***2D Materials* Roadmap**

# The 2025 2D Materials Roadmap

**Wencai Ren[1*], Peter Bøggild[2*], Joan Redwing[3*], Kostya Novoselov[4,5], Luzhao Sun[6], Yue Qi[6], Kaicheng Jia[6], Zhongfan Liu[6], Oliver Burton[7], Jack Alexander-Webber[7], Stephan Hofmann[7], Yang Cao[8], Yu Long[9], Quan-Hong Yang[9], Dan Li[8, 10], Soo Ho Choi[11,12], Ki Kang Kim[11,12], Young Hee Lee[11,12], Mian Li[13], Qing Huang[13], Yury Gogotsi[14], Nicholas Clark[4], Amy Carl4, Roman Gorbachev[4], Thomas Olsen[2], Johanna Rosen[15], Kristian Sommer Thygesen[2], Dmitri K. Efetov[16], Bjarke S. Jessen[2], Matthew Yankowitz[17], Julien Barrier[18], Roshan Krishna Kumar[18], Frank HL Koppens[18], Hui Deng[19], Xiaoqin Li[20], Siyuan Dai[21], D.N. Basov[22], Xinran Wang[25,31], Saptarshi Das[3], Xiangfeng Duan[24], Zhihao Yu[25, 26], Markus Borsch[27], Andrea C. Ferrari[7], Rupert Huber[28], Mackillo Kira[27], Fengnian Xia[29], Xiao Wang[25,30,31], Zhong-Shuai Wu[23], Xinliang Feng[32,33], Patrice Simon[34,35], Hui-Ming Cheng[1,36,37], Bilu Liu[38], Yi Xie[39], Wanqin Jin[40], Rahul Raveendran Nair[8], Yan Xu[41], Qing Zhang[1], Ajit K. Katiyar[42], Jong-Hyun Ahn[42], Igor Aharonovich[43], Mark C. Hersam[44], Stephan Roche[45], Qilin Hua[46], Guozhen Shen[46], Tianling Ren[38], Hao-Bin Zhang[47], Chong Min Koo[12], Nikhil Koratkar[48], Vittorio Pellegrini[49], Robert J Young[4], Bill Qu[50], Max Lemme[51] and Andrew J. Pollard[52]**

[1] Shenyang National Laboratory for Materials Science, Institute of Metal Research, Chinese Academy of Sciences, 72 Wenhua Road, Shenyang 110016, P. R. China.
[2] Technical University of Denmark, Denmark
[3] The Pennsylvania State University, USA
[4] University of Manchester, UK
[5] Institute for Functional Intelligent Materials, National University of Singapore, Singapore
[6] Beijing Graphene Institute, China
[7] University of Cambridge, UK
[8] Department of Chemical Engineering, The University of Melbourne, Victoria, Australia
[9] Nanoyang Group, Tianjin Key Laboratory of Advanced Carbon and Electrochemical Energy Storage, School of Chemical Engineering and Technology, Tianjin University, Tianjin 300072, China
[10] The Hong Kong University of Science and Technology, Hong Kong, China
[11] Center for Integrated Nanostructure Physics, Institute for Basic Science, Suwon 16419, Republic of Korea
[12] Sungkyunkwan University, Suwon 16419, Republic of Korea
[13] Zhejiang Key Laboratory of Data-Driven High-Safety Energy Materials and Applications, Ningbo Institute of Materials Technology and Engineering, Chinese Academy of Sciences
[14] Drexel University, USA
[15] Linköping University, Sweden
[16] Ludwig-Maximilians-Universität München, Germany
[17] University of Washington, USA
[18] ICFO – The Institute of Photonic Sciences, Castelldefels 08860, Barcelona, Spain
[19] University of Michigan, USA
[20] University of Texas at Austin, USA
[21] Auburn University, USA
[22] Columbia University, USA
[23] State Key Laboratory of Catalysis, Dalian Institute of Chemical Physics, Chinese Academy of Sciences, Dalian, China
[24] University of California, Los Angeles, USA
[25] Suzhou Laboratory, Suzhou, China
[26] School of Integrated Circuit Science and Engineering, Nanjing University of Posts and Telecommunications; Nanjing 210023, China
[27] Department of Electrical Engineering and Computer Science, University of Michigan, Ann Arbor, MI, USA
[28] University of Regensburg, Germany
[29] Department of Electrical and Computer Engineering, Yale University, New Haven, CT, USA
[30] School of Integrated Circuits, Nanjing University, Suzhou, China

[31] Interdisciplinary Research Center for Future Intelligent Chips (Chip-X), Nanjing University, Suzhou, China
[32] Center for Advancing Electronics Dresden (cfaed), Faculty of Chemistry and Food Chemistry, echnische Universität Dresden, Dresden 01062, Germany
[33] Max Planck Institute of Microstructure Physics, Halle (Saale) 06120, Germany
[34] CIRIMAT, UMR CNRS 5085, UniversitéPaul Sabatier Toulouse III, Toulouse 31062, France
[35] RS2E, Réseau Français sur le Stockage Electrochimique de l'Energie,FR CNRS 3459, Amiens Cedex 80039, France
[36] Shenzhen Key Laboratory of Energy Materials for Carbon Neutrality, Shenzhen Institute of Advanced Technology, Chinese Academy of Sciences, Shenzhen, China
[37] Faculty of Materials Science and Energy Engineering, Shenzhen University of Advanced Technology, Shenzhen, China
[38] Tsinghua University, P. R. China
[39] University of Science and Technology of China, P. R. China
[40] Nanjing Technical University, China
[41] Huawei Technologies Co., Ltd
[42] School of Electrical and Electronic Engineering, Yonsei University, Seoul 03722, Republic of Korea
[43] University of Technology Sydney, Australia
[44] Northwestern University, USA
[45] ICREA and Catalan Institute of Nanoscience and Nanotechnology, Spain
[46] School of Integrated Circuits and Electronics, Beijing Institute of Technology, China
[47] State Key Laboratory of Organic-Inorganic Composites, Beijing University of Chemical Technology, Beijing 100029, China
[48] Rensselaer Polytechnic Institute, USA
[49] BeDimensional, Italy
[50] The Sixth Element Inc., China
[51] RWTH Aachen University & AMO GmbH, Germany
[52] National Physical Laboratory, UK

*Guest Editors of the Roadmap:
Emails: Wencai Ren wcren@imr.ac.cn; Peter Bøggild pbog@dtu.dk and Joan Marie Redwing jmr31@psu.edu

## Abstract

Over the past two decades, 2D materials have rapidly evolved into a diverse and expanding family of material platforms. Many members of this materials class have demonstrated their potential to deliver transformative impact on fundamental research and technological applications across different fields. In this roadmap, we provide an overview of the key aspects of 2D material research and development, spanning synthesis, properties and commercial applications. We specifically present roadmaps for high impact 2D materials, including graphene and its derivatives, transition metal dichalcogenides, MXenes as well as their heterostructures and moiré systems. The discussions are organized into thematic sections covering emerging research areas (e.g., twisted electronics, moiré nano-optoelectronics, polaritronics, quantum photonics, and neuromorphic computing), breakthrough applications in key technologies (e.g., 2D transistors, energy storage, electrocatalysis, filtration and separation, thermal management, flexible electronics, sensing, electromagnetic interference shielding, and composites) and other important topics (computational discovery of novel materials, commercialization and standardization). This roadmap focuses on the current research landscape, future challenges and scientific and technological advances required to address, with the intent to provide useful references for promoting the development of 2D materials.

# Contents

# Introduction

**Wencai Ren[1], Peter Bøggild[2] and Joan M. Redwing[3]**

[1] Shenyang National Laboratory for Materials Science, Institute of Metal Research, Chinese Academy of Sciences, China
[2] Technical University of Denmark, Denmark
[3] Department of Materials Science and Engineering, The Pennsylvania State University, USA

The advent of 2D materials has revolutionized condensed matter physics and materials science, offering unprecedented opportunities to explore exotic physical phenomena, engineer novel functionalities, and address critical technological challenges across diverse fields. Over the past two decades, the exploration of 2D materials has expanded beyond graphene, encompassing a vast library of atomically thin crystals and their heterostructures. These materials exhibit extraordinary electronic, optical, thermal, mechanical, and chemical properties, and hold promise for breakthroughs in electronics, optoelectronics, quantum technologies, energy storage, catalysis, thermal management, filtration and separation, and beyond. Many exciting new physics and phenomena continue to emerge, while select 2D materials, such as graphene, h-BN, and the semiconducting transition metal dichalcogenides (TMDCs), are transitioning from laboratory-scale demonstrations to industrial applications. In this context, a holistic understanding of synthesis, structure-property relationships, integration, and performance optimization is essential. This roadmap reviews the multifaceted challenges and opportunities in 2D materials research, focusing on the synthesis, properties and applications of representative systems including graphene and its derivatives, TMDCs, MXenes as well as their heterostructures and moiré systems.

**Scalable and reliable synthesis**

As research on many 2D materials progress beyond the "laboratory demonstration" phase, the consistent and scalable production of high-quality materials has become a critical bottleneck. This challenge is equally relevant to fundamental research, where precise control over structure, doping, strain, and contamination is required, and commercial applications, which demand large-scale production while maintaining material integrity. To address this challenge, we see two pathways. The first is the robotic automation of mechanical exfoliation and stacking to reproducibly produce microscopic single-crystal flakes. This versatile approach offers an unprecedented ability to rationally design and fabricate 2D material assemblies with customized configurations. Provided that transfer-induced inhomogeneities can be minimized, this approach would provide sufficient throughput for fundamental research and some customized high-end devices.

The second approach is wafer-scale synthesis of 2D single-crystal films (e.g. graphene, TMDCs and Moiré materials) or large-scale cost-effective production of bulk micro-flakes (e.g. graphene oxides and MXenes), representing a mainstream solution to the industrial production of materials that inherit the superlative properties demonstrated with microscale flakes. Technological progress requires not only the precise growth and transfer of large-area high-quality crystals, but also the development of complete pathways to heterogeneous integration, quality control and cost reduction at every process stage. Another perspective is to combine the two approaches towards automated wafer- or even roll-to-roll based lamination systems, creating large-scale heterostructures on demand.

New directions in synthesis research are emerging to refine growth control and deepen our scientific understanding. Site selective synthesis of microscale 2D crystals via direct growth on patterned substrates is a growing topic of interest as a pathway to avoid complications of layer transfer. To transition from empirical, trial-and-error methods to scalable and application-specific synthesis, a shift toward data driven strategies is underway. Artificial intelligence (AI)-assisted simulations of growth dynamics in combination with operando characterizations and real-time process control will substantially improve the yield, consistency and repeatability.

**Discovery of novel 2D materials**

In addition to the controllable synthesis of established materials, the discovery of novel 2D materials is an exciting frontier. Density functional theory (DFT) and first-principles calculations have become powerful tools for predicting the stability, properties, and potential applications of hypothetical 2D materials — especially those without bulk counterparts. AI-driven approaches will undoubtedly play a crucial role in this field by integrating machine learning algorithms with high throughput computational screening. However significant challenges persist. Realistic assessment of synthesizability and stability, accurate modelling of moiré materials and reliable predictions of electronic/magnetic properties remain open questions. This necessitates further and close collaboration between computational and experimental scientists.

**Van der Waals and 3D assemblies**

Moiré materials are at the forefront of research in strongly correlated electron systems, including unconventional superconductivity and topological phases, offering immense potential for applications in quantum electronics, photonics, and polaritonics. However, many challenges persist in understanding and manipulating these exotic quantum states, including establishing direct correlation between the atomic-scale properties and emergent mesoscopic phenomena, developing robust sources and detectors for non-classical states of light, new theoretical frameworks for modeling strongly correlated and topological states, and advancing computational methods for predicting multi-electron interactions.

Similarly, the fabrication of 3D assemblies significantly expands the application of 2D materials in areas such as energy storage and thermal management by enhancing their functionality and compatibility. However, a lack of comprehensive structure-property mappings hampers precise engineering for tailored properties and the realization of new functionalities. The complex hierarchical structures and dynamic operational behaviours of 3D assemblies necessitate multimodal characterization techniques to decode their intricate architecture and correlate them with macroscopic properties. To advance this field towards consistent and scalable production of high-quality materials, it is necessary to develop new techniques for controlling their properties as well as advancing theoretical modeling capabilities for understanding and predicting new properties in emerging materials and systems. The integration of AI with advanced characterization and modelling tools will be pivotal for addressing challenges in microstructural recognition and property prediction.

**Real-world applications**

2D materials have demonstrated their potential to deliver transformative impact on diverse applications including, but not limited to, electronics, optoelectronics, energy storage, catalysis, filtration and separation, thermal management, flexible electronics, sensing, electromagnetic interference shielding, and composites. A key consensus within the 2D materials community is that scalable production of 2D materials with consistent properties is essential for realizing real-world applications. Furthermore, field-specific challenges must be addressed to bridge the gap between laboratory-scale demonstrations and industrial implementation. In electronics, the integration of 2D materials into 3D device architectures has emerged as a groundbreaking approach to overcome the limitations of traditional silicon-based technologies. This enables the development of task-specific, energy-efficient, and versatile electronic systems, making it a promising candidate for next generation device components for storage and computation. To realize the vision of industrial-scale adoption, research and development efforts must prioritize scalable and reliable fabrication processes, which require holistic solutions across material, process, and architecture dimensions. For interdisciplinary fields such as flexible electronics, neuromorphic computing, quantum technologies, spintronics, and sensing, advancing material property engineering, improving mechanistic understanding, scaling up high-performance device fabrication, and integrating with mainstream architectures are essential for commercialization. For energy storage, electrocatalysis, filtration and separation, thermal management, electromagnetic interference shielding, and composites, future efforts should be devoted to improving performances at relatively low cost to compete with existing technological solutions. Application-driven design and long-term stability should be prioritized in optimization efforts, which require precise structural control of both the materials and their assemblies. This structural and process optimization should be based on in-depth mechanistic insights, which will benefit from the combination of atomic-scale characterizations, real-time simulations and progress in theory and modeling. The development of hybrid systems and novel 2D materials may offer pathways to resolving fundamental trade-offs in performance, scalability and selectivity while operating under realistic conditions. In this process, identifying critical and irreplaceable applications for 2D materials remains a high-priority task for researchers in academia and industry alike. In addition to addressing the previously mentioned technical challenges with an increased focus on reproducibility and consistency, a coherent, hierarchical structure of international standards is needed in the future to finally achieve commercialization of 2D materials.

This 2025 roadmap underscores a shift in 2D materials research: from individual material, phenomenon and process exploration to holistic, application-driven development strategies. By leveraging breakthroughs in synthesis, theory, and system integration, the community stands poised to unlock the full potential of 2D materials paving the way for an era of advanced and sustainable technologies. The journey ahead demands interdisciplinary collaboration and relentless effort on translating scientific discoveries into commercial impact.

## 2. Graphene

**K. S. Novoselov[1,2]**

[1]National Graphene Institute, University of Manchester, UK
[2]Institute for Functional Intelligent Materials, National University of Singapore, Singapore

Graphene[1] is probably one of the simplest possible materials – monolayer of carbon atoms (and carbon is one of the lightest atoms) arranged in a honeycomb lattice, **Figure 1**. And yet, despite its simplicity, it is one of the most popular materials to study. One reason is the very high quality of the crystals widely available – carbon-carbon bonds are very strong, so it is relatively easy to obtain samples without any defects. Electron mobility in modern graphene devices can be as high as tens of millions $cm^2/V{\cdot}s$ with the mean free path of a few tens of microns at low temperatures. Another – is the linear electronic spectrum. Apart of the zero bandgap, which ensures possible applications in low-energy photonics (THz and far infra-red), such linear dispersion relation provides chiral properties to the quasiparticles.

### The basics

The linear dispersion relation is not unique to graphene, but graphene is the first material where the linear dispersion was demonstrated to determine the majority of the electronic and optical properties. Thus, for the large part of the spectrum (ranging from THz, practically to ultra-violet, where the deviation from the linear dispersion is not too large), the optical adsorption (and the optical conductivity) is given simply by the combination of the fundamental constants: πα=2.3% (here $\alpha = e^2/4\pi\varepsilon_0\hbar c$ is the fine structure constant)[2]. More importantly, the linear dispersion relation implies the chiral properties of electrons in graphene – the specific relation between the electron's momentum and its pseudospin (the phase between the components of the electron wavefunction which reside at different sublattices). For electrons in graphene their pseudospin is always either collinear or anti-collinear with the momentum (depending on the valley), and it has more complex relations for bilayer graphene. This has an immediate effect on the scattering properties of electrons in graphene, as any changes of the electrons momentum would imply a change in the electrons pseudospin (sublattice composition), which would require very special symmetry of the scattering potential. This partly explains the Klein paradox (electron penetration through any rectangular barrier is always 100%) and, as a consequence, the high mobility of the charge carriers. Other consequences of the linear dispersion relation are the half-integer quantum Hall effect in graphene[3,4] and the chiral quantum Hall effect in bilayer graphene.

**The chemistry**

Honeycomb lattice of graphene implies that each of the carbon atoms has three neighbours. Thus, out of its four valence electrons, three are involved in forming the σ-bonds, leaving one electron in the delocalised π-states, which determines the electronic and optical properties of this material. However, the same electron can be utilised for the formation of yet another chemical link, thus creating new chemical compounds. Historically, the first chemical derivative of graphene was graphane[5], when one hydrogen atom is attached to each of the carbon atom (the most stable configuration being when the two sublattices of graphene are hydrogenated from two different sides), **Figure 1**. Unlike graphene – graphane is an insulator with a few eV bandgap. Interestingly, the process is completely reversible – hydrogen can be removed (for instance, by annealing), bringing the material back to the pristine, zero-bandgap state of graphene. Later, other chemical derivatives have been demonstrated, including fluorographene (the two-dimensional analogue of Teflon, where one fluorine atom is attached to each carbon).

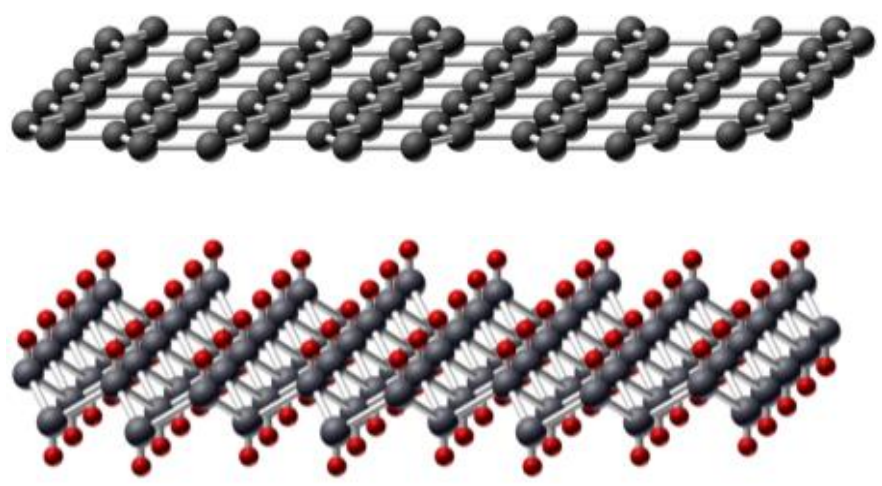

**Figure 1.** Atomic structure of graphene (top) and graphane (bottom). Grey spheres – carbon atoms; red spheres – hydrogen.

One of the popular derivatives, which also offers a viable production route for graphene, is graphene oxide. Obtained via exfoliation of oxidised graphite, this material is hydrophilic (unlike graphene, which is hydrophobic, but lipophilic) and thus can form suspensions in water, which significantly simplifies its processing. Graphene oxide is a rather disordered material, with a variety of hydroxyl, epoxy and carboxyl groups. The typical carbon-to-oxygen ratio is around 1.5-2.5, and can be varied to balance the conductivity and the hydrophilicity. The hydrogen bonds allow the formation of stable laminated multilayer structures, known as graphene oxide paper[6] (or graphene paper, if a succeeding reduction has been used). Such structures are very efficient for a variety of water and ion filtration applications.

The possibility of functionalisation of graphene offers multiple opportunities for the use of this 2D crystal for a variety of sensing applications. Graphene can be functionalised with various receptors, which can selectively bind certain chemicals. The process of binding is usually accompanied with the changes in the electrical environment, which influences the electronic transport in graphene (effectively the layer of molecules on the surface act as a floating gate on the graphene channel). Since the distance between graphene and the absorbed molecules is very small, the sensitivity achieved can be significantly higher than that for similar structures based on silicon.

**The use**

Clearly, graphene applications are not limited only to the chemical sensors. One niche, though very important, application is the use of graphene for quantum metrology as a resistance standard. For years, III-V GaAs/AlGaAs heterostructures have been used for the measurements of the von Klitzing constant in the quantum Hall regime. Such structures, however, require milliKelvin temperatures and significant magnetic fields. Graphene, grown on silicon carbide, on the other hand, demonstrates pinning of the zero Landau level, due to charge exchange with the substrate, which, in conjunction with very large cyclotron gap for massless quasiparticles, allows observation of a very high-quality quantised resistance plateau at helium temperatures and at very modest magnetic fields.

Apart from this, graphene has been following the very usual pathway of any new material in terms of applications. Applications in composites to improve strength and toughness and reduce weight become very popular. Interestingly, graphene does the same to concrete: the addition of a fraction of a percent of graphene to concrete improves its strength by up to 100% (the numbers in the literature vary from 20% to 100%). This result is extremely important for our sustainability goals, as concrete industry is one of the largest $CO_2$ emitters. Functional composites are being utilised as well, where graphene improves their conductivity or fire retardancy. Large amounts of graphene is now used for batteries applications as a conductive material for electrodes as well as a host for nanosilicon particles.

Graphene ink is used for printable electronics. Graphene's high thermal conductivity is now utilised for thermal management in electronic devices (in this case reduce graphene oxide paper is usually used). A number of optoelectronic applications are in the developing stage, where graphene is being adopted as an optical modulator in silicon photonic circuits. In terms of electronic applications, graphene is being tried for a variety of roles as well, from the active channel to interconnects. The recent progress in chemical vapour deposition of graphene[7] makes such applications more and more realistic.

**The stack**

The isolation of graphene confirmed the very possibility of the existence of single atom thin two-dimensional materials. This gave rise to many other 2D crystals to be isolated and studied[8]. Today we are talking about several hundred of those investigated experimentally and many more being predicted theoretically. Having access to such 2D crystals means that one can stack them into heterostructures, by placing one crystal on top of another. Interestingly, very high-quality interfaces, practically free of contaminants, could be achieved this way, so the crystals then occur in the very close proximity to each other and mainly interact via van der Waals forces. Sandwiching graphene with crystals that contain heavy atoms (for instance tungsten, as in $WS_2$) has been demonstrated to promote spin-orbit interaction in graphene. High-quality hexagonal boron nitride (hBN) is being used as a protective layer to screen graphene from the environment, as well as a very thin, high-quality dielectric for applying the gate. As the lattice constants of graphene and hexagonal boron nitride are very close to each other (2.460 Å for graphene and 2.504 Å for hBN), if aligned, these crystals form a moiré structure (corresponding to the periodic variations of the atomic stacking between the two crystals) with a period of up to 14nm (and even larger if hBN is allowed to be compressed, as, for instance, happens in the case of monolayer hBN). The moiré structure acts as an additional scattering potential for electrons, which results in strong spectrum reconstruction at the corresponding wavevectors. In fact, the reconstruction results in the formation of the replicas of the linear Dirac spectrum with locally linear dispersion relations as well. Furthermore, as the period of the moiré is large enough so that one can fit a flux quantum in experimentally achievable magnetic fields – it opened up the possibility of studying the Hofstadter physics in such devices[9-11].

**The twist**

A very special type of heterostructures with peculiar and highly unusual properties has been obtained by simply stacking two graphene layers under a small angle. The interplay between the Bragg scattering in the moiré potential and the van der Waals interaction between the layers leads to the formation of the flat electronic bands[12] (the specific angle at which it happens has been dubbed the magic angle and is about 1.1°). These flat bands trigger a number of electron-electron correlation effects, including superconductivity, the Mott insulator and orbital ferromagnetism.

**The future**

It is amazing that even after twenty years of intensive research, graphene still brings us surprises. At the same time, it is already becoming one of the platform materials, similar to silicon, being used in many areas of scientific research due to its unique properties: as a transparent electrode; as a substrate in transmission electron microscopy; as nanoreactor, *etc*. It is impossible to forecast where the next breakthrough in graphene research will happen – in the study of its electronic, chemical, or mechanical properties – but one thing it is possible to predict with high accuracy: it *will* happen. A similar situation exists for the applications of this material. It looks like we have lived the full cycle: from the denial (graphene is too difficult to produce, so it is unlikely that it will be used in applications) – to euphory (in the future everything will be made from graphene) – to disappointment (why we don't see graphene everywhere). In fact, the penetration of graphene into our life is gradual but steady, and we expect it to continue at the ever-increasing pace for the years to come.

**The read**

# 3. Graphene synthesis: a route driven by industrialization and markets demands)

**Luzhao Sun[1], Yue Qi[1], Kaicheng Jia[1], Zhongfan Liu[1,2,*], Oliver Burton[3], Jack Alexander-Webber[3], Stephan Hofmann[3,*]**

[1] Beijing Graphene Institute, China
[2] Peking University, China
[3]University of Cambridge, UK

Email: * zfliu@pku.edu.cn , sh315@cam.ac.uk

## 1. Introduction

Graphene has spearheaded research into 2D materials, accelerating the development of various material systems in its wake. The term 'graphene and related materials' has since evolved into a broad label that encompasses a wide range of materials. However, grouping these distinct materials together can obscure important factors. Even the simple concept of the 'cost of graphene' is often expressed in USD per ton to match bulk applications like inks and composites. This can overshadow the unique challenges and opportunities specific to high-quality graphene films and wafers. Starting from a brief review of the progress on the synthesis of continuous films, this section discusses the challenges ahead for continuous graphene to fulfill its potential as a core material in future technology.

In 2007, small, exfoliated graphene domains (~1000 $\mu m^2$) had become commercially available. The development of chemical vapor deposition (CVD) graphene and silicon carbide (SiC) growth methods soon enabled larger films, with today's global providers offering large-area CVD graphene films at costs <1 $EUR/cm^2$, a factor of $10^8$ lower than the exfoliated, microscopic flakes initially traded on a one-by-one basis. The production of graphene films has advanced to the foundry pilot line stage [1], with industrially compatible infrastructure for manufacturing of basic opto-electronic device architectures at technology readiness level (TRL) approaching 8. Technological progress thereby requires not only the capability to grow large, high-quality crystals but development of complete pathways to heterogeneous integration, and adequate characterisation and quality control at every process stage.

We here review the progress in the view of holistic graphene integration pathways beyond individual process steps, as integration represents an Achilles heel of all current 2D material-based roadmaps and translation of their widely hailed unique properties to scalable technology. We structure our reflections accordingly, starting from the currently dominant “general purpose” approach of CVD growth on Cu support and subsequent graphene transfer, to highlight a range of developments and discoveries that can serve the needs of future application-specific integration. Addressing yield and reproducibility thereby arguably requires a deeper scientific understanding and overcoming the “black-box” that such processing often presents by new characterisation approaches.

## 2. Current and future challenges

### 2.1 General-purpose graphene films

The strong in-plane bonding that gives graphene its excellent mechanical properties and chemical stability comes at the cost of requiring high energy input during synthesis to achieve high crystallinity. This motivates the use of transition metal growth substrates, i.e. a catalytic CVD approach [2]. Cu was found to not only present a broad process window for monolayer film CVD [3], but subsequently also enable effective graphene release, helped by interfacial decoupling or its dissolution in etching solution. Subsequent graphene transfer allows general-purpose utilization compatible with a wide range of substrates, from silicon to plastics. Cu-based CVD has become the most widespread approach, driving a large fraction of the current graphene commercialisation, be it batch-to-batch (B2B) or roll-to-roll (R2R) manufacturing processes. While much of the early work utilised polycrystalline Cu foils, substrate design has progressed to epitaxial metallisation and surface alloying, such as single-crystal Cu(111) and CuNi(111) on sapphire wafers, in order to promote planarity and lateral graphene domain orientation control for their effective merging to a single-crystal film.

Compared with the growth step itself, the subsequent transfer and handling of large, continuous atomically thin films remains a critical bottleneck. While progress has been made in polymer handling layers [4, 5], it still faces challenges of cracking and wrinkling and strain, especially when scaling up. The polymer and liquid solvents inevitably lead to contamination effects, compromising e.g. achievable graphene channel mobilities, and complicating further processing and functionalisation. The decoupling relies on diffusion of oxygen or water species at the interface between the graphene and the growth substrate, which significantly slows down the transfer process for high-quality, defect-free graphene. This makes dry transfer of industrially relevant graphene particularly challenging. This epitomises the need for combined process optimisation [6], which comes with a vast parameter space, where many important parameters remain hidden. Despite progress towards streamlining and processes and robotic automation, successful transfer generally depends on skilled individuals. The economics and sustainability of transfer-based routes is also influenced by the reusability or recovery of the growth substrate and the transfer handle.

### 2.2 Application specific routes

While general-purpose metal-catalysed CVD and wet transfer techniques have introduced graphene films to commercial markets, numerous integration challenges remain unresolved. This situation has driven alternative approaches. CVD on Ge, sapphire, SiC or $SiO_2$ substrates offers metal-free processing. However, such approaches require higher temperatures and/or result in poorer crystallinity of the graphene due to the reduced catalytic activity of the substrate. To meet these challenges, numerous efforts have been made such as confined space CVD and co-catalysis CVD. For semiconductor/dielectric substrates the complexity and importance of surface preparation typically increases, as studies on sapphire surface orientation/termination and miscuts have shown. On select stepped sapphire or e.g. Ge(110) surfaces graphene domains have been shown to align in select orientation, i.e. highly crystalline graphene can be achieved by merging of such domains.

Processing without transfer necessitates incorporation of the substrate into the device design and demands precise interfacial control, due to the high sensitivity of the exposed single-layer films to numerous external influences and imperfections. The many discoveries that contributed to the successful development of graphene growth directly on SiC via surface thermal decomposition over the last decades highlight the importance of such interface control and engineering [7]. Motivated by CMOS back-end integration, low temperature processing enabled by e.g. plasma or hot filament enhanced CVD or ALD type processes has been explored. This may lead to a significant reduction of graphene crystallinity and layer control. Localised heating is another strategy for specific applications, such as sensor platforms.

Graphene CVD synthesis generally affects the substrate and its surface, i.e. process steps should not be thought as being just additive. Rather, system design is required, and holistic approaches need to be tailored and matched to application specific requirements. The ‘Catch 22’ is that more specific development tends to spring from more specific application pull. While for general purpose growth the graphene quality is typically seen in “absolute” terms, *e.g.*, single crystallinity/defect density, for application specific growth it is the functionality, robustness and consistency achieved with graphene in the concrete application scenario that should be taken as a measure. Transfer-free integration with direct graphene CVD on metals that form part of the device structure, such as on ferromagnets for spintronics, ferroelectrics in memristors or as diffusion barrier on contacts, but also graphene-skinned materials that cultivating on engineering materials with various morphologies, such as glass fibres, powders, 3D metal foams, and zeolites [8], present diverse examples where CVD process integration pathways have started to be explored.

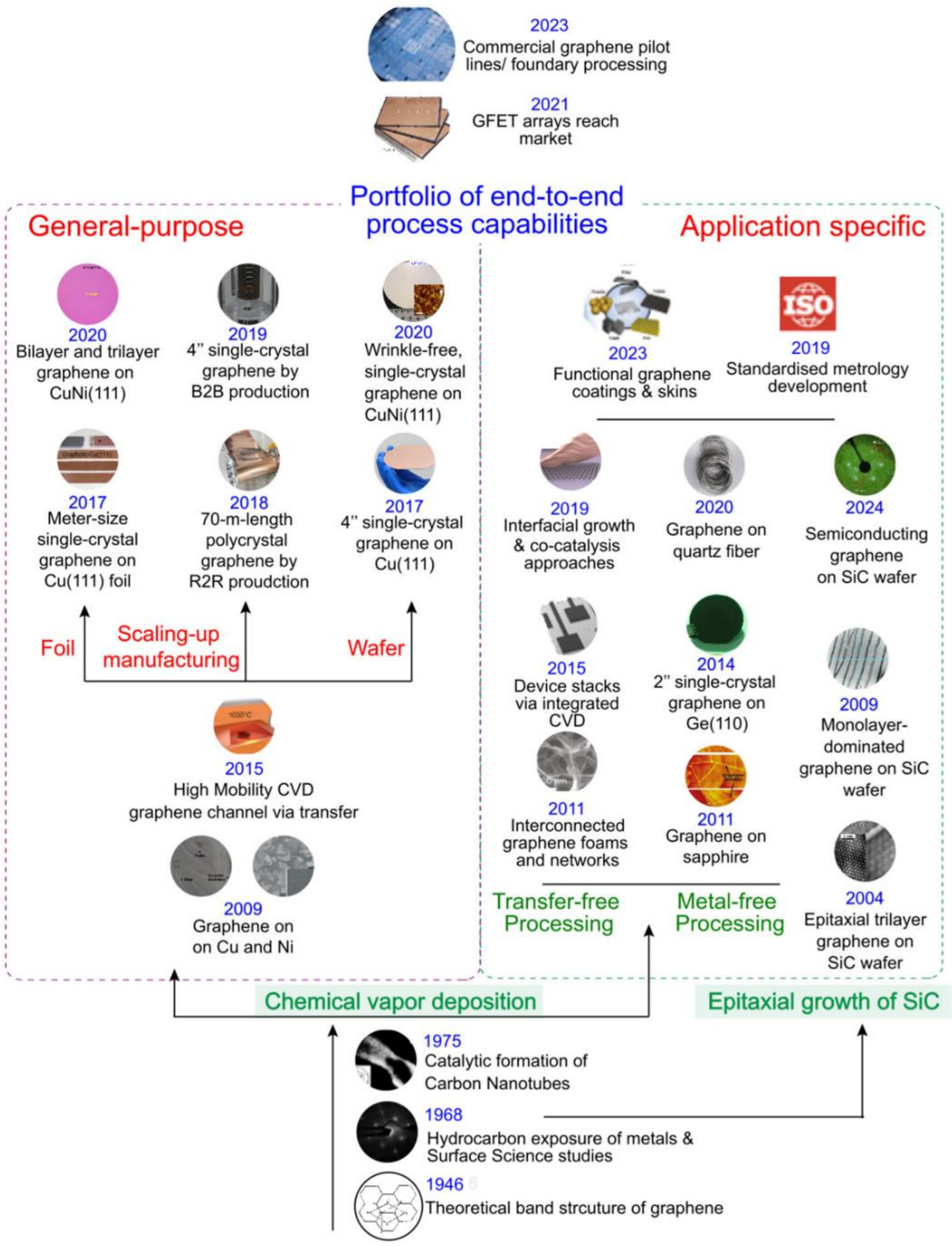

Figure 1 Development of graphene CVD processes & technology

### 3. Advances in Science and Technology to Meet Challenges

***Industrial production*** of graphene materials is the prerequisite for the commercialization of graphene products and motivated scientific research in this field. Improving quality control and reducing production costs are long-term critical concerns in the materials industry. Currently, the expense associated with producing continuous graphene films and wafers exceeds market demand, limiting their industrial applications. The field is still far from being able to scale-up all the superlative properties demonstrated at small flake level for individual “hero” devices, using complex, often manual assembly strategies. Increasing automation of R2R or B2B will be crucial for production with high reproducibility [9].

Arguably, graphene is the major model system for process development at atomically thin level. The vast parameter space, intrinsic and extrinsic disorder and distinct reaction kinetics need more scientific underpinning. The story of synthetic graphene clearly highlights that the empirical, trial-and-error development is inadequate in terms of reaching stable, cost-effective application scenarios. Today, growth control is mainly limited to monolayers. Even for graphene bilayers, controlling stacking order and microstructure remains highly challenging. This underscores the need for ***new approaches*** to exploit rather than combat strong anisotropy. Technological maturing requires standardisation that also feeds into health and safety protocols. This reflects a need for scalable screening methods not only for the general-purpose route but supporting each tailored process development step.

Different applications also require varying graphene properties. To proceed towards commercially viable graphene applications, synthesis needs to be seen in a holistic perspective as part of an increasing portfolio of process capabilities at monolayer level, including film handling and transfer, which need to tie in with diverse existing manufacturing pathways across targeted application areas. While, for instance, large-area, transparent conductive films may not demand single-crystal graphene, they still require high-quality, homogenous films; electronic applications require significantly more uniform and clean graphene, as even minor inconsistencies can lead to issues with performance and variability. Modelling of yield, ***driven by end-application performance***, will guide the optimization of graphene production processes, ensuring the highest possible device yields within specific performance thresholds. This approach will allow graphene manufacturers to ***customize production*** for different industries, enhancing both yield and performance across diverse applications. We also anticipate that graphene will more frequently serve as a key supporting component in an evolving portfolio of materials and applications, rather than being the primary focus or centerpiece material.

Many graphene growth processes are optimised and communicated without much concern regarding subsequent integration and performances, which limits the ability to optimize yield and reproducibility of end devices at an industrial scale. To address this, the field is shifting towards ***data-driven synthesis and assessment***, where automation and improved process control will make a substantial improvement in repeatability and application specific process understanding. Incorporating real-time, *in operando* methods [10], and artificial intelligence (AI) will allow for continuous feedback, analysis, and adjustment, enabling increasingly automated optimisation through statistical methodologies already prevalent in main-stream manufacturing. Such a transition is crucial for aligning the quality of graphene with the performance demands of a wide range of industries, moving from trial-and-error methods to a more systematic, scientific approach.

Improving the environmental footprint often remains an after-thought after a material/application reached mass market. ***Sustainable product life cycle design*** has driven the exploration of carbon source alternatives to highly purified gases, from waste gases to organic waste. There are many carbon waste streams and graphene growth can offer new routes to up-cycling. Therefore, eco-friendly process is another aspect that should be judged by much more holistic metrics.

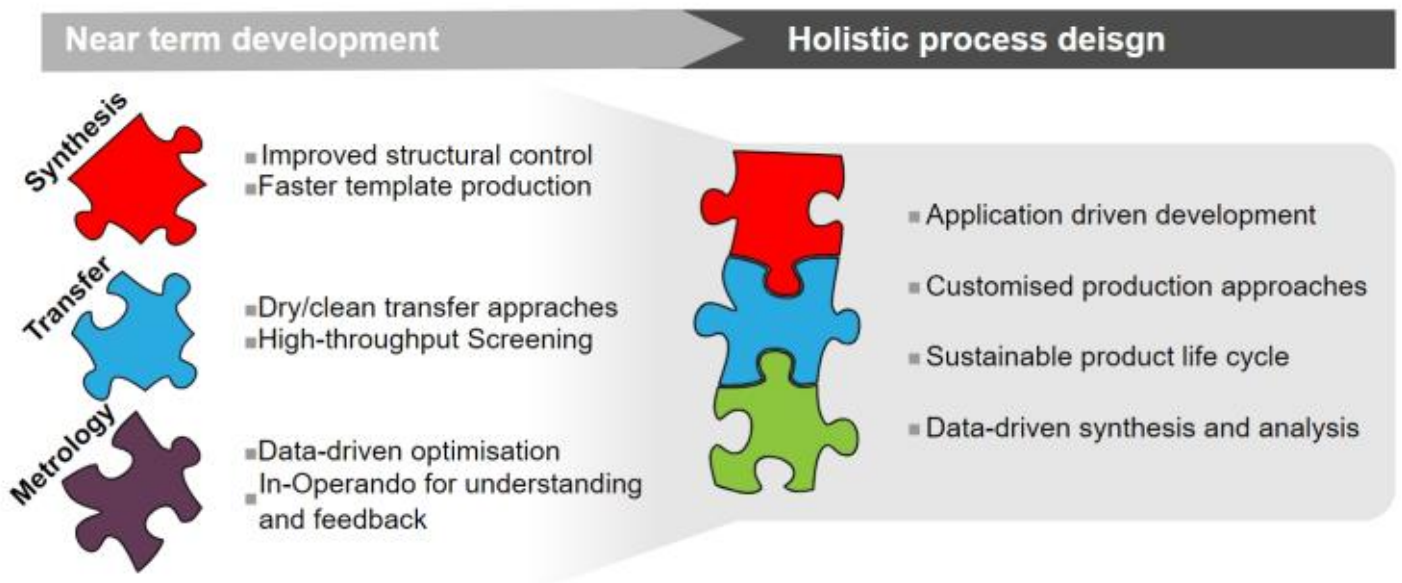

Figure 2 Schematic of the technology roadmap for graphene industry in synthesis and transfer.

## 4. Concluding Remarks

At this 20-year anniversary, the celebrations often focus on the fundamental science discoveries in properties and functionalities. If the 2D material research and innovation communities are serious about impactful commercialization, much more appreciation and attention should be given to the science of crystal growth and scale-up, keeping integration and the needs of the end-application in mind. Automation, high-throughput and *operando* characterization approaches now allow us to go beyond 'black-box' reactor and process calibrations, allowing us to make a large leap from 'black magic' to systematic, data-driven science.

**Acknowledgements**

This work was financially supported by National Natural Science Foundation of China (NSFC, No. T2188101), EPSRC (EP/K016636/1, EP/P005152/1, EP/T001038/1) and the European Union's Horizon 2020 research and innovation program (Grant Agreement No number 785219). J.A.-W. acknowledges support of his Royal Society Dorothy Hodgkin Research Fellowship. O.J.B. acknowledges the support of the Oppenheimer research fellowship.

## 4. Graphene derivatives

**Yang Cao[1], Yu Long[2], Quan-Hong Yang[2,*] and Dan Li[1,3,*]**

[1]Department of Chemical Engineering, The University of Melbourne, Victoria, Australia
[2]Nanoyang Group, Tianjin Key Laboratory of Advanced Carbon and Electrochemical Energy Storage, School of Chemical Engineering and Technology, Tianjin University, Tianjin 300072, China
[3]Department of Chemical and Biological Engineering, The Hong Kong University of Science and Technology, Hong Kong, China.

Email*: qhyangcn@tju.edu.cn and dan.li@ust.hk

**Status**

Graphene derivatives, chemically modified or structurally altered forms of graphenes, are designed to enhance or diversify the already remarkable properties of graphenes for various applications. Their exploration dates back to 1859, with the discovery of graphite oxide, long before pristine graphene was isolated in 2004 by Andre Geim and Konstantin Novoselov. These derivatives are especially valuable due to their chemical versatility and ease of production. For instance, graphene oxide (GO) is notable for its dispersibility in water and potential for chemical modifications, making it ideal for use in energy storage, flexible electronics, and biomedical applications. Reduced graphene oxide (rGO), which is produced by deoxygenation of GO, restores part of the $sp^2$ carbon network, providing a balance between conductivity and functionality for electronic devices and sensors.

As graphene derivatives continue to evolve, researchers are exploring new structural forms that enhance their functionality and open new application areas. One important advancement in this field is the development of 3D graphene derivatives with the graphene nanosheets, usually are GO and rGO, being arranged into 3D forms, such as fibres, membranes and aerogels. The added structural variety introduces new features to these 3D graphene assemblies. For example, the nanoporous network generated during assembly, particularly the nanochannels formed between two adjacent layers, creates a transport pathway for ions and molecules, thus enabling the applications such as ion/molecules sieving and separation.[1] Moreover, their flexible carbon skeleton allows rapid electron transfer and electromechanical responses, making them highly beneficial for the development of wearable electronics, where flexibility, lightweight, and quick response times are crucial.[2]

The ease of processing and diversity in properties make graphene derivatives integral to many large-scale commercialised products, including composites, anticorrosion coatings, separation processes, and energy storage. Over the next decade, various commercial graphene products are expected to emerge, underscoring the critical importance of ongoing research in graphene derivatives.

**Current and Future Challenges**

Over the past years, a range of graphene derivatives has been developed, each presenting unique challenges. Here, we specifically focus on the challenges associated with graphene derivatives used in large quantities for practical applications, particularly GO, rGO, and their assemblies. While our review centres on these specific materials, the insights and challenges discussed may also be relevant to other graphene derivatives.

*Large-Scale, Cost-Effective Production with Controlled and Consistent Quality*. The variability in production quality arises from the complexities in controlling the synthesis and functionalization processes of graphene materials. For instance, the nonstoichiometric nature of GO and rGO results in varying ratios of carbon, hydrogen, and oxygen. These variations pose fundamental issues in determining the reproducibility of rGO by just comparing the atomic ratios of each sample, making it difficult to evaluate the consistency of product quality of graphene derivatives. Additionally, diverse synthesis methodologies for rGO and its precursors lead to significant variations in their defects, clustering of functional groups, sheet size and overall composition. These variations complicate efforts to achieve good uniformity in one batch and reproducibility across different batches.

*Structural Characterization and Control at Multiple Length Scales.* The challenges of structural characterization and control ranges from atomic structure of building blocks to the assembled stacking structure. In specific, GO, as the commonly used building blocks of graphene assembly, has an overall chemical composition that is generally understood. However, its detailed local and long-range chemical functional groups remain unclear. The surface chemistry and interfacial effects of functional groups on GO—particularly their interaction with solvents—require further research to fully understand their role in regulating their assembly process. Moreover, defects, in-plane pores and corrugations of nanosheets complicate the interactions between the building blocks and the resultant stacking patterns. Consequently, structure irregularities and disorder occur at different length scales, including the variation in the interlayer features, formation of intermediate structure unites, and the existence of large voids.[3] The difficulty to accurately characterize and describing those structural irregularities complicates efforts to precisely engineer the hierarchical structure of those assemblies to achieve optimized performances for given applications.

*Establishing Proper Structural-Property Relationship.* Even though advanced characterization techniques can provide insights into the local atomic structure and hierarchical organization, understanding how these structures influence macroscopic properties across multiple scales remains incomplete. For example, studies report two diverse trends of sheet size effects on the mechanical properties of graphene membranes.[4] These seemingly controversial discoveries indicate how this lack of comprehensive structure-property mapping hampers the precise engineering of graphene derivatives for targeted applications.

*Exploring New Properties and Translating it into Real-World Applications.* Studies have shown that graphene derivatives hold great promises in many advanced applications including flexible electronics, energy storage, and ion separation. However, in most studies, only principal concepts have been demonstrated. Gap exists in the practical applicability of these materials in real-world applications. For example, graphene assemblies with controlled nanochannels have been theoretically predicted and experimentally demonstrated to exhibit ultra-fast ion transport, remarkable ion selectivity and storage properties.[1] Although these properties are promising in catalysis, energy storage and filtration field, their integration into commercially viable products remains limited. Bridging this divide requires not only technological advancements but also interdisciplinary collaborations to identify critical application domains where these properties can deliver transformative impacts.

**Advances in Science and Technology to Meet Challenges**

Graphene derivatives are now being produced at industrial scales, marking a pivotal era for designing graphene-based materials with broader commercial applications in the coming years. While many technical advancements and applications are discussed elsewhere in this roadmap, three often-overlooked aspects are critical for advancing graphene derivative research (Figure 1): (i) the development of integrated multiscale characterization techniques for structural analysis, (ii) harnessing the hierarchical structures of graphene assemblies to unlock new properties and applications, and (iii) leveraging artificial intelligence (AI) for graphene material design and optimization.

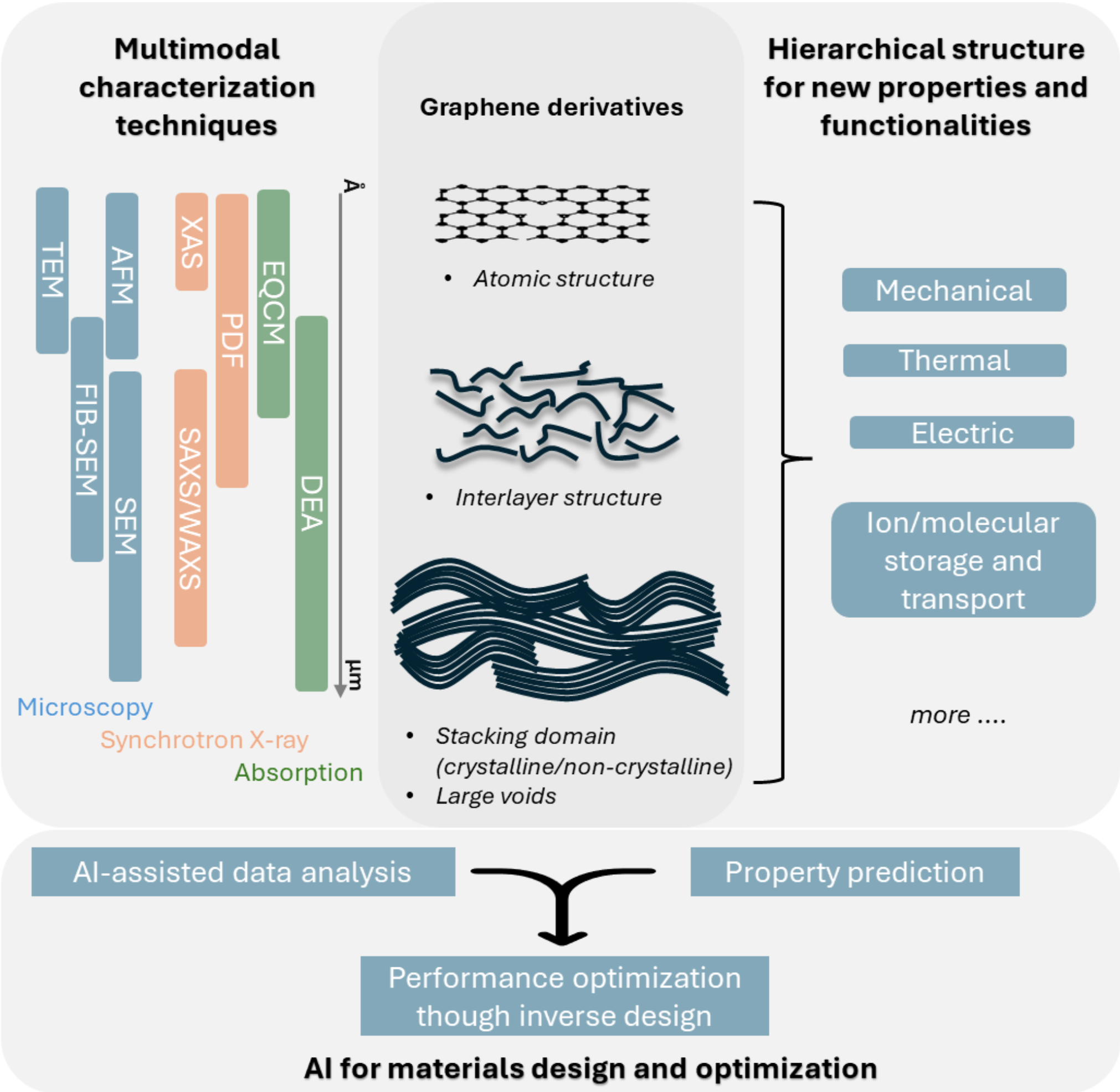

Figure 1. Some Key considerations to advancing graphene derivative research in the coming years: (i) integrating multimodal characterization techniques to decode complex hierarchical structures, (ii) leveraging these hierarchical assemblies to unlock novel properties and applications, and (iii) employing AI to optimize material design.

*Integrating Multimodal Characterization Techniques.* Recent advancements in characterization techniques have significantly enhanced our understanding of graphene assemblies. Microscopic tools such as SEM, AFM, and focused ion beam SEM (FIB-SEM) provide nanoscale insights into atomic and nanoscale structure of graphene derivatives. For example, FIB-SEM nanotomography has recently been employed to reveal intricate morphological details in printed graphene networks, including porosity, tortuosity, surface area and sheet orientation. [5] Synchrotron-based methods, including X-ray absorption spectroscopy (XAS), pair distribution function analysis (PDF), and small-angle/wide-angle X-ray scattering (SAXS/WAXS), address intrinsic structural complexities of graphene assemblies by offering exceptional spatial and spectral resolution for multiscale characterizations from sub-nm to μm and enabling in situ and operando observations.[3]

Despite these advances, the complex hierarchical structures and dynamic operational behaviors of graphene assemblies demand more integrated approaches. Fully understanding these assemblies also requires complementary methods, such as dynamic electroadsorption analysis (DEA) and electrochemical quartz crystal microbalance (EQCM), which use small ions and molecules as probes to obtain pore surface area, surface chemistry and active sites.[3], [6] Together, these techniques provide structural information at different dimensions to systematically decode the intricate architectures and correlate it with collective properties of graphene derivatives.

*Harnessing Hierarchical Structures of Graphene Assemblies.* Research on graphene-based bulk materials has traditionally focused on leveraging the intrinsic properties of individual graphene sheets, such as mechanical strength and thermal or electrical conductivity. However, increasing research has found that the performance of graphene assemblies is influenced more by the hierarchical organization than by the properties of individual sheets. For example, Mechanical properties and thermal conductivity of assemblies depends heavily on the inter-sheet bonding and connectivity.[2] Moreover, controlled stacking of graphene sheets creates nanochannels structures showing unique ion transport behaviors, critical for energy storage and filtration applications.[7] Particularly, the hierarchical porous structures are shown to remarkably enhance molecules/ion transport efficiency, demonstrating the importance of architectural design.[8]

Structural features within assemblies can sometimes conflict, necessitating trade-offs and precise multiscale characterization and structural control. For instance, while maximizing surface area and porosity benefits ion transport efficiency, it can also reduce volumetric energy density in energy storage electrodes and ion selectivity in separation applications.[9] Advances in understanding interfacial interactions during graphene oxide assembly have enabled tailored internal microstructures, balancing surface area and packing density. Techniques such as solvent-controlled removal and capillary compression have produced reduced graphene oxide (rGO) hydrogels with superior volumetric performance while maintaining high gravimetric capacitance.[9] These developments highlight the need for deeper exploration of hierarchical assembly processes to unlock new applications.

*Leveraging Artificial Intelligence for Graphene Derivative Research.* The structural complexity and diverse property requirements of graphene derivatives present significant challenges for material design and optimization. Artificial intelligence (AI) offers transformative potential in this domain through microstructural recognition, property prediction and performance optimization. [10] As the field evolves, combining AI with advanced characterization and modelling tools will be pivotal in addressing the multifaceted challenges of graphene research. Specifically, by integrating AI with advanced characterization techniques and computational models, researchers can correlate hierarchical structures—such as nanochannels and defect distributions—with macroscopic properties like conductivity and ion transport behavior. This approach uncovers intricate structure-property relationships and guide the optimization of graphene assemblies, from sub-nanometer channels for high-efficiency filtration to dense nanosheet packing for next-generation batteries.

**Concluding Remarks**

The discovery of graphene has revolutionized materials science, while its derivatives continue to expand the field's horizons with their ease of production and diverse properties. From GO's dispersibility to the newly introduced functionalities of graphene assemblies, these materials have reshaped the landscape of technological applications of graphene. However, challenges remain in ensuring consistent production quality, precise structural characterization, hierarchical stacking controlling and real-world applications integration. Addressing these with advanced characterization, new structural control strategies, and AI-driven design will be key. As graphene derivatives and their applications transition from conceptual innovations to practical solutions, they hold the promise of revolutionizing industries and driving transformative progress in addressing global challenges.

**Acknowledgements**

The author acknowledges the financial support from the Australian Research Council (FL180100029, D.L.) and National Natural Science Foundation of China (Nos. 52432005, Q.-H. Y.).

## 5. Towards large-scale synthesis of transition metal dichalcogenides

**Soo Ho Choi[1,2], Ki Kang Kim[1,2], and Young Hee Lee[1,2]**

[1]Center for Integrated Nanostructure Physics, Institute for Basic Science, Suwon 16419, Republic of Korea
[2]Department of Energy Science, Sungkyunkwan University, Suwon 16419, Republic of Korea

### Status

Despite extensive research on the remarkable physical and chemical properties of transition metal dichalcogenides (TMDs), there are still significant challenges in large-area synthesis techniques toward practical applications. In this section, we will provide an overview of the historical development of large-area synthesis techniques for TMD materials.

To achieve large-area TMD films, various techniques have been applied, including pulsed laser deposition, molecular beam epitaxy, atomic layer deposition, and chemical vapor deposition (CVD). Among them, CVD is the most extensively studied technique for economically viable large-area TMD film synthesis. In 2012, powder-CVD method was developed to synthesize TMD grains of a few tens micrometres in size by vaporizing or supplying transition metal oxide powder and chalcogen powder precursors [1]. At the same time, the different approach was proposed to synthesize large-area poly-crystal (PC) TMD films with uniform thickness [2]. This approach involves the deposition of transition metal films onto substrates, followed by subsequent heat treatment in a sulphur or selenium atmosphere. However, it is of note that TMD films exhibit poor crystallinity, leading to a substantial reduction in charge carrier mobility in the field effect transistor (FET). In 2015, monolayer PC TMD films were successfully synthesized on 4-inch wafers using metal-organic chemical vapor deposition (MOCVD) [3]. This approach enables precise control over the supply of volatile powder or liquid precursors, facilitating the successful synthesis of high-quality TMD films at a large scale. Furthermore, liquid-precursor-intermediated (LPI) CVD method was developed to synthesize the large-area polycrystalline TMD films by spin-coating the liquid transition metal precursor onto the substrate, followed by controlled thermal annealing in chalcogen atmosphere [4]. This method offers precise control of the film coverage by adjusting the precursor concentration and spin coating speed, ensuring reproducibility and versatility in the growth. Moreover, different transition metal precursors are incorporated to result in the doping in TMD materials. This significantly broadens the range of potential applications. Meanwhile, recent research has progressed on the synthesis of single-crystal (SC) TMD films to address the degradation of their intrinsic properties such as mechanical strength, chemical stability, and carrier mobility, caused by grain boundaries (GBs) in PC TMD film. As a result, single crystalline monolayer TMD films have been successfully synthesized with sizes of up to 2 inches using substrates including atomic sawtooth Au and miscut sapphire substrate [5,6].

Lateral and vertical TMD heterostructures offer unique physical phenomena for electronic and optoelectronic devices including Coulomb drag, Bose-Einstein condensation, band renormalization, interlayer excitons, etc. Initially, diverse physical phenomena and applications from TMD heterostructures have been investigated through the transfer of each TMD layers. The presence of residuals at the heterostructure interface during the transfer process often obscures intrinsic interface physics. Therefore, developing direct synthesis techniques of heterostructures has been highly desired to achieve clean interfaces. In 2014, TMD lateral heterostructure grains were synthesized by sequentially supplying powder precursors in a two-step process [7]. Furthermore, by utilizing the MOCVD technique, the width of the heterostructure can be precisely controlled up to a few tens nanometres [8]. On the other hand, several problems still remain in the synthesis of vertical heterostructure. Among various challenges, most important issue is the nucleation on dangling bond-free surface. To address this, artificial defects were introduced on TMD surface to enable controlled nucleation for the patterned synthesis of vertical heterostructures [9]. However, the utilization of micrometre-scale defects concerns regarding potential degradation in the intrinsic properties of TMD layers. In 2021, large-scale vertical TMD superlattice has been successfully synthesized through the nucleation at GBs within polycrystalline TMD films [10]. Nevertheless, the small grain sizes prevail and inevitably shrinks as the film thickness increases, consequently constraining the potential applications.

To date, the synthesis techniques for achieving large-area SC TMD films at the monolayer scale have been successfully developed. However, the techniques for synthesizing lateral and vertical heterostructures using TMD materials are predominantly limited to grain growth methods and PC films. With further advancement of CVD techniques, the performance of current electronic and optoelectronic devices will be significantly improved, and the scaling issues faced by silicon-based devices could be addressed.

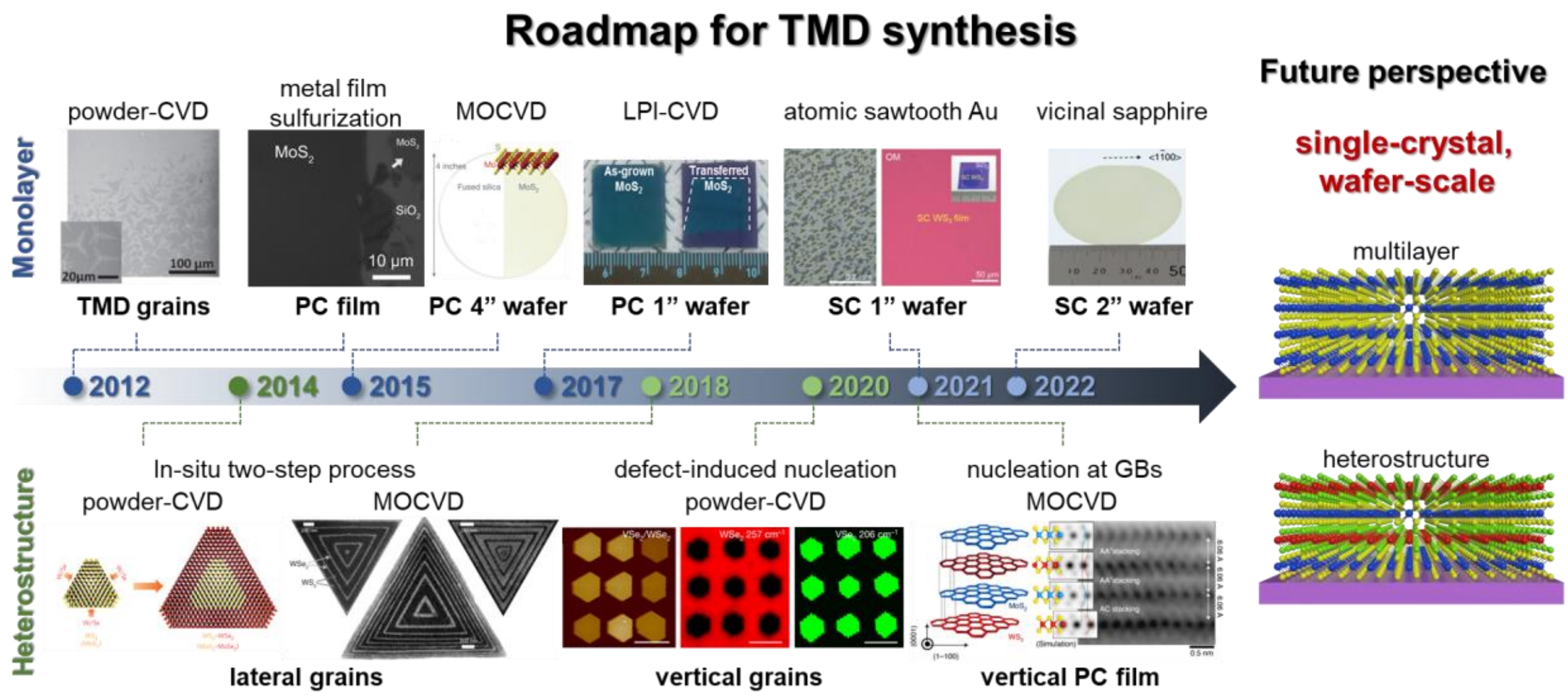

**Figure 1.** Roadmap for the synthesis of large-area TMD monolayers and heterostructures. Reproduced from [1-10], permission from John Wiley & Sons. WILEY-VCH Verlag GmbH & Co. KGaA, Weinheim, Nature Publishing Group, and AAAS.

## Current and Future Challenges

Progresses of TMD materials synthesis have been remarkable in the past decade for large-area TMD films. However, despite these notable achievements, several critical challenges persist for the commercialization of large-area TMD film. In this section, we will discuss the key issues.

i) Synthesis of SC multilayer TMD film

Multilayer TMDs are often needed, for example, for enhanced charge carrier mobility in FET devices compared to monolayers. While the four-layer SC graphene film has been successfully synthesized on SC Si-Cu alloy, only two-layer SC $MoS_2$ film has been synthesized on the miscut sapphire substrate to date. Generalization to thicker films beyond two-layer film together with wafer-scale size, single-crystal, and uniformity of the film is necessary in the field of electronic devices.

ii) Nucleation on dangling-bond-free surface for TMD heterostructures

To date, the synthesis techniques for TMD heterostructures have been primarily focused on the limited number of materials, and the synthesis of large-area heterostructure films has been predominantly achieved in the polycrystalline materials. Homo/heterostructures with single crystallinity and well-defined stacking order and orientation are deemed necessary. One important key for TMD heterostructure is the nucleation on the dangling-bond-free surface of TMD materials. While defect-induced nucleation has been developed, the recovery of film damage resulting from defect formation still remains to be solved.

iii) Other issues

Various dopants could be doped in TMD materials using LPI-CVD and MOCVD techniques. Yet, the dopant uniformity is of primary concern. In addition, the current synthesis techniques require a transfer process of large-area TMD films onto another substrate prior to the fabrication of devices. In this transfer process, wrinkles and tearing on TMD films and contaminations caused by the chemical residues inevitably arise. Thus, the development of residual-free transfer techniques is highly desired.

## Advances in Science and Technology to Meet Challenges

To synthesize large-area SC multilayer TMD films, two possible ways could be suggested: 1) thickness-controlled synthesis of few-layer TMD islands followed by their coalescence, or 2) layer-by-layer epitaxy. Recently, bi-layer SC $MoS_2$ film has been successfully synthesized via thickness-controlled nucleation on miscut sapphire substrates. Due to the higher thickness of step edges on the sapphire substrate, coherently aligned bi-layer $MoS_2$ islands predominantly synthesized, resulting in the formation of bi-layer SC $MoS_2$ film. Furthermore, 3-5 layers of SC hBN film also have been synthesized on Ni (111) substrate through the coalescence of thickness-controlled hBN islands. Thus, similar approaches could enable the synthesis of few-layer SC TMD films. The layer-by-layer homo- and heteroepitaxy is an ideal approach to synthesize few-layer SC TMD films as well as SC TMD heterostructures. Therefore, the development of techniques to precisely control the surface energy of TMD materials is highly desired, while maintaining a defect-free surface. This could offer the synthesis of various TMD materials with the desired thicknesses. In addition, from the perspective of modulating the electronic structure of TMD materials, the development of novel doping techniques or post-processing strategies for dopant redistribution could lead to the uniform distribution of dopant elements within the film.

**Concluding Remarks**

For the practical application of TMDs in various fields, significant advancements have been made in the development of CVD techniques for the synthesis of large-area TMD films. Nevertheless, further progress is still required for layer-by-layer epitaxy or transfer techniques that enable precise control of TMD film thickness and stacking orders for next-generation electronic/optoelectronic devices.

**Acknowledgements**

This work was supported by the institute for Basic Science of Korea (IBS-R036-D1) and Hubei University of Technology, China. K.K.K. acknowledges support from the Basic Science Research through the National Research Foundation of Korea (NRF), which was funded by the Ministry of Science, ICT and Future Planning, and the Korean government (MSIT) (2022R1A2C2091475 and RS-2024-00439520).

## 6. The roadmap of MXenes

**Mian Li[1], Qing Huang[1], Yury Gogotsi[2]**

1. Zhejiang Key Laboratory of Data-Driven High-Safety Energy Materials and Applications, Ningbo Institute of Materials Technology and Engineering, Chinese Academy of Sciences
2. Department of Materials Science and Engineering, A. J. Drexel Nanomaterials Institute, Drexel University.

Since the first report on unconventional physical properties in graphene, materials with two-dimensional (2D) morphology and one or a few atoms in thickness have been attracting increasing interest. This led to the discovery of many new 2D materials, including a family of 2D transition metal carbides/nitrides (MXenes), which continue growing and show remarkable potential in many applications. MXenes have a general formula of $M_{n+1}X_nT_x$, where M is an early transition metal, X is carbon or nitrogen, $T_x$ is the surface terminations that bonded to the outer M layers, *n*+1 and *n* is the respective numbers of layers of M and X atoms. Thus, the thinnest MXenes have 3 atomic layers (without terminations), similar to transition metal dichalcogenides. The thickest members of the family ($M_5C_4T_x$) approach 1.5 nm, having 11 atomic layers, which provide them stiffness and bending rigidity exceeding other 2D materials. The diverse compositions and versatile structures endow MXenes with tunable physical and chemical properties, which are useful in various applications. This article provides a brief perspective on the latest advances in synthetic methods, surface chemistry, and MXene applications. We further give an insight into the future opportunities for the MXene materials.

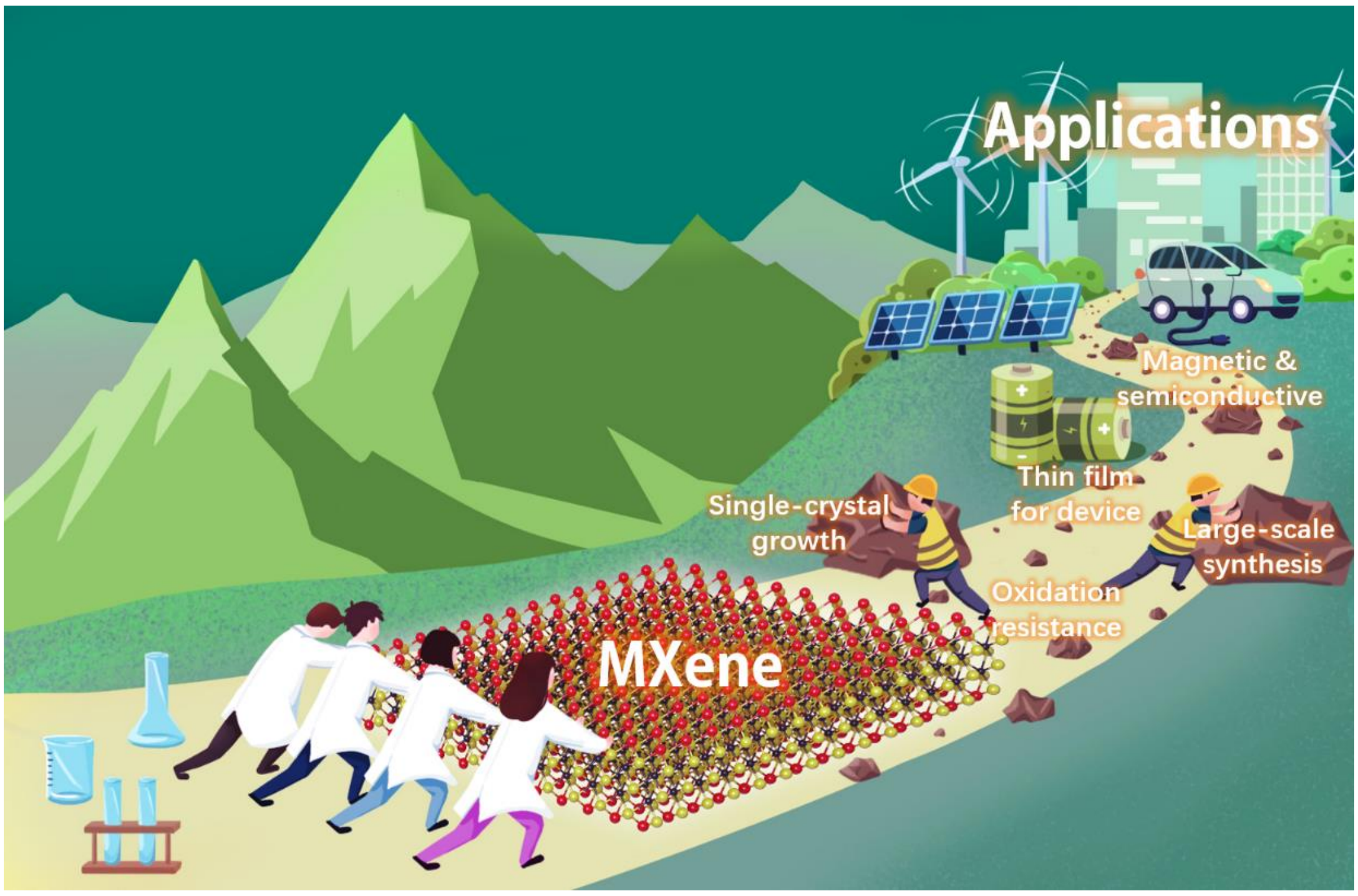

Figure 1. The challenges and opportunities for the MXene materials

**Synthesis methods**
The synthetic route determines surface terminations and allows one to tune the properties and expand the application scopes of MXene materials. Since the first discovered MXene $Ti_3C_2T_x$ was obtained by using HF to etch the $Ti_3AlC_2$ MAX phase in 2011, selective etching of MAX in fluorine-containing aqueous solutions has become the most widely used method to prepare MXenes[1]. Besides HF, fluorine-containing etchants such as LiF+HCl (MILD method), bifluoride salts (e.g., $NH_4HF_2$), and fluorine-containing molten salts are used for the synthesis of MXenes[2]. Considering the limitations of fluorine-ion etching methods, fluorine-free etching methods such as electrochemical etching, alkaline hydrothermal etching, HCl-hydrothermal etching, halogen etching, and Lewis acidic molten salts have been developed[3]. Gaseous halogen etching produces halogen-terminated MXenes with uniform surface terminations[4]. Among the above methods, Lewis acidic molten salts received the most attention because of the ability to etch different MAX phases and precisely control the surface terminations of MXenes [5], benefitting from the versatile constituents of molten salts and a wide range of etching temperatures.

A main limitation of the selective etching (“top-down”) synthetic route is that the structure and composition of the MXenes rely on the availability of precursors, such as regular MAX phases or double A layer counterparts like $Mo_2Ga_2C$. Such “top-down” route is always energy-consuming and expensive, and not practical for synthesizing high-quality MXenes on a large scale. A direct synthetic route of MXenes through the reactions of metals and metal halides with graphite, methane, or nitrogen was proposed in 2023[6]. It enables chemical vapor deposition (CVD) growth of MXene carpets and spherulite-like morphologies. Considering the enormous variety of elemental stoichiometries in the MXene family of materials, direct synthetic routes could substantially expand the MXene family and the range of accessible properties. It is important to mention that the fluidized bed CVD synthesis of $Ti_2CCl_2$ MXene from titanium chloride (an intermediate product of $TiO_2$ synthesis) and methane (natural gas) can provide large amounts of chlorine-terminated MXenes at the price level between titania and multiwall carbon nanotubes[7], which are produced using similar processes.

**Surface chemistry**

In a 2D flake of MXene, the terminations $T_x$ are bonded to the outer M layers and exposed on the flake's surface. This fact makes the terminations not only influence the intrinsic properties of the MXenes but also play a dominant role in determining the interaction between MXenes and the ambient environment. Numerous theoretical studies have predicted that specific terminations of MXenes can lead to remarkable properties, such as ultra-high electron mobility, widely tunable work functions, half-metallicity, and 2D ferromagnetism. Therefore, surface chemistry is one of the most important issues, and it should be controlled along with the structure and composition research of MXenes. Thus, control over the surface chemistry of MXenes has drawn many attentions. Generally, the terminations of MXenes originate from the etchant during the etching of MAX phases or gaseous reactants in CVD synthesis. Fluorine-containing aqueous etching methods inevitably result mixed -O, -OH, and -F terminations, whose effect on the properties of MXenes has been addressed by both theoretical and experimental studies. Fluorine-free etching methods enable a wide range of uniform surface terminations. For example, the recently developed Lewis acidic molten salt etching methods can form MXenes with pure terminations of -Cl, -Br, -I and their combinations[8]. Substitution and assembly of the MXene terminations in molten salts can further transform MXenes with terminations such as $-NH_2$, -S, -Se, and bare MXenes (no surface termination)[9]. Furthermore, chemical scissors–mediated structural editing, based on the molten salt chemistry, can form MXene with -P and -Sb terminations[10]. Recently, a flux-assisted eutectic molten salts etching approach enables the synthesis of MXenes with triatomic-layer borate polyanion terminations (OBO terminations)[11]. Some of those terminations have been demonstrated to induce superconductivity[9] and extraordinary electrochemical properties[11]. The versatile surface chemistry provides large room to explore properties and applications the of MXenes.

**Properties & applications**

The 2D structure and diverse elemental constituents endow MXenes with a unique combination of properties, including a large exposed surface area, high electronic and ionic conductivity, outstanding mechanical properties inherited from bulk carbides and nitrides, and a hydrophilic nature, particularly when terminated with oxygen or hydroxyl groups. These characteristics are vital for a multitude of applications, such as electrochemical energy storage, sensing, microwave absorption, electromagnetic interference (EMI) shielding, catalysis, and more.

MXenes exhibit excellent performance as supercapacitor electrodes due to their pseudocapacitive (redox) charge storage mechanism coupled with electric double-layer behavior. Engineering the surface and interlayer chemistry of MXene electrodes can significantly enhance the capacity and charge-discharge performance of batteries. Moreover, MXenes demonstrate their advantages by enabling supercapacitors and batteries to extend into unconventional fields, such as micro-supercapacitors, hybrid capacitors, and beyond Li-ion batteries. This remarkable performance in electrochemical energy storage can be attributed to MXenes' high electrical conductivity, 2D morphology, and adaptability to various customizations[12].

In gas sensing applications, MXenes can operate effectively at room temperature through the strategic design of their constituents and structure. Notably, MXenes have shown significant progress in flexible sensors, including wearable devices, soft robotics, and smart medical equipment. A wide array of flexible sensors can be developed based on the properties of MXenes and various substrate materials, encompassing pressure, gas, electrochemical, and biosensors. However, several challenges remain to be addressed to further enhance the performance of MXenes in sensor applications. For instance, improving the environmental stability of MXenes is crucial for the effective construction of flexible sensors. Additionally, most current research focuses on $Ti_3C_2T_x$ MXenes, and exploring other MXene compositions could unveil new possibilities for designing and fabricating high-performance MXene-based sensors[13].

MXenes exhibit significant potential for microwave absorption and electromagnetic interference (EMI) shielding due to their high specific surface areas, abundant functional groups and defects, high electronic conductivity, and numerous interfaces within assembled films and coatings. Various MXenes have been investigated for microwave absorption and EMI shielding, highlighting the advantage of fabricating micrometre-thin, free-standing films[14]. However, further efforts are required to fully understand and optimize the applications of MXenes in these fields. Detailed elucidation of the microwave absorption and EMI shielding mechanisms is essential to customize the design of MXene constituents and structures for effective interaction with infrared, terahertz, and gigahertz waves [15].

MXene-based materials are also garnering attention in catalysis, particularly for the hydrogen evolution reaction, oxygen evolution reaction, and carbon dioxide reduction reaction. In electrocatalysis, MXenes can function as both catalysts and supports. Their unique features provide ample opportunities for designing and preparing electrocatalysts with high activity, selectivity, and durability, positioning MXenes as promising alternatives to platinum-based catalysts. Moreover, due to their low Fermi level, MXenes can serve as photo-generated electron acceptors in photocatalysis, facilitating fast charge carrier separation and enhancing photoconversion efficiency. Considering their large specific surface area and diverse surface terminations, MXenes play a multifaceted role in improving photocatalytic activity beyond acting merely as electron acceptors[16].

The application of MXenes in biomedical and environmental domains is also noteworthy. Their remarkable mechanical properties, flexibility, high specific surface area, and hydrophilicity make MXenes promising candidates for high-performance water treatment membranes. Their electrical and ionic conductivity, coupled with significant photothermal properties, further enhance their utility in water treatment processes. However, improvements in their environmental stability are necessary for commercial applications. In the biomedical field, the biocompatibility and low cytotoxicity of certain MXenes, along with their plasmonic resonance and high photothermal conversion efficiency in the near-infrared and infrared ranges, make them suitable for cancer therapy. Additionally, MXenes hold potential for various biomedical applications, including antibacterial coatings, regenerative medicine, medical imaging, drug delivery, diagnostics, and biosensing.

**Outlook**

The research on MXene materials is rapidly evolving, with significant advancements made in recent years. The development of new synthetic routes has greatly expanded the elemental composition of MXenes, endowing them with novel properties and enhancing their applicability. These advancements also allow for better control over surface chemistry and have the potential to substantially reduce production costs. We anticipate further expansion in MXene research in the near future, with the following key issues warranting attention:

1. Expansion of Elemental Composition: The incorporation of lanthanide and actinide elements at the M-site, and non-metallic elements such as oxygen, boron, nitrogen, and phosphorus at the X-site, with multi-element terminations as $T_x$ groups.
2. Structural stability in varied environments and working conditions: Strategies to resist oxidation and performance degradation must be implemented.
3. Synthesis of Large-sized Single-crystal MXene Nanosheets: Techniques to derive large-size single-crystal MXene nanosheets directly from MAX phase single crystals or via bottom-up methods (such as CVD) are crucial for studying their fundamental physical properties.
4. Heteroepitaxial CVD Growth Techniques: The development of heteroepitaxial chemical vapor deposition (CVD) growth methods will provide a new route for producing high-quality MXenes or complex layered nano-architectures with tailored properties.
5. Exploration of Intrinsic Physical Properties of MXenes: It is crucial to investigate the physical properties of MXenes, focusing on semiconductive and magnetic behaviors that have never been reported.
6. Large-scale Production: Scaling up the synthesis of MXenes through etching approaches or CVD techniques will be essential for industrial applications.
7. Emergence of MXene-analogous Layered Materials: The layered transition metal carbo-chalcogenides (TMCCs), such as $Nb_2S_2C$, which can be considered analogous to $Nb_2CS_x$ MXene but with doubled sulfur terminations, are promising to combine the merits of both transition metal dichalcogenides and MXenes.
8. Applications of MXenes: Given their unique properties, MXenes promise breakthroughs in various fields. Potential applications include conductive transparent, flexible, and wearable electronic and optoelectronic devices, energy harvesting and storage, electromagnetic shielding, epidermal and implantable electrodes, tissue engineering, soft robotics, thermal insulation, and beyond. The ongoing and future research efforts will undoubtedly unlock new potentials for MXenes, making them pivotal in addressing technological challenges across multiple domains.

## 7. From promise to progress: transfer induced-inhomogeneity in 2D heterostructures

**Nicholas Clark[1], Amy Carl[1], Roman Gorbachev[1]**

[1]University of Manchester, UK

**Status**

The stacking of individual 2D crystals into heterostructures offers the unprecedented ability to rationally design functional materials at the atomic level with nearly limitless configurations. The large library of 2D materials (2DM) with diverse electronic properties can be used to assemble complex devices layer-by-atomic-layer, allowing for novel material combinations and interfaces that cannot be realised through direct growth methods. Over the past decade, numerous proof-of-principle devices have been demonstrated, however for many the prospect for commercialisation is currently remote. Transfer-induced inhomogeneity and interfacial contamination drastically limit the clean area of heterostructures as the number of layers and interfaces increase, impeding the development of high end-applications.

This problem is especially acute for those applications where high electronic quality and large-scale uniformity is essential, such as in optoelectronics, quantum technologies and aerospace. While 2DM can substantially impact these fields, the transition from micrometre-scale prototypes to large-scale manufacturing has become a critical bottleneck. Despite this, advances in 2D material growth and transfer techniques are starting to yield viable commercial products, albeit only those with less stringent requirements for electronic grade materials quality. This discussion outlines the state of the field and the necessary developments to enable the commercial applications where tailored vdW heterostructures can really shine.

**Current and Future Challenges**

The major challenge for both academic investigation into van der Waals (vdW) heterostructures and their industrial utilisation is the fabrication of repeatable and homogenous 2D crystal structures. While layer-by-layer growth techniques show promise[1], [2], they are limited to specific systems with compatible chemistry, demonstrated over small areas, and incapable of generalizing to more complex arrangements, such as controllable crystalline misalignment[3]. This makes transfer of individually grown 2D materials the next best option, although it presents its own set of problems.

Considerable progress has been made in thin film growth of 2D materials in the last 5 years, despite its thermodynamic disadvantage compared to bulk crystal formation. However, challenges remain. For instance, few-layer grown hexagonal boron nitride (hBN), the key dielectric crystal employed in the majority of vdW heterostructure devices, exhibits limited uniformity and poor crystalline quality compared to exfoliated crystals. The rapid development in this field instils optimism that the quality of few-layer grown hBN will match that of exfoliated crystals within a few years[4], although it currently falls short of the necessary standards.

The primary issue for 2DM devices occurs as individual crystals are stacked to form a heterostructure. Currently, this process is performed in air or inert gases, resulting in the adsorption of a wide range of volatile atmospheric species on the crystal surfaces. These contaminants subsequently become trapped at the interfaces of the heterostructure. This contamination remains mobile between the layers and can diffuse forming isolated pockets (also called “bubbles” or “blisters”) leaving micrometre-scale regions of atomically clean interfaces behind. This phenomenon, typically referred as “segregation”, has been both a blessing and a curse for the field. Whilst it provides small clean areas perfect for device prototyping, it prevents scale-up with dense patterning or large area devices. Segregation occurs regardless of whether the 2DM is exfoliated or grown, however it does not occur if one of the surfaces is not atomically flat or has high defect density. In such cases the contamination is pinned, forming a continuous film at the interface.

The second problem stems from the use of polymeric transfer methods. Polymer carrier films are widely used to delaminate and support fragile atomically thin crystals. These soft layers have drastically different thermal expansion coefficients compared to those of 2DM. In combination with the difference in topography between the growth and target wafer, this leads to formation of wrinkles and cracks in transferred 2D materials.

Due to these issues, even with 12 years of intensive development, all currently demonstrated assembly techniques suffer from large scale interfacial inhomogeneities. Atomically clean areas are limited to tens of microns at most, even in the simplest 3-layer heterostructures. The homogeneous area decreases with each additional layer in the vdW stack, compounding the issue in more complex devices such as LEDs, where the optically active area is typically limited to a few micrometres due to contaminants at the numerous interfaces. Another issue relates to the exciting advances in studies of the rotational alignment, or so-called “twist”, between each crystal. This relative twist enables the formation of novel optoelectronic states and fine tuning of material properties. Such structures are extremely sensitive to small local strain variations, resulting in a complex and unpredictable moiré superlattice. Even using state of the art optimised transfer processes, significant twist angle disorder is observed over ~100 nm length scales. This, together with the sensitive dependence on twist angle of the electronic behaviour, means that measurements on nominally identical stacks display considerable variation. This variation prevents repeatability, external verification of results, and targeted testing of hypotheses, let alone the development of applications. While certain techniques can partially remove most of the trapped contamination after transfer, these are time consuming, limited to small areas, and do not necessarily restore the interface to a pristine state[5].

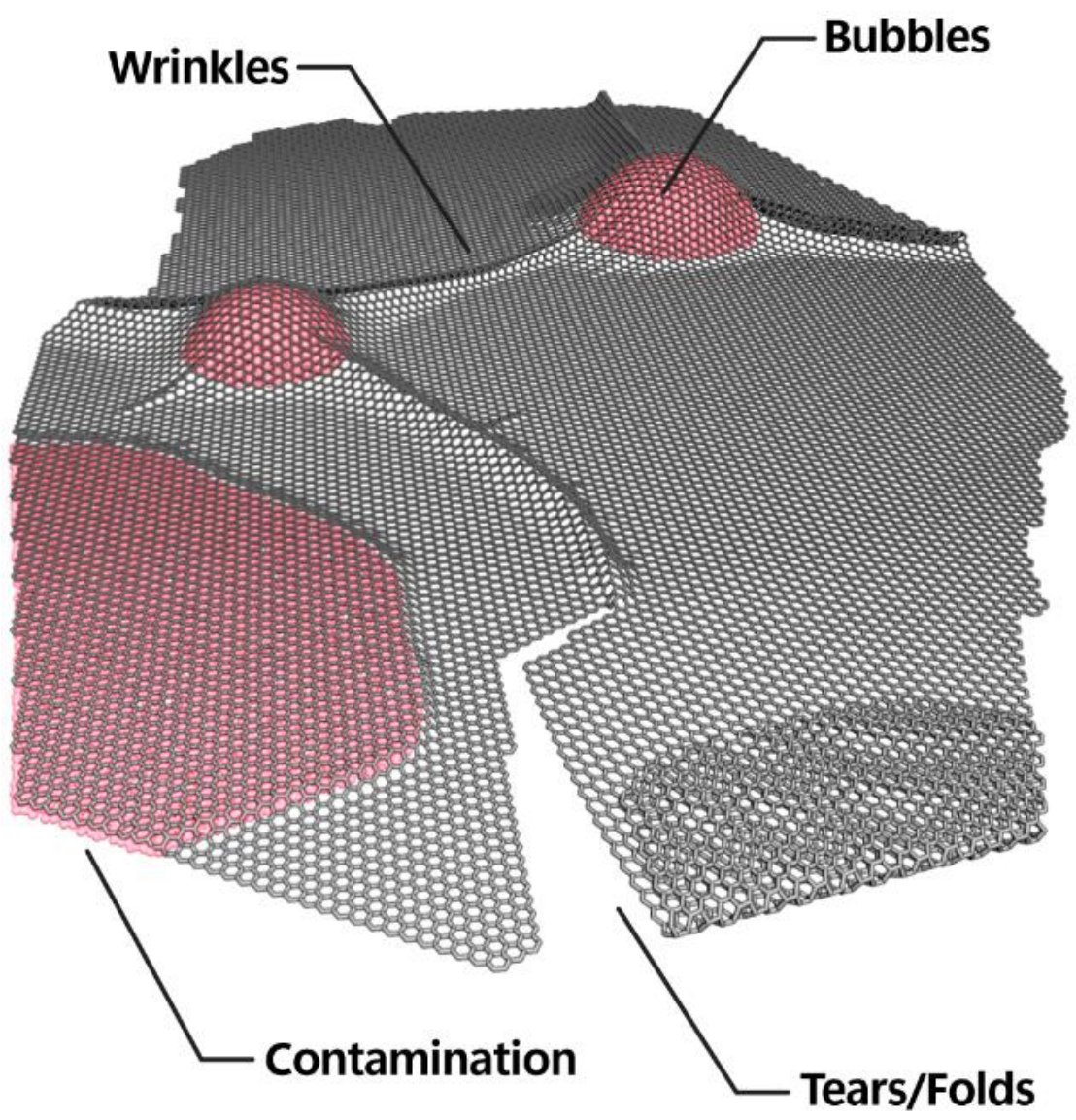

*Figure 1 Sources of inhomogeneity in vdW heterostructures*

**Advances in Science and Technology to Meet Challenges**

An obvious solution to address these challenges is to modify the operational conditions during layer assembly. Alternatives include operating in high or ultra-high vacuum (preceded by surface conditioning, e.g. annealing) or performing assembly with interfaces submerged in solvents capable of stripping away surface contamination. However, such alternative environments remain unattainable due to the use of polymeric carriers in all 2DM transfers. These polymers exhibit low melting temperatures (100-200C), a loose molecular matrix which is prone to outgas in vacuum, and poor chemical resistance against solvents capable of removing hydrocarbon surface contamination. In initial experiments contamination persists even in high vacuum environment due to the use of these polymeric carrier layers. Turning away from the polymers in favour of inorganic materials to transfer 2DM (e.g. metal foils, nitride or oxide films) should significantly improve the situation, allowing for higher processing temperatures, and providing more closely matched thermal expansion rates, low outgassing in vacuum and the ability to employ more aggressive chemical treatments.

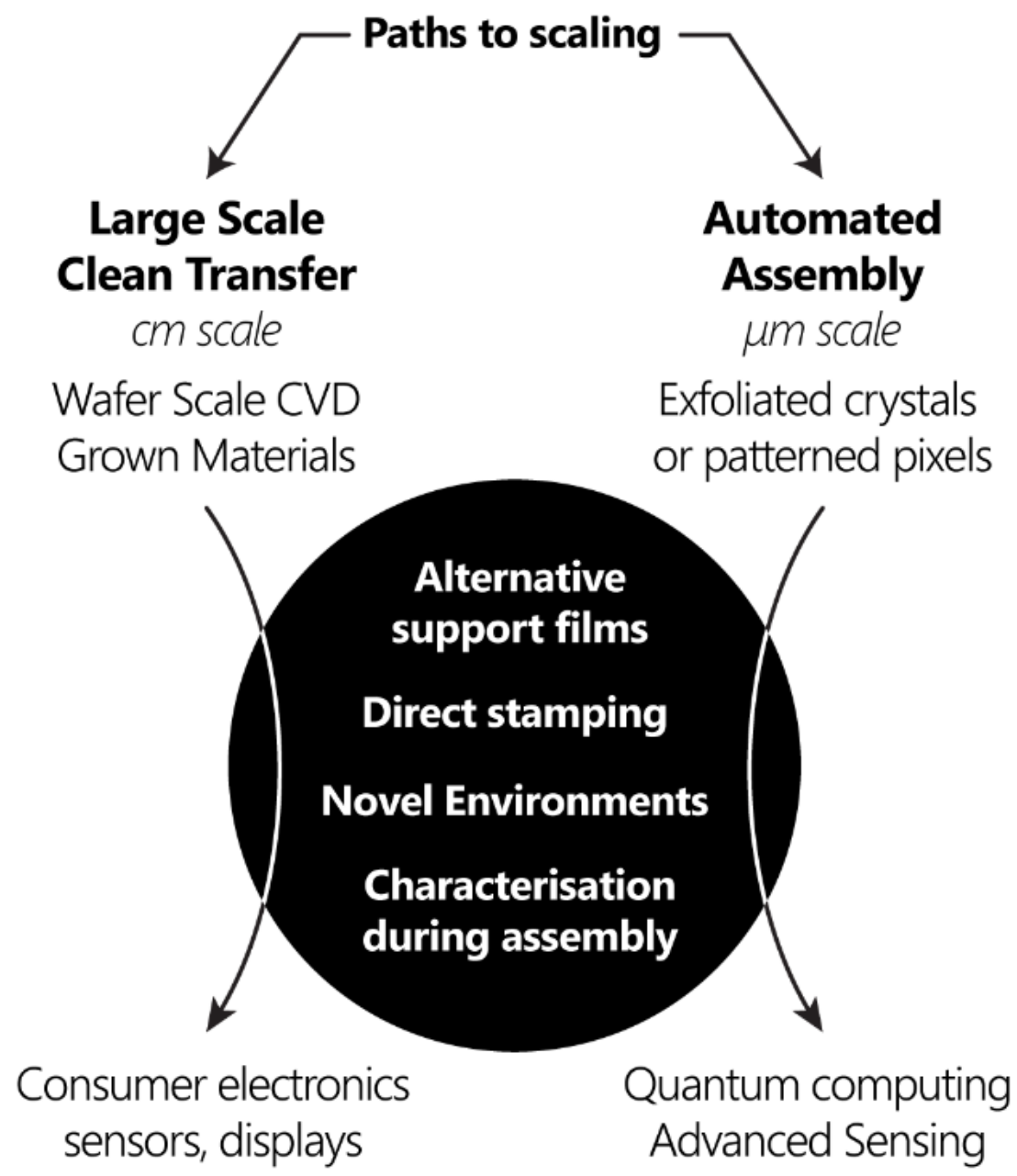

*Figure 2 Routes to scalable, clean heterostructure fabrication*

Regarding compatible scale-up methods, we foresee two major pathways. One obvious and widely explored route of development is to apply the current research best-practice techniques at a large scale to CVD-grown materials. These are polymer based "dry" delamination techniques, where 2D materials are transferred from low adhesion growth substrates to higher adhesion atomically flat 2DM. These avoid any contact with polymers or liquids on the vdW heterostructure interfaces, confining most contamination to an outer encapsulation layer. As almost any potential inorganic replacements for the carrier would be less pliable than polymers, techniques will also require extremely low roughness of the 2D materials (and thus growth substrates) at the large scale to ensure conformal contact with the pickup layer, as well as a relatively weak bond with the growth substrate. Various pathways to achieve this at scale have recently been demonstrated for multiple materials, for example growth of CVD graphene on ultra-flat Cu and subsequent oxidation to enable stamp transfer[6], or direct growth of weakly bonded graphene on polished $SiO_2$[7]. Such processes will need to be optimised and applied to a larger section of the 2DM library for commercial applications.

An alternative approach for both research and commercialisation is the automation of the manual stacking of small-area crystals. VdW heterostructures with unprecedented numbers of layers have already been fabricated by automated transfer systems using both automatically identified mechanically exfoliated crystals[8], and lithographically defined sections of grown thin materials[9]. Although interlayer contamination remains an issue, if the aforementioned transfer problems are reduced, automated stacking may provide sufficient throughput for specialised segments of the market, where extremely high quality but low batch sizes are required. Indeed, one automated system can output hundreds of heterostructures per year at a fraction of the cost required to run a silicon foundry, which is enough to fulfil the need for specific high-end optoelectronic devices in aerospace, military, quantum computing and medical fields and should not be overlooked.

**Concluding Remarks**

At present, the technological side of the field is stagnating. Eliminating transfer-induced inhomogeneity (both strain and interlayer contamination) is essential to progress and access the full potential of 2D materials. When such problems plague even bespoke research devices fabricated by leading experts in sophisticated cleanroom environments using 'clean' mechanically exfoliated 2D crystals, it is superficially challenging to see how these may be eliminated in an industrial setting.
A paradigm shift in the way 2D crystals are transferred is therefore required, with a transition away from polymeric carriers towards more mechanically, thermally, and chemically stable carrier surfaces. Ultra-high vacuum or fluid environments are a promising way forward [10], but the current reliance on polymeric support films must be overcome to enable their full potential.

**Acknowledgements**

*We acknowledge support from Royal Society, ERC Consolidator grant QTWIST (101001515), EPSRC grant numbers EP/V007033/1, EP/S030719/1 and EP/V026496/1, the European Graphene Flagship Project (881603) and the European Quantum Technology Flagship Project 2DSIPC (820378).*

## 8. Computational discovery and novel 2D materials

**Thomas Olsen[1], Johanna Rosen[2] and Kristian Sommer Thygesen[1]**

[1] *Technical University of Denmark, Denmark*
[2]*Linköping University, Sweden*

**Status**
First-principles calculations based on density functional theory (DFT) are increasingly used to guide and accelerate the search for new materials. In the field of 2D materials, a number of computational databases have been created to support this development and to provide an overview of known as well as hypothetical (potentially synthesizable) 2D monolayers. For example, the Computational 2D Materials Database (C2DB) [1] contains about 16k monolayers with a large variety of calculated properties (mechanical, electronic, magnetic, topological, linear- and nonlinear optical) while the 2D Materials Cloud [2] and the 2D Materials Encyclopedia [3] contain respectively 3k and 6k monolayers and their electronic band structures.

The strategies used to generate new 2D materials can be categorized as 'top-down' or 'bottom-up' approaches [XXX: We are working on a figure to illustrate this]. In the top-down approach one starts with a 3D bulk crystal and calculates whether single layers can be exfoliated from it either mechanically [2,4] (relevant for weakly van der Waals bonded layers) or chemically [5] (relevant when specific elements binding the layers together can be selectively etched away). The mechanical exfoliation approach facilitates predictions based on a straightforward calculation of the interlayer binding energy using a van der Waals (vdW) compliant exchange-correlation functional. The chemical exfoliation approach is more involved as it is necessary to have a representation of the chemical processes involved, including how to passivate the obtained surfaces by, e.g., O, F, or OH terminations. Large-scale computational studies of chemical exfoliation, such as MAX to MXene conversion, have identified stable 3D precursors [5b] by evaluating exfoliability based on Pourbaix diagrams (for electrochemical etching) and assessment of the exfoliation free energy (for conventional etching with imposed electroneutrality) [5c].

Several different 'bottom-up' approaches have also been employed. The simplest uses a known 2D monolayer as scaffold structure and generates new ones by substituting the atoms by chemically similar ones [1]. The crystal symmetry approach starts from an abstract space group (or more precisely layer group) and systematically decorates the Wyckoff sites by different elements [6]. In the intercalation approach one starts with a known homobilayer and inserts atoms into the vdW gap to produce new covalently bonded 2D crystals [7]. Recently, AI-based generative models trained on existing 2D materials databases have been successfully used to produce new stable monolayers [8]. A key question concerns the stability of the resulting structures. Here an important indicator is the *energy above the convex hull*, which is defined as the energy of the given structure relative to all other competing phases (including mixed phases) of the same chemical elements and stoichiometry. A hull energy close to zero indicates thermodynamic stability at low temperature and in the absence of any other chemical species. Considering that freestanding layers of 2D materials are typically not thermodynamically stable with respect to their 3D counterpart, the important convex hull indicator can be complemented by evaluation of properties and processes leading from 3D to 2D, in combination with assessment of the dynamical stability of the 2D derivatives.

Computational approaches have been pivotal for a number of new classes of 2D materials that have recently emerged or attracted renewed interest. These include high-entropy alloys (HEAs) of TMDs [9] and MXene [10] containing four or more metal elements with low mixing energy. The HEAs exhibit enhanced stability at elevated temperatures and show promise as e.g. (electro)catalysts. Natural van der Waals (vdW) bilayers, i.e. homobilayers in naturally occurring stacking configurations, are attractive due to their highly tunable and switchable properties arising from the presence of the vdW gap and their stacking/sliding degrees of freedom. Compared to twisted homobilayers and general vdW heterostructures, natural bilayers have better stability and are easier to produce. A systematic computational exploration of all natural bilayers resulting from stacking of 1k different monolayers was recently reported [11]. TMD homobilayers with native metal atoms filling a periodic pattern of sites in the vdW gap comprise a class of ultrathin, covalently bonded 2D crystals with stoichiometry dependent properties [12]. Computations show that self-intercalation generally promotes magnetism, metallicity, and basal plane reactivity [7]. For classes of 2D materials from non-vdW structures, theoretical predictions have facilitated development of 2D materials from selective etching, beyond MXenes. Through identified reduced interlayer interaction via alloying, and screening procedures targeting optimal compositions for chemical exfoliation, both a 2D metal boride, *boridene* [12b], and 2D sheets based on primarily Ru and O have been realized [5]. Both materials show promise as catalysts.

The discovery of new 2D materials with optimal properties for specific applications has largely been driven by high throughput computational screening studies. Such approaches usually take existing databases as a starting point [1,2,3] and apply workflows to compute properties systematically for thousands of materials [13a, 13b, 13c]. Magnetic order in 2D materials is arguably one of the most interesting properties with rather direct prospects of applications for 2D spintronics [14]. In this regard the most pertinent challenge is to find a compound that remains magnetic at room temperature [15a] and high throughput computations have been applied to compile lists of promising 2D magnets [15b, 15c] for room-temperature applications. Such studies comprise important seeds for further analysis of physical properties – for example multiferroic order [16] or altermagnetic band splitting [17]. Other classes of 2D materials that have been characterized systematically by first principles calculations include topological insulators [18], ferroelectrics [19a], superconductors [19b], thermoelectrics [19c], high mobility semiconductors [19d], catalysis [19e] and non-linear optical properties [20a, 20b].

**Current and Future Challenges**

It is evident that first-principles electronic structure codes in combination with big data analytics and AI models comprise a powerful tool for accelerated digital discovery of novel materials. Yet, a number of outstanding challenges must be overcome before its full potential can be leveraged and fully benefit the field of 2D materials. One is related to the development of deep generative models to autonomously create materials with prescribed properties. While methods like crystal diffusion, with proper training data, can generate stable and chemically meaningful structures [8], the problem of property conditional structure generation is still in its infancy. As computer power continues to increase and automated workflow management software and generative AI tools become more available, the computational throughput will grow tremendously. This situation entails a risk of the field being flooded by potentially interesting hypothetical materials. To mitigate this risk, it is crucial that computational discovery studies invoke more careful assessment of candidate materials. This includes a more serious evaluation of the synthesizability of proposed structures and their stability under realistic conditions, i.e. beyond the 'low formation energy' paradigm, as well as a characterization of the materials most basic properties. Another long-standing challenge in the field concerns the accurate modeling of twisted 2D materials and general van der Waals heterostructures because large number of atoms per unit cell of such structures (many thousands) simply precludes conventional first principles descriptions.

**Advances in Science and Technology to Meet Challenges**

The upcoming exascale supercomputers provide an opportunity for expanding computational studies to larger structures and more complex processes. However, as these supercomputers are GPU-based, the current simulation codes must be ported, which is a rather demanding endeavor. Emerging machine learning interatomic potentials will enable more realistic simulations of 2D materials including dynamics and growth processes. They may also permit determination of the detailed atomic structure of materials with thousands of atoms per unit cell, e.g. twisted Moiré structures. However, the potentials are currently limited to the prediction of total energies and forces and cannot provide electronic or magnetic properties, which are essential for many envisioned applications of these materials.

**Concluding Remarks**

The computational development of novel 2D materials is an exciting frontier in materials science, driven by advancements in computational techniques and high-throughput screening methods. By employing DFT and machine learning algorithms, one may predict the stability, properties, and potential applications of hypothetical 2D materials before experimental synthesis. However, challenges remain in accurately modelling the synthesizability, necessitating further collaboration between computational and experimental scientists.

Machine learning will undoubtedly play a crucial role for future materials discoveries, leveraging the vast amounts of computed data available in publications and databases.

**Acknowledgements**

T. O. was supported by VILLUM FONDEN (grant no. 29378). K. S. T. is a Villum Investigator supported by VILLUM FONDEN (grant no. 37789).

## 9. Electronics of twisted 2D materials

**Dmitri K. Efetov[1], Bjarke S. Jessen[2], Matthew Yankowitz[3]**

[1] Ludwig-Maximilians-Universität München, Germany
[2] Technical University of Denmark
[3] University of Washington, USA

### Status

Some of the most exciting and enigmatic phases of matter result from strong electron-electron interactions or topological electronic bands. Recently, moiré systems have become key platforms for studying the interplay between these effects, drawing from a wide range of two-dimensional (2D) van der Waals (vdW) materials, which include single and multilayers of graphene and various semiconducting transition metal dichalcogenides.

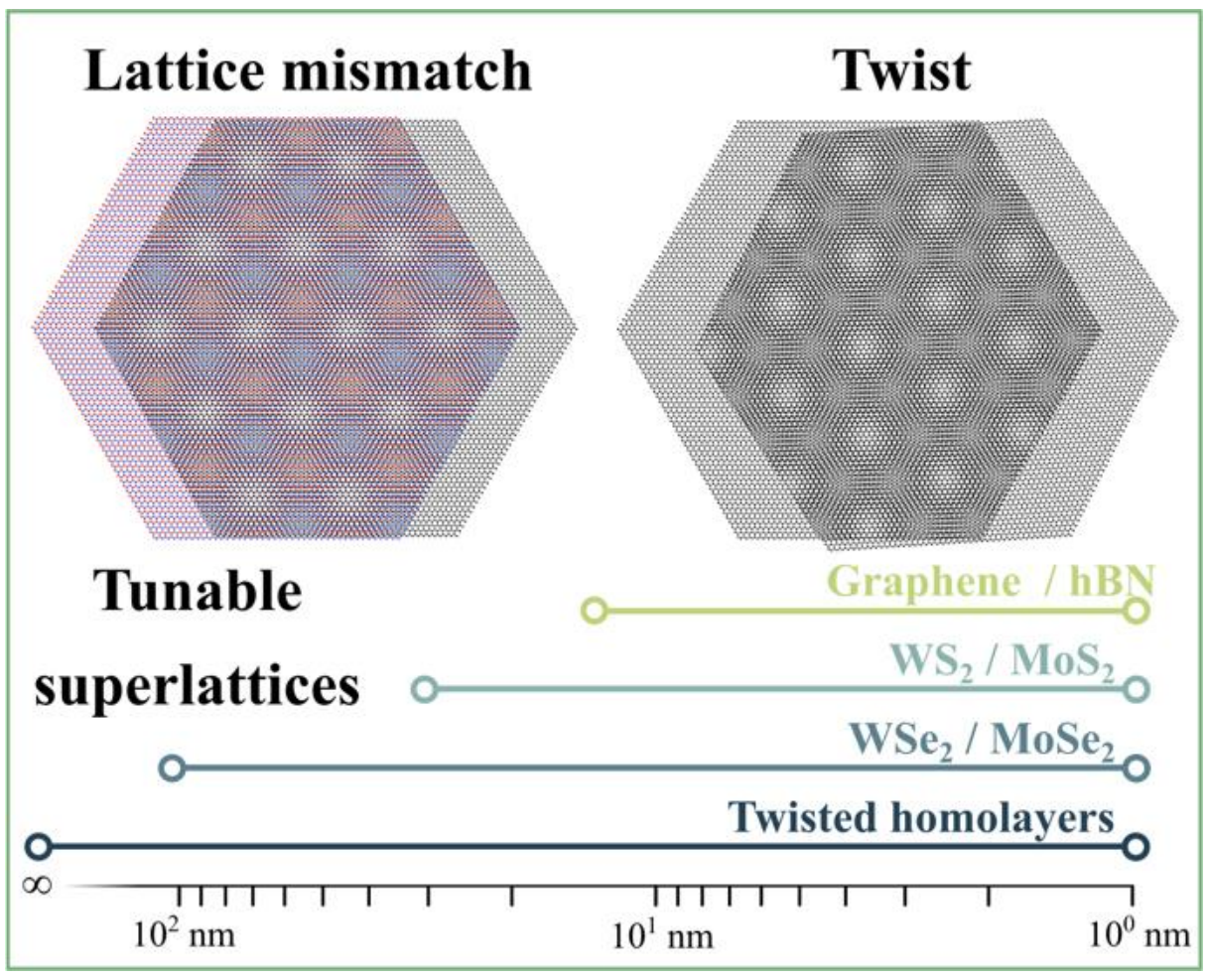

*Figure 3- Cartoon schematic of moiré pattern formed by stacking atomically thin crystals with either a lattice mismatch or a relative twist angle. The characteristic length scales of different types of moiré materials are shown below, varying depending on the relative twist angle. "Homolayers" indicate structures in which the same material is twisted upon itself (e.g., twisted bilayer graphene).*

A unique aspect of 2D materials lies in the ability to directly modify their electronic properties by layering one atomically thin crystal atop another, which allows for their surface-exposed electronic wavefunctions to strongly overlap. When stacked vdW crystals have a small lattice mismatch or relative twist angle, a "moiré pattern" arises, creating a tunable artificial superlattice potential many times larger than the constituent atomic lattices (Fig. 1). Under suitable conditions this can generate extremely flat electronic bands with quenched kinetic energy, leading to the emergence of strongly correlated phases with broken symmetries including correlated insulators, generalized ferromagnets, and unconventional superconductors. In addition, the moiré-induced mini-Brillouin zone can have inherently altered Berry curvature compared to its constituent materials, giving rise to nontrivial topology and quantum geometry.

Early work relied on scanning tunneling microscopy to establish the existence of moiré patterns in twisted graphene structures and angle-aligned graphene on hexagonal boron nitride (hBN) [1]. In 2013, the graphene/hBN moiré lattice was shown to host Hofstadter butterfly physics in a magnetic field, arising due to the interplay between comparable electronic and magnetic length scales [2]. The field was subsequently supercharged by the 2018 discovery of unconventional superconductivity and correlated insulating states in “magic-angle” twisted bilayer graphene (MA-TBG) [3], [4], comprising two sheets of monolayer graphene twisted by the so-called “magic” angle of approximately 1.1° (resulting in a superlattice period of ≈10 nm). Since then, interest in moiré materials has grown rapidly. Researchers from around the world remain fascinated by the countless new quantum states of matter arising in moiré materials, and the field has grown to encompass well over a dozen distinct moiré platforms featuring a wide range of exciting physics and unexpected new discoveries each year.

**Current and Future Challenges**

A vast array of exotic quantum states are possible within the framework of moiré quantum matter, and many current and future challenges lie in discovering, cataloging, understanding, and manipulating these states. A notable example is superconductivity, which was first discovered in magic-angle twisted bilayer graphene but has since been found in a larger family of twisted graphene multilayers, and in twisted bilayer $WSe_2$. Numerous experiments have probed these superconducting phases, including studies of potential rotational symmetry breaking, electrostatic Coulomb interaction screening, macroscopic quantum coherence in Josephson junctions and SQUID devices, and microscopic studies of the superconducting gap [5]. These experiments collectively point to unconventional superconductivity in twisted graphene, with signatures of nematicity and anisotropic pairing symmetries. However, key open questions remain, including the roles of electron-phonon and electron-electron interactions in Cooper pair formation, and whether common mechanisms underlie superconductivity across various twisted graphene and TMD moiré systems.

Many other fingerprints of strong correlations have been observed, including ‘strange-metal’ behavior in magic-angle graphene and signatures of tunable Kondo lattice formation in aligned $MoTe_2/WSe_2$ heterostructures. Gapped states at commensurate moiré band dopings, typically associated with a spontaneous lifting of spin/valley degeneracies or broken discrete translational symmetry, are also common. Accurately predicting these phases across different moiré platforms, and unambiguously determining their ground state ordering, often remain as outstanding challenges. For example, many of the features seen in MA-TBG are captured within a topological heavy fermion description, characterized by the coexistence of itinerant and localized charge carriers within the flat band. However, developing a complete understanding of the resulting correlated states and their connection to superconductivity is currently work in progress.

The combination of intrinsic Berry curvature and spontaneously broken time-reversal symmetry leads to new forms of electronic ordering with intertwined topological and correlated properties. Numerous twisted graphene and TMD moiré systems support the integer quantum anomalous Hall (QAH) effect, driven by interaction-induced spin and orbital magnetism that opens a gap between otherwise degenerate Chern bands. Remarkably, the fractional QAH effect has recently been discovered in twisted $MoTe_2$ and in moiré lattices of rhombohedral-stacked multilayer graphene aligned with hBN https://www.zotero.org/google-docs/?bEWH3l[6], [7]. The nature of these fractional QAH states and their connection to fractional quantum Hall states in Landau levels is a topic of intense interest.

Alongside the emerging physics, several overarching challenges face the field. A major experimental hurdle is reliably reproducing states sometimes seen in only a single sample https://www.zotero.org/google-docs/?DdauSu[8], and scaling these effects to wafer-scale device arrays. Key issues include the extreme sensitivity of many-body ground states to precise twist angles, dielectric environments, and unintentional strains formed during the mechanical device assembly [5]. Theoretically, several challenges persist in modeling strongly correlated systems. These include accurately calculating band structures for moiré materials with thousands of atoms per unit cell, distinguishing between nearly degenerate ground states of strongly interacting systems, and developing new paradigms for understanding correlation-driven topological phases of matter.

**Advances in Science and Technology to Meet Challenges**

Currently, state-of-the-art moiré devices are made by exfoliating bulk crystals down to atomically thin layers with Scotch tape, then stacking flakes at controlled twist angles using a dry transfer technique. However, this strategy yields only a single micrometer-scale sample at a time, and is susceptible to introducing large in-plane strains that can substantially change the desired physics of the structure. The development of robotic exfoliation and stacking systems equipped with machine-learning–enabled flake detection and control may provide a pathway towards increasing the yield of this fabrication technique. Additionally, further development of techniques to directly manipulate the stacking of 2D vdW flakes by pushing them with an atomic force microscopic (AFM) tip may be useful for constructing strain-free devices with on-demand twist angles [9]. Efforts to scale these systems to wafer scale will likely require new synthesis and transfer techniques based on chemical vapor transport or molecular beam epitaxial growth of large-area single crystals, paired with advanced robotic transfer and device assembly technology.

In parallel, the ongoing development of advanced characterization and modeling techniques will be critical for future progress. Experimentally, the field would benefit from further development of cryogenic nano-ARPES to probe the evolution of the spectral weight of moiré systems as a function of doping and other key tuning parameters. Advanced terahertz spectroscopy techniques are also valuable for probing the low-energy physics of the flat moiré bands. Microscopic probes operating at millikelvin temperatures will be informative, enabling direct connections between the atomic- and moiré-scale properties of these materials with their mesoscopic phenomena. These will likely include advances in scanning techniques including tunneling, compressibility, magnetization, and optical spectroscopy. The quantum twisting microscope represents an emerging frontier in the study of moiré quantum matter [10], in which vdW layers adhered to an AFM tip can be deterministically rastered across and twisted upon a separate vdW stack, all while various quantities are measured.

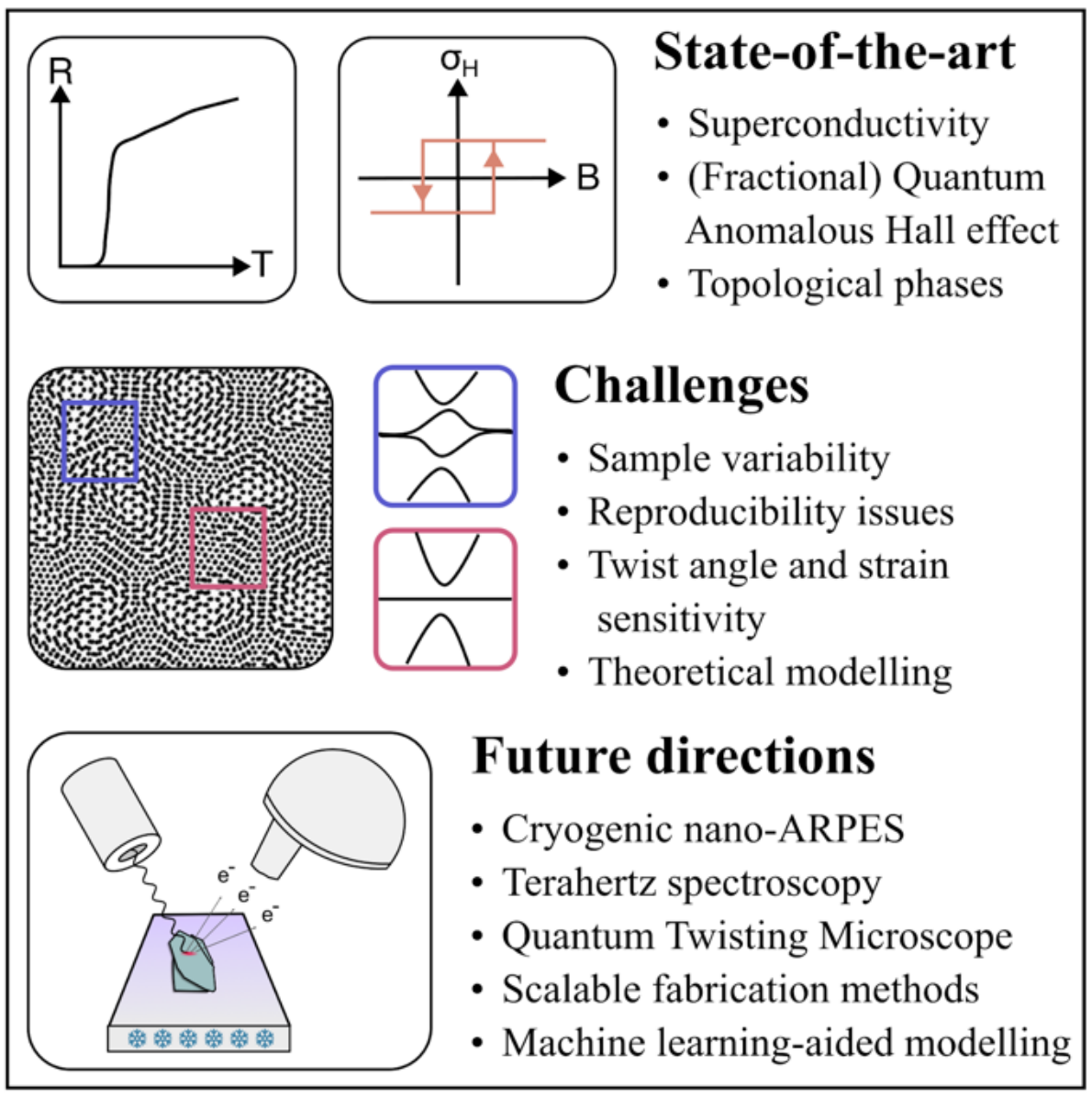

*Figure 4 - Cartoon illustrations summarizing the state-of-the-art of the field of electronic twisted materials, key challenges facing the field, and likely future directions.*

Additionally, it will be fruitful to further develop techniques for directly controlling the properties of moiré materials, including the use of pressure and uniaxial strain to control the effective Hamiltonian of the system, and the integration of metallic screening layers near the moiré interface to control the strength of Coulomb interactions. Fabricating lateral junctions between different emergent phases presents new opportunities for deterministically engineering exotic quantum states of matter. For example, interfacing superconductivity with integer or fractional QAH states is predicted to generate Majorana or parafermion modes carrying non-Abelian excitations, which may be useful for the development of topological qubits. Further manipulating the resulting topological boundary modes with electrostatic gates and quantum point contacts will be needed in order to detect and braid these exotic quasiparticles.

Lastly, new theoretical paradigms are needed for improving the modeling of strongly correlated and topological states of matter, and for answering key open questions facing the field. Understanding the critical role of quantum geometry in setting the properties of strongly correlated and unconventional superconducting phases lies at the frontier of condensed matter theory. Additionally, fractional Chern insulator states arising at zero magnetic field present new opportunities to understand the microscopic properties of quasiparticles with fractionalized charge excitations and anyonic quantum exchange statistics.

**Concluding Remarks**

Moiré materials are currently at the forefront of the study of intermixed correlations, topology, and superconductivity in condensed matter physics owing to their extraordinary versatility. Future progress in this field necessitates a multifaceted approach to improving and scaling sample fabrication, developing new techniques for characterizing and directly controlling their properties, and advancing theoretical capabilities for better understanding and more efficiently predicting new physics in existing and emerging moiré materials (Fig. 2). Beyond fundamental discoveries, these designer quantum materials hold promise for the development of future electronics, including as ultra-sensitive quantum detectors and in various forms of quantum information storage and processing devices.

**Acknowledgements**

M.Y. acknowledges support from the Department of Energy, Basic Energy Science Programs under award DE-SC0023062. B.S.J. acknowledges support from the Novo Nordisk Foundation grant NNF23OC0084494. D.K.E. acknowledges funding from the European Research Council (ERC) under the European Union's Horizon 2020 research and innovation program (grant agreement No. 852927), the German Research Foundation (DFG) under the priority program SPP2244 (project No. 535146365), the EU EIC Pathfinder Grant "FLATS" (grant agreement No. 101099139) and the Keele Foundation.

## 10. Moiré nano-optoelectronics

**Julien Barrier[1], Roshan Krishna Kumar[1] and Frank HL Koppens[1]**

[1]ICFO – The Institute of Photonic Sciences, Castelldefels 08860, Barcelona, Spain

### Status

*Advent of moiré materials*. The possibility to stack 2D crystalline layers with nearly perfect interfaces has revolutionised the design of electronic and nanophotonic materials. Van der Waals heterostructures offer virtually limitless possibilities for materials creation as, unlike traditional growth techniques, material selection is free from constraints imposed by lattice mismatch. Such freedom allows for the creation of electronic and photonic phases through modulation of spin-orbit coupling, exciton amplitudes, energy shifts or scattering rates. The range of achievable quantum phases expands with additional tuning routes offered by electrostatic gating, electric or magnetic fields, temperature, strain engineering, etc. As 2D heterostructures fabrication grew, rotational stages were developed to align crystallographic axes of two different crystals, resulting in superlattice potentials with length scales of 10-20 nm and observations of phenomena such as replica of Dirac fermions, Hofstadter gaps and Brown-Zak fermions. This technique could later be used for precise control over the twist angle between two or more homolayers. In that case, the twist angle defines the size of the band-folding Brillouin zone and possible inter-layer hybridization, enabling control of the ratio between interaction strength and electronic bandwidth. Such 'moiré' materials stand out for their exceptionally tuneable band structures and their ability to host a wealth of novel physical phases, including correlated insulators, superconductors, orbital ferromagnets, Wigner crystal states, strip phases, topological multiferroic orders, moiré excitons, and bosonic exciton crystals to name a few [1], [2], [3]. These unique properties make twisted moiré materials ideal candidates for easy-to-fabricate quantum simulators, with all known correlation-induced symmetry-broken phases identified in these materials. However, their exact nature and underlying mechanism remain elusive.

*Optoelectronic probes of moiré materials*. Since the early days of graphene research, optical probes have been used for characterization: Raman spectroscopy, for example, allows for identification of stacking order and layer number of various 2D materials. Over time, both free space and near-field techniques were incorporated to gain insights into e.g., carrier energy relaxation dynamics, carrier transport, bandgap, exciton energy, Berry curvature and heat transport [3], [4]. For example, scanning photocurrent experiments, achieved using near-field optical microscopes (SNOM), allow for the observation of the plasmonic response in moiré systems, for the extraction of parameters such as the local Seebeck coefficient or for visualization of plasmons. For example, near-field photocurrent studies of minimally twisted graphene bilayers revealed that plasmons reflect at domain boundaries, shedding light on the local strain profile and displacement of the graphene layers [4]. Similarly, photocurrent nanoscopy experiments revealed inter-band plasmons [5] in graphene bilayers or correlated insulating states in twisted trilayers [4]. Notably, the plasmons in graphene can also be leveraged to observe phenomena in proximity layers such as moiré ferroelectricity [6].

By probing the current response of a moiré material exposed to light, photocurrent measurements are another versatile tool to sense the properties of electronic states, Bloch band quantum geometry, quantum kinetic processes and device characteristics [4]. Two prominent quantum geometric photocurrents arise from inter-band transitions: injection and shift photocurrents. These exhibit a sensitive dependence on the crystalline symmetries, parameters that are readily controlled within moiré heterostructures. The bulk photo-galvanic effect (BPGE) is an example of such phenomenon emerging from broken symmetries, which offers valuable insights into the dependence of the quantum geometry on e.g., the displacement field, Fermi level. With their exceptional tunability, large unit cells, topological properties and large density of states, moiré systems represent an ideal system to study giant bulk photocurrents. The complex structure of photocurrent as a function of polarization-helicity and gate voltage arising from moiré minibands allows probing symmetry broken states and quantum geometry, particularly when assisted by strong electron-electron interactions [7].

*Moiré optoelectronic devices.* In terms of optoelectronic device design, moiré materials present strong light-matter interactions and large photo-response, positioning them as prime candidates for photodetection and photoconversion applications. For these applications, the exceptional tunability of their spectrum as well as the energy scale of relevant optical transitions present advantages. By simply adjusting the twist angle, it becomes possible to control the wavelength from ultraviolet down to the THz range. For twist angles above 1°, bandgaps of 50-100meV facilitate photo-induced transitions throughout the infrared (IR) range, enabling broadband light detection [8] and advanced spectroscopy [9]. For smaller twist angles, the formation of flat minibands with a high density of states significantly enhances the dynamical conductivity, leading to a strong photo-response. In this regime, correlated insulating gaps and bandwidths are typically around 10 meV, leading to resonant inter-band transitions with detection capabilities in the far-IR and THz energy ranges. Equally, moiré materials hold promise for the exploration of previously unattainable ranges in single-photon detection, owing to the emergence of superconductors at ultra-low carrier densities, which allow low-energy single photons to cause a sufficient temperature increase to induce phase slips. Such phenomenon is expected to offer the possibility for single-photon detection in the mid-IR and THz ranges with exceptionally fast response times and high energy resolution [10]. In addition, the ability to engineer symmetry breaking in twisted 2D materials opens doors for symmetry breaking optoelectronics, where groundbreaking effects like the BPGE and opto-valley Hall effects can be engineered with potential applications in chiral light-emitting diodes or moiré quantum emitters [3], [4].

*Beyond optoelectronics: moiré nanophotonics*. Besides their potential for optoelectronic devices, moiré materials are sparking significant interest due to the possibility of engineering entirely new types of photonic excitations, including non-reciprocal plasmons, quantum geometric plasmons, moiré excitons, resonantly hybridised excitons, and other reconstructed collective excitations [2], [3]. Near-field techniques, in particular, offer promising avenues for exploring various light-matter coupling mechanisms within these materials. Since the properties of excitons are directly linked to the underlying Bloch states, they can serve as sensitive probes for the potential landscape of twisted materials. For example, in twisted transition metal dichalcogenides, the spatially displaced wave functions of electron-hole pairs forming excitons (at K and K' moiré sites) allow for optical transitions with properties similar to conventional bilayer excitons but tuneable energies, and spin-layer selectivity [2]. Moiré materials can also host more complex excitonic states such as moiré-trapped interlayer trions, satellite moiré exciton peaks and non-trivial many-body states involving interlayer charge transfer excitons and interlayer moiré excitons with, e.g., a hole wavefunction surrounded by three electron wavefunctions. The long lifetimes and strong dipolar interactions in moiré materials can also lead to the formation of correlated excitonic states. These states hold promise as sensitive probes for the correlated phases observed in neighbouring systems. They also offer a tuneable platform to access various exotic quantum phases, including non-interacting Bose gases, superfluid phases, quadrupole and dipolar excitations, and electron-hole plasmas.

**Current and Future Challenges**

Despite significant research, fundamental challenges remain in moiré optoelectronics. Understanding the nature of moiré correlated phases is one of them, which requires the combination on various techniques. Optoelectronic techniques hold promise to reveal properties invisible to conventional electron transport and scanning probes. However, extending purely optical methods across a wide frequency range would face hurdles like low signal-to-noise ratio and substrate contributions as moiré materials are smaller than IR and THz wavelengths. It then becomes crucial to upscaling moiré materials for practical photodetection applications. In basic science, photocurrent measurements offer a potential alternative to circumvent these limitations.

Additional challenges in practical devices originate from unwanted reflections from gates and need for careful integration with existing technology to avoid compromising performance. Going further, integration with waveguides, cavities, plasmonic structures or ring resonators could create a vast array of high-performance optoelectronic devices such as single-photon emitters or optical modulators.

Unlocking new functionalities also presents exciting possibilities. Trapping interlayer excitons using periodic strain profiles or electrostatics could lead to tuneable quantum light sources, photon-induced gating and Purcell enhancement for enhanced light-matter interactions. Promising tuning knobs could include ultrafast optical excitations, electric and magnetic fields, pressure and Floquet engineering, although not yet demonstrated.

**Advances in Science and Technology to Meet Challenges**

An important challenge lies in the improved fabrication methods. Both device applications and fundamental research calls for the necessity to fabricate large, clean and reproducible moiré devices.

Assembling atomically thin crystals on top of one another in unfavourable stacking configurations has proved a challenging task, and naturally places severe doubts on the upscaling potential. Nonetheless, researchers have already demonstrated the capabilities to grow twisted heterostructures using chemical vapour deposition (CVD), which marks an important milestone towards device applications based on moiré materials [11]. However, single-crystal domains so far remain small (10 µm). In addition, one should be able to stack different layers similar to tandem solar cells to improve low-power photodetectors while maintaining the desired properties.

**Concluding Remarks**

The study of moiré materials at the interface between condensed matter physics and photonic techniques presents a fertile ground for fundamental advances. Probing the band structure, their interactions with valleys and quantum metrics through advanced techniques hold immense potential to unlocking novel functionalities. The unique combination of ultra-strong optoelectronic response and precise control over energy scales in moiré materials represents a paradigm shift and promises a future brimming with groundbreaking applications and transformative technologies.

**Acknowledgements**

J.B. acknowledges support from the European Union's Horizon Europe program under grant agreement 101105218. R.K.K. acknowledges the FLAG-ERA program (PhotoTBG).

## 11. Polaritonics of 2D materials

**Hui Deng[1], Xiaoqin Li[2], Siyuan Dai[3] and D.N. Basov[4]**

[1] University of Michigan, USA
[2] University of Texas at Austin, USA
[3] Auburn University, USA
[4] Columbia University, USA

### Status

Polaritons are omnipresent in van der Waals (vdW) materials. A polariton is a quantum mechanical superposition of a photon with a matter excitation: plasmons, phonons, excitons, magnons, to name a few [1, 2]. Polaritons can be readily implemented in vdW monolayers, bulk crystals; vdW materials can be readily integrated in different forms of electromagnetic/optical cavities and meta-structures promoting strong and even ultra-strong light matter interaction. A combination of the unique physical properties and reduced dimensionality enable a number of nanophotonic virtues including: the atomic-scale localization, exotic nano-light propagation, dynamic and versatile tunability. In addition, a few transition metal dichalcogenide monolayers, heterostructures, and cavities (Fig. 1a) have been used to create exciton-polaritons at room temperature with interesting valley properties (Fig. 1b-c).

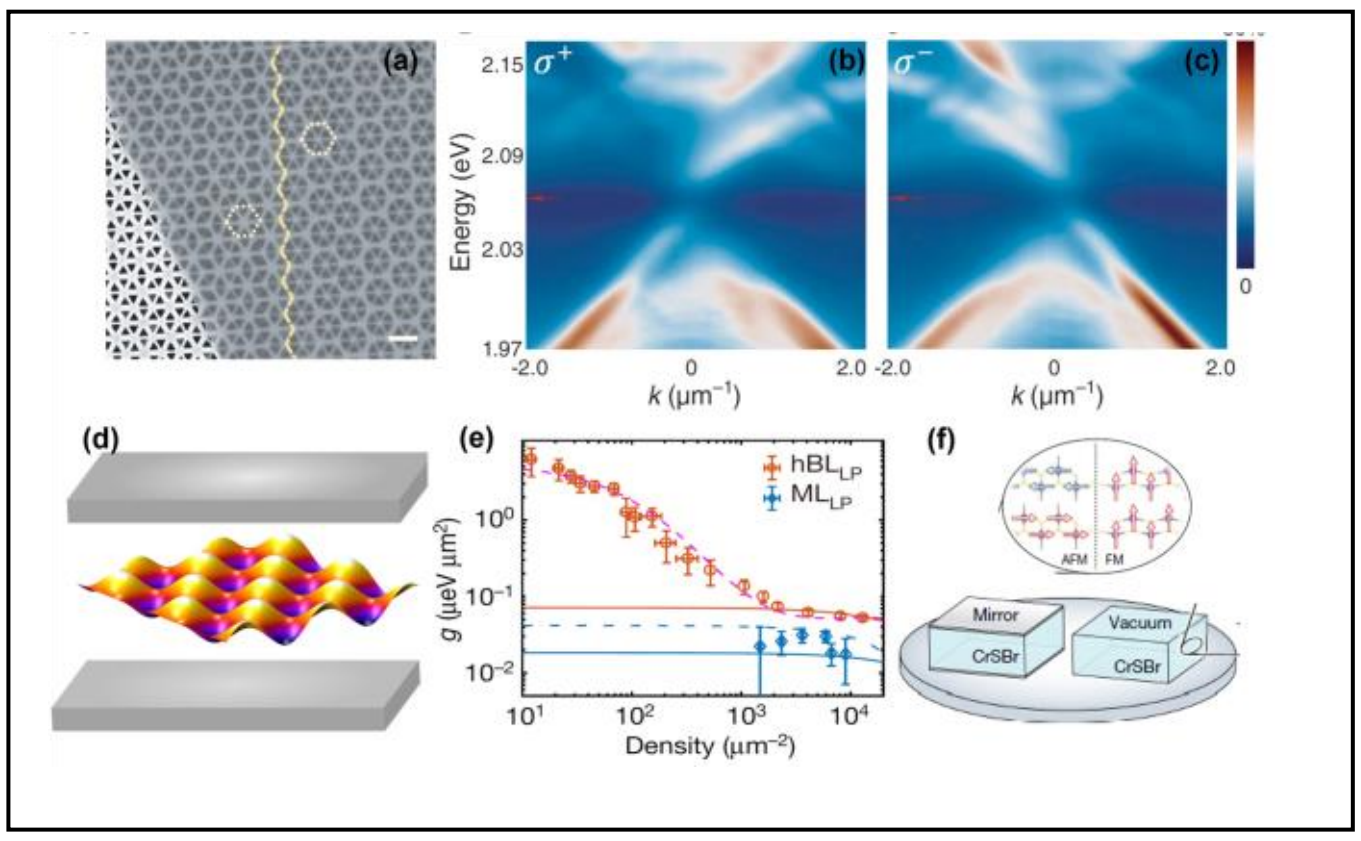

**Figure 1.** (a) Two spin Hall photonic crystals supporting edge states in-between. (b-c) Polariton dispersions showing helical edge states. Adapted from [3] (d-e) Moiré lattice excitons in a cavity, forming polaritons with enhanced nonlinearity. Adapted from [4]. (f) Schematics of bulk CrSBr crystals supporting polariton modes, with polariton energies correlated with the antiferromagnetic and ferromagnetic orders

One prominent advance of 2D materials is that different atomic layers can be easily stacked into heterostructures for exotic yet desired properties not existing in nature. These advances include low-loss, hybrid nano-light, phase-changing behaviors, charge-transfer nano-light, and nano-light photonic crystal and bandgap, etc. In addition, various 2D materials can be easily transferred onto desired substrates, including patterned ones without degradation of their inherent properties. It is possible to form exciton-polaritons by directly placing 2D materials on 2D photonic crystals[3] and to tailor the properties of the polaritons by design of the photonic crystals, such as forming helical edge states of polaritons on a spin Hall photonic crystal[4]. Another emerging frontier is related to 2D magnets, where exciton-polaritons have been reported in 2D magnets NiPS3 and CrSBr [5], and the polariton properties become correlated with the magnetization in the crystals as illustrated in Fig. 1f.

In addition to stacking, heterostructure component layers can be twisted and form moiré superlattices. Twisted heterostructures produce plasmonic nano-light crystal, wavefront engineering, nano-light canalization, and topological transitions. Furthermore, moiré supercell sizes and electronic band structures are readily tunable by the twist angle between two stacked layers. Each moiré cell confines the excitons on the nanometer scale, effectively forming a quantum dot, giving rise to a quantum dot lattice with homogeneity and density beyond the reach of any other existing methods. When the moiré superlattice is integrated into a cavity, the resulting moiré polaritons retain the large oscillator strength of 2D excitons while acquiring strong nonlinearity typical of 0D excitons (Fig. 1d-e), circumventing the conditional trade-off between light-matter coupling strength and polariton nonlinearity [5].

It is worth noting that a series of propagating polaritons in 2D materials have been imaged in real space using scattering-type scanning near-field optical microscopy (s-SNOM, Fig. 2). The nanoscale sharp metallic tip in the s-SNOM experiments (Fig. 2a-b) transfer the energy and bridge the momentum mismatch between photons (straight arrow) and polaritons (wave arrow), thereby launching the polariton waves whose standing wave interferences can be imaged as fringes. Representative works include the imaging of phonon polaritons in hBN (Fig. 2c), surface plasmon polaritons in graphene (Fig. 2d), exciton waveguides dynamics in TMDs (Fig. 2e), and moiré plasmon-phonon polaritons in hBN encapsulated twisted bilayer graphene (Fig. 2f).

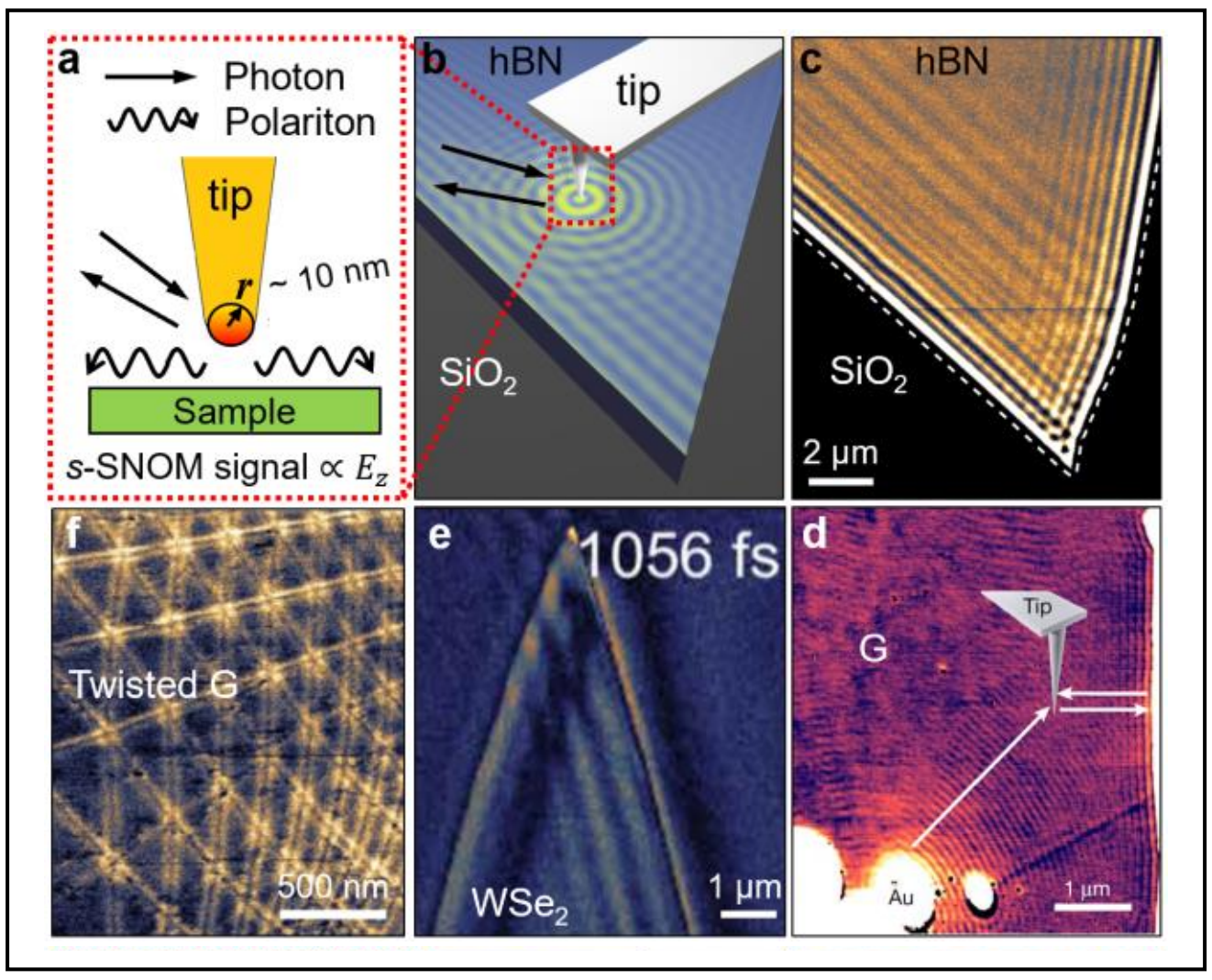

Figure 2. Real-space Imaging of polaritons in 2D materials using the scattering-type scanning near-field optical microscopy (s-SNOM). (a), The schematic of s-SNOM. (b) The tip and edge can launch polaritons whose standing wave fringes will be imaged in the s-SNOM scan. (c) – (f), Nano-optical images of phonon polaritons in hBN (c), plasmon polaritons in graphene (d), exciton waveguides in TMD (e), and moiré plasmon-phonon polaritons in hBN encapsulated twisted bilayer graphene (f).

### Current and Future Challenges

Integrating advanced photonic structures with high-quality 2D optical materials and heterostructures in a scalable fashion is a major challenge. Demonstrations of high optical quantum yield, long coherence length or coherence time of the matter excitations integrated with cavities have mainly been achieved with manually prepared 2D heterostructures with lateral dimensions of a few microns.

Additional theoretical developments are needed to describe vacuum properties in the ultra- and deep-strong coupling regime, to describe collective phenomena when long-range Coulomb interactions exist in hybrid polaritonic systems, or to describe strong coupling of light with different types of excitations simultaneously.

A unique feature of vdW materials is that their thickness can be easily controlled, reaching the ultimate quantum confinement limit of a monolayer. Photonic cavities, on the other hand, typically have a mode volume that far exceeds the thickness of atomically thin vdW materials. This mismatch in dimensions is a commonly encountered challenge.

**Advances in Science and Technology to Meet Challenges**

The current challenges can be addressed through iterative advancements in the growth and fabrication technologies of 2D materials, the development and measurement of novel polariton devices incorporating advanced materials and structures, and the theoretical prediction and analysis of properties associated with these continuously emerging devices and phenomena. For example, additional theoretical developments are essential to accurately describe vacuum properties within the ultra-strong and deep-strong coupling regimes, to model collective phenomena arising from long-range Coulomb interactions in hybrid polaritonic systems, and to explore the simultaneous strong coupling of light with various types of excitations.

**Concluding Remarks**

Recent progresses are the first steps to create new types of polaritons based on advanced vdW materials and heterostructures. The field will rapidly evolve with improved understanding of the new vdW heterostructures and superlattices. Future research with these and new 2D material polariton systems will leverage their novel properties to explore quantum and collective phenomena previously difficult to access, such as polariton quantum blockade, fractional quantum hall states of polaritons, and integrated polariton circuits of polaritons—in device configurations and at operating temperatures that are suitable for practical technological applications.

**Acknowledgements**

*Li gratefully acknowledges funding from NSF EECS-2130552.*

## 12. 2D semiconductor field-effect transistor and 3D integration

**Xinran Wang[1,2], Saptarshi Das[3,4,5,6], Xiangfeng Duan[7,8], Zhihao Yu[2,9]**

[1]School of Integrated Circuits and Interdisciplinary Research Center for Future Intelligent Chips (Chip-X), Nanjing University, Suzhou, China
[2]Suzhou Laboratory, Suzhou, China.
[3]Department of Engineering Science and Mechanics, The Pennsylvania State University, University Park, PA, USA
[4]Department of Electrical Engineering, The Pennsylvania State University, University Park, PA, USA
[5]Department of Materials Science and Engineering, The Pennsylvania State University, University Park, PA, USA
[6]Materials Research Institute, The Pennsylvania State University, University Park, PA, USA
[7]Department of Chemistry and Biochemistry, University of California, Los Angeles, Los Angeles, CA, USA
[8]California NanoSystems Institute, University of California, Los Angeles, Los Angeles, CA, USA
[9]School of Integrated Circuit Science and Engineering, Nanjing University of Posts and Telecommunications; Nanjing 210023, China.

Email: Xinran Wang (xrwang@nju.edu.cn), Saptarshi Das (sud70@psu.edu), Xiangfeng Duan (xduan@chem.ucla.edu), Zhihao Yu (zhihao@njupt.edu.cn)

The atomically thin channels of 2D semiconductors (in particular, transition-metal dichalcogenides) enable ideal switching behavior at nanometer channel length and provide the ultimate material solution for device miniaturization. The weak van der Waals interaction allows for seamless monolithic three-dimensional (M3D) integration with Si and other materials to address the limitation silicon-based approaches in terms of density, interconnection and multi-function. The ultimate device scaling and M3D integration are the unique advantages for 2D semiconductors in reshaping future transistor technology.

Since the first $MoS_2$ field-effect transistor (FET), 2D semiconductor transistors have achieved remarkable progress over the past decade.[1-3] At device level, developing technologies that simultaneously satisfy performance, power, and area (PPA) requirements is crucial, which demands co-optimizations of contacts, gate dielectrics, doping and parasitic capacitance while scaling down device dimensions. Effective switching behavior at gate lengths as short as 0.3 nm has been demonstrated[4], [5] showcasing the scaling potential of 2D semiconductor. However, advanced node transistor scales contacted gate pitch (CGP) all together, where both channel and contact length need to meet the stringent requirements. Recently, the semimetal has been shown to suppress metal-induced gap states (MIGS) and the Fermi level pinning in 2D semiconductor contacts.[6, 7] Building on this and orbital hybridization, state-of-the-art semi-metallic Sb ($01\bar{1}2$) contact now achieves $R_c$ = 85 Ω·µm at $L_c$ = 20 nm, lower than Si transistors. [8], [9] By simultaneous scaling channel and contact, the first 1 nm node FET with CGP = 40 nm was successfully demonstrated whose key PPA metrics meet the IRDS target.[9]

Achieving small effective oxide thickness (EOT) in a gate-all-around (GAA) geometry is essential to suppress short-channel effects in scaled transistors. Due to the absence of dangling bonds, direct atomic layer deposition (ALD) is challenging. By introducing organic or inorganic buffer layers, sub-1 nm EOT and 60 mV/dec switching have been demonstrated in 2D semiconductor FETs by ALD.[10-14]

van der Waals stacking is another interesting approach. By transferring high-k dielectric nanosheets - such as single-crystal $Al_2O_3$, $SrTiO_3$, and $GaF_2$ - on to 2D semiconductor channel, transistors with scaled EOT and pristine interface have also been demonstrated.[15-17]. Beyond planar devices, the industry has been placing significant emphasis on advancing technologies for 2D semiconductor GAA devices. Leading organizations such as IMEC, TSMC, and Intel have unveiled their manufacturing process flows for GAA devices and conducted systematic simulations.[18-20] A critical challenge remains achieving uniform deposition of the surrounding dielectric layer. TSMC presented the first Demonstration of a GAA Monolayer-$MoS_2$ Nanosheet nFET, achieving 410 μA/μm $I_D$ at 1V $V_{DD}$ with a 40nm gate length.[19]

The integration of 2D materials into 3D architectures has emerged as a groundbreaking advancement in semiconductor technology, addressing the limitations of traditional silicon-based approaches. Furthermore, monolithic 3D integration (M3D) offers significant advantages over traditional packaging approaches by enabling seamless vertical stacking of devices on a single substrate without the use of through-silicon vias (TSVs) or microbumps[21]. Unlike packaging methods, which rely on stacking pre-fabricated chips and introduce parasitic capacitances and interconnect delays, monolithic integration employs ultra-dense inter-tier vias and low-temperature processes to achieve higher vertical interconnect density and reduced signal delay. This approach not only minimizes the physical footprint but also enhances energy efficiency and performance by allowing transistor-level partitioning and closer proximity of stacked layers. Furthermore, monolithic integration supports the incorporation of diverse materials and functionalities—such as logic, memory, and sensing—within a unified 3D architecture, making it ideal for advanced computing applications and energy-efficient systems.
Recent advancements in M3D have achieved remarkable progress, including wafer-scale fabrication of 2D material-based FETs using $MoS_2$ and $WSe_2$.[22, 23] These breakthroughs feature multi-tier stacking with low thermal budgets (<200°C), preserving the integrity of lower layers while enabling diverse functionalities such as logic operations, memory, and sensing. CMOS circuits, such as inverters and NAND gates, have been successfully realized using complementary n- and p-type FETs in vertically stacked configurations.[24] Additionally, heterogeneous M3D integration of graphene-based sensors with $MoS_2$ memtransistors has enabled near-sensor computing, supported by ultra-dense interconnects with 62,500 I/O per $mm^2$. Expanding this paradigm further, a six-layer 3D nanosystem integrating transistors and memristors has demonstrated AI task execution.[25] Collectively, these advancements highlight the scalability and versatility of 2D materials in redefining semiconductor architectures and capabilities.

Despite these promising developments in transistor technology and M3D, several challenges must be addressed to transition from experimental demonstrations to industrial-scale adoption. One of the most pressing issues is the development of high-performance p-type transistor, which currently lag significantly behind their n-type counterparts. Ohmic contact stability remains a key bottleneck, as demonstrated n-type contacts, despite achieving contact resistance lower than silicon-based devices, still lack verification of long-term reliability. Developing high-quality, ultra-thin dielectric interfaces with minimal EOT and low interface states for GAA and CFET architecture is another hurdle. Similarly, transfer processes can introduce contamination, structural defects or strain, further compromising device reliability.[26] System-level challenges, including self-aligned processes, reliability, and design-technology co-optimization (DTCO) are still in their infancy, necessitating increased research focus and investment. [27]

For monolithic 3D integration, the inter-tier challenges such as electrical isolation, topography management, and thermal dissipation become increasingly critical as the number of stacked layers grows.[28] Heat buildup in dense 3D architectures poses a significant risk to performance and reliability, necessitating innovative solutions like heat spreaders, thermal vias, and materials with high thermal conductivity. Additionally, achieving a smooth planarized surface is essential to reduce topography-related impacts on top-tier device fabrication, ensuring layer uniformity and alignment. Device-to-device variability further exacerbates these issues, particularly in high-density circuits where precision is paramount. Finally, achieving dense and reliable interconnects while maintaining structural integrity and ensuring compatibility with back-end-of-line (BEOL) processes adds to the complexity of scaling efforts, requiring holistic solutions across material, process, and architectural dimensions.

On the design side, reliable technology computer aided design (TCAD) model, process design kit (PDK) and electronic design automation (EDA) tools for 2D semiconductors are needed to transition from individual device and small-scale systems to very-large-scale integration.[29-31] We need to better control material defects and process variations and come up with variation-aware design methodology to bridge this gap. The EDA tools for 3D integration of 2D materials include multi-PDK support for hybrid CMOS and 2D material processes, advanced multi-physics simulation for electrical, thermal, and mechanical effects, and precise interface modeling to optimize 2D material-silicon contacts. For 3D architectures, tools must handle high-aspect-ratio structures, ensure uniform deposition, and extract parasitic in complex interconnects. System-level co-simulation is critical to integrate analog, digital, and RF components with 2D material-based devices. Additionally, tools must enable modular design, efficient thermal management, and robust reliability analysis to support scalable, high-performance 3D heterogeneous integration.

To realize the vision of commercial-scale M3D integration of 2D materials, research and development efforts must prioritize scalable and reliable fabrication processes. Immediate goals include refining synthesis methods to produce defect-free, uniform 2D materials at 6-8 inch wafer scale and developing transfer techniques that preserve material quality and ensure clean interfaces and low strain across the entire wafer. Innovations in BEOL-compatible processes will be crucial for seamless integration into existing silicon-based manufacturing workflows. Mid-term objectives involve the realization of unified 3D architecture that integrate diverse functionalities, such as logic, memory, and sensing, within a single chip. Achieving inter-tier electrical isolation and mitigating thermal effects will be essential for robust, high-performance systems. Long-term aspirations include the commercialization of multifunctional 3D chips that leverage the unique properties of 2D materials for applications in neuromorphic, quantum, and artificial intelligence computing. The integration of 2D materials into 3D architecture offers transformative potential, enabling task-specific, energy-efficient, and versatile electronics. By addressing these challenges, the semiconductor industry stands poised to usher in a new era of sustainable and advanced computing technologies.

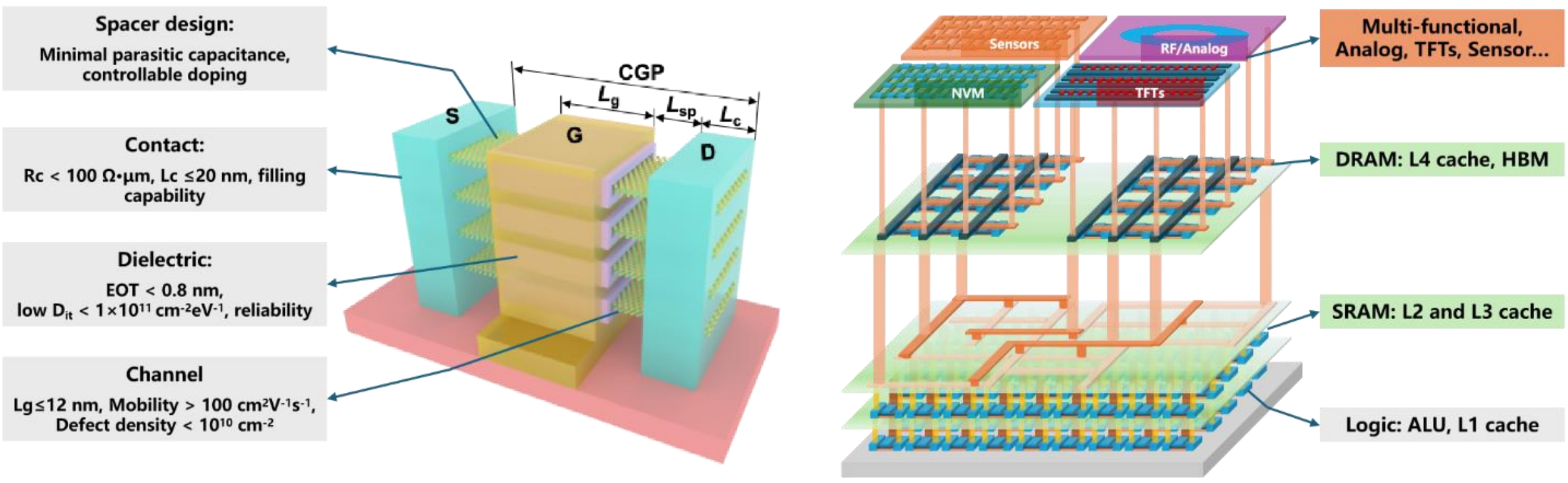

Figure 1. Advanced 2D semiconductor FET (left) and 3D heterogeneously integrated system (right)

## 13. Quantum photonics and lightwave electronics

Markus Borsch[1], Andrea C. Ferrari[2], Rupert Huber[3], Mackillo Kira[1] and Fengnian Xia[4]

[1]Department of Electrical Engineering and Computer Science, University of Michigan, Ann Arbor, MI, USA
[2]Cambridge Graphene Centre, University of Cambridge, Cambridge, CB3 OFA, UK.
[3]Department of Physics, University of Regensburg, Regensburg, Germany.
[4]Department of Electrical and Computer Engineering, Yale University, New Haven, CT, USA.

**Status** Optoelectronics stands on the cusp of a major transformation through the integration of layered materials (LMs) [1-3]. These herald a new era of solutions to pressing issues, including the quest for energy sustainability. They can significantly boost the efficiency of both information and communication technology (ICT) and light harvesting, as well as providing energy-efficient resources for artificial intelligence (AI). Concurrently, LMs are poised to thrust quantum information science and engineering (QISE) into a realm of practical usability, offering unprecedented performance advantages over classical options. Here, we highlight recent ideas and advances for energy solutions, AI, and QISE based on graphene, semiconducting LMs, and their heterostructures (LMHs).
Graphene can provide advanced functionalities in key technological devices and is the most mature of the LMH technologies. It already has delivered competitive energy efficiency and speed for devices such as low-power and hyperspectral photodetectors useful for autonomous driving, and modulators integrated in silicon-photonics, outperforming the state of the art [2]. Emerging applications include photosynthetic devices, LiDAR, security, and ultrasensitive physical and chemical sensors for industrial, environmental, and medical technologies.

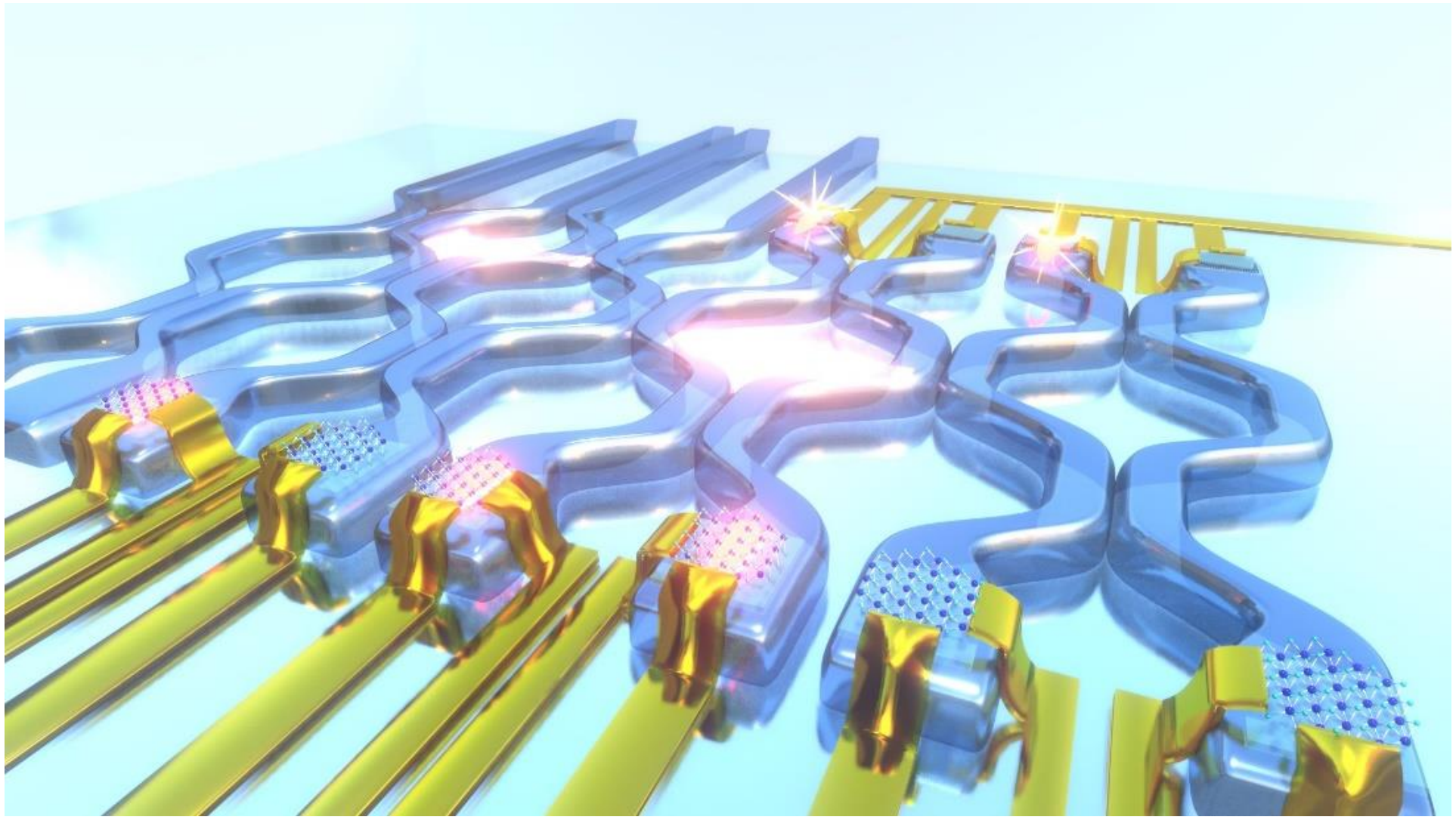

**Figure 1 | Scalable integrated photonics.** Scalable integrated photonic circuits built from on-chip quantum devices using critically coupled LMHs, adapted from A. R.-P. Montblanch et al., Nat. Nanotechnol. 16, 555–571 (2023).

Twisted LMs have provided promising new directions for AI devices. Asymmetric potential and ferromagnetic switching in twisted bilayer graphene (t-BLG) were used to achieve synaptic behaviours for neuromorphic computing functions [4,5]. By leveraging tuneable quantum geometric properties of t-BLG and advanced AI algorithms, intelligent sensing of high-dimensional optical information with a single on-chip sensor was demonstrated [6]. Recently, LMs have been configured to mimic the dynamics of biological synapses [7], heralding a new generation of energy-efficient AI hardware.

Semiconducting LMs are currently evolving fast as a new platform for quantum technology. They have already shown potential as scalable components, such as quantum light sources, photon detectors and nanoscale sensors, and enabled new materials discovery within the broader field of quantum simulations. Yet, harnessing quantum information *within* semiconducting LMs has remained challenging when operating at GHz clock speeds because each electron scattering event—occurring roughly $10^4$ times per GHz cycle—terminates quantum operations.

Lightwave electronics addresses this critical QISE barrier by using the oscillating electric field of intense optical carrier waves as ultrafast biasing fields, enabling quantum operations that outpace the oscillation cycle of light [3]. As illustrated in Fig. 2, lightwave electronics can transport electrons coherently at petahertz (PHz) clock speeds—one million times faster than conventional electronics and 100 times faster than scattering occurs—and flip quantum states within a few fs [8,9] under ambient conditions. Utilizing a quasiparticle-collider setup shown in Fig. 2, such quantum operations have been timed with attosecond precision to access delicate multi-electron interactions and correlations within LMs, providing direct access to electronic entanglement states. These breakthroughs have set the stage for meaningful QISE advancements that will enable quantum sensing, communication, and transduction applications, and augment classical electronic and photonic devices with entanglement capabilities.

**Current and Future Challenges**

The intrinsic confinement of electrons within LMs leads to minimally screened interactions that synergistically enhance the coupling between light, charge, lattice vibrations, and spin states. By crafting LMHs, electrons can be confined, transported, and topologically altered across desired layers—unlocking new functionalities and broadening the design landscape. However, many demonstrations are still based on manually constructed, micro-scale materials. Wafer-scale growth of LMs and LMHs is the next challenge.

Realizing PHz and entanglement electronics (Fig. 2) requires overcoming several escalating challenges: conducting single operations swiftly enough (1-fs timescale) to surpass electron scattering events, generating a sequence of such operations, ensuring high-precision timing and control (100-as timescale), and chaining multiple operations to leverage coherent, correlated, or entangled electrons for QISE applications. Although significant progress has been made, further innovations, particularly in sequencing and chaining operations, are necessary. To achieve these stringent goals, lightwave electronics could utilize short, asymmetric lightwaves that impart net, directional momentum onto electrons. This necessitates the development of flexible and stable methods to generate arbitrary waveforms, akin to DC pulses. Integrating lightwave sources into the devices themselves is critical for improving efficiency and reducing costs, a goal that could be facilitated using photonic structures that guide and focus light and plasmonic nanostructures to enhance electric fields.

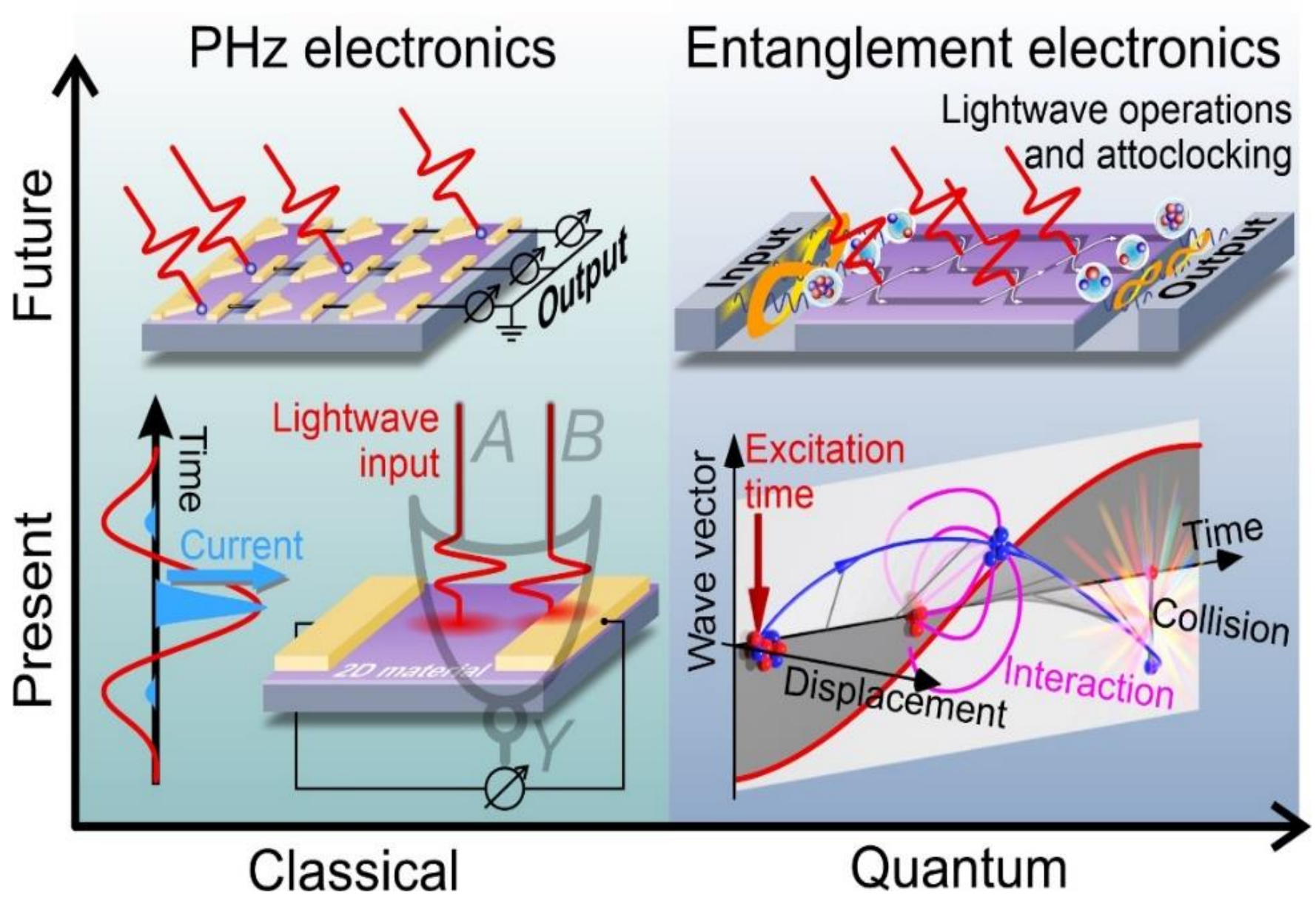

**Figure 2 | Vision of lightwave electronics.** Strong lightwaves can *nonlinearly* and coherently excite directional currents, following the main peak of few-cycle pulses (inset, lower left). This principle has enabled basic logic-gate operations (lower left, schematic; adapted from Ref. [9]) using the carrier-envelope phase of two pulses as inputs (A and B) and the output (Y) encoded in the current between electrodes, laying the groundwork for PHz electronics. Future directions could include building an entire PHz processor (top left) by integrating hybrid multi-electrode–lightwave arrays. Lightwave-operated quasiparticle colliders can directly detect and drive multi-electron correlations (spheres) by attoclocking excitation-to-collision timing (lower right; adapted from Ref. J. Freudenstein et al. Nature 610, 290–295 (2022)). Advances could improve excitation, control, timing, and readout of multi-electron correlations in LMs (top right; adapted from Ref. [3]) by chaining multiple quantum-state operations within the lifetime of coherences for entanglement electronics.

A combination of theoretical investigations and advanced quantum spectroscopic tools will be essential for directly accessing QISE-relevant correlations. At present, both theory and experiment face significant challenges. Multi-electron predictions are computationally prohibitive, and flexible and robust sources and detectors for complex quantum states of light remain elusive. Developing such predictive tools and technologies would be crucial for leveraging quantum states for various applications.

**Advances in Science and Technology to Meet Challenges**

While promising progress has been made for wafer-scale synthesis of graphene, hBN, and transition metal dichalcogenides (TMDs), many applications in energy, AI, and QISE demand LMHs consisting of a variety of LMs with diverse physical properties. Since different LMs can require distinctive synthesis conditions, it is challenging to directly synthesize the desirable LMHs on demand. As a result, future technologies will require both wafer-scale synthesis and autonomous/robotic and AI-assisted transfer techniques [10] for the preparations of wafer-scale LMHs on demand. Successful integration of LMs into existing electronic and silicon-photonics platforms is also paramount.

Lightwave electronics has also been extended to the videography of single-electron dynamics [3] by combining near-field techniques with attosecond science. Extending this capability to visualize multi-electron correlation dynamics—capturing the temporal evolution of interactions among multiple electrons—remains an elusive goal. Understanding how many-body correlations and decoherence influence emergent phenomena is equally challenging, yet addressing these challenges could lead to fascinating quantum effects, with implications ranging from entanglement electronics to exploring the fundamental properties of quantum phases.

For QISE advances, achieving a reliable readout via quantum transduction—which converts quantum information in solids to different modes—is essential. Current developments are promising but require significant improvements to minimize losses detrimental to quantum-information applications. This underscores the need for further exploration into LMHs, digital alloys, highly efficient high-harmonic generation sources, and sophisticated quantum transduction techniques.

**Concluding Remarks**

With their unique electronic properties and reconfigurability, LMs and their heterostructures offer numerous opportunities in energy efficient ICT, neuromorphic computing, intelligent sensing, and QISE. Future AI systems will greatly benefit from large-scale synaptic circuits based on LMHs. As one of the fastest-growing electricity consumers globally, ICT can markedly improve its energy efficiency through integrating LMH optoelectronics and silicon-photonics. By integrating sensing systems with synaptic circuits, it becomes possible to directly process sensed information with AI algorithms, substantially mitigating information transmission bottlenecks.

This ongoing integration promises reduced production costs through simpler designs, while delivering higher data rates at lower power consumption, critical for advancing 5G, 6G, ubiquitous sensor networks, and AI. The capability of LMs to transmit and receive multiple wavelengths provides a cost-effective solution to achieve necessary data rates beyond the year 2025. Additionally, LMs used as broadband photodetectors (PDs) are pivotal for mid-infrared detection—the fingerprint region for many gases and molecules—where conventional silicon-photonics PDs are unsuitable. This is crucial for a range of applications in healthcare, security, and environmental monitoring.

For QISE, lightwave electronics presents unique opportunities to control and measure the dynamics of electronic quantum states and their correlations at natural time and length scales. The integration of LMs is essential in enhancing existing optoelectronics with entanglement processing capabilities.

This could revolutionize classical technologies by increasing clock rates into the PHz regime, and quantum technologies by unlocking the full quantum potential of semiconducting LMs.

**Acknowledgements**
We acknowledge funding from ARO through Award W911NF1810299, W.M. Keck Foundation, the National Science Foundation through grants 2118809, 1741693, and 2150561, the Deutsche Forschungsgemeinschaft through Project ID 422 314695032-SFB 1277 and Research Grant HU1598/8, the Yale Raymond John Wean Foundation, EU Graphene Flagship, ERC Grants Hetero2D and GSYNCOR, EPSRC Grants EP/K01711X/1, EP/K017144/1, EP/N010345/1, EP/L016087/1.

## 14. 2D materials for next-generation energy storage

**Xiao Wang[1], Zhong-Shuai Wu[1,*], Xinliang Feng[2,3,*], Patrice Simon[4,*] and Hui-Ming Cheng[5,6,7,*]**

[1] State Key Laboratory of Catalysis, Dalian Institute of Chemical Physics, Chinese Academy of Sciences, Dalian, China
[2] Center for Advancing Electronics Dresden (cfaed), Faculty of Chemistry and Food Chemistry, echnische Universität Dresden, Dresden 01062, Germany
[3] Max Planck Institute of Microstructure Physics, Halle (Saale) 06120, Germany
[4] CIRIMAT, UMR CNRS 5085, UniversitéPaul Sabatier Toulouse III, Toulouse 31062, France; RS2E, Réseau Français sur le Stockage Electrochimique de l'Energie,FR CNRS 3459, Amiens Cedex 80039, France
[5] Shenzhen Key Laboratory of Energy Materials for Carbon Neutrality, Shenzhen Institute of Advanced Technology, Chinese Academy of Sciences, Shenzhen, China
[6] Faculty of Materials Science and Energy Engineering, Shenzhen University of Advanced Technology, Shenzhen, China
[7] Shenyang National Laboratory for Materials Science, Institute of Metal Research, Chinese Academy of Sciences, Shenyang, China

Email: wuzs@dicp.ac.cn, xinliang.feng@tu-dresden.de; patrice.simon@univ-tlse3.fr; hm.cheng@siat.ac.cn.

**Status**

Energy storage, particularly batteries and supercapacitors, has garnered overwhelming interest in the last few decades to overcome the issue of intermittency of sustainable green energy due to the geographical and seasonal dependence for impending energy crisis and $CO_2$ emission reduction.[1-4] As essential components of batteries or supercapacitors, electrode materials greatly determine the key electrochemical performance and practicality for applications. Unlike bulk materials, which suffer from limited ionic adsorption sites and violent volume changes, 2D materials have emerged as promising electrode materials, having made a monumental leap since the discovery of graphene in 2004 (Figure 1). For energy storage devices, most researchers focus on inorganic 2D materials, such as graphene and their derivates, transition metal dichalcogenides, transition metal oxides, layered double hydroxides, metal carbides/nitrides, hexagonal boron nitride, black phosphorus, etc. Presently, 2D materials could perform as different components in energy storage devices with fast ionic transport and adjustable physicochemical characteristics. (ⅰ) Electrodes materials. With tuning of morphology, framework and charge transport properties through doping, alloy composition, hybridization, and surface functionalization, heterostructures through van der Waals, especially for combination of conductive and active components, are prone to high capacity, high power/energy density and long cyclability. (ⅱ) Inactive components. Some 2D materials could perform as conductive additives for constructing conductive cross-linked network, depending on the full contact among points, lines and surfaces. Remarkably, holey 2D materials with adjustable and customized framework hold promise for filtration or selective ion transport as separators or electrode|electrolyte interface. For improving the potential of 2D materials in present and future applications, it is important to tune the properties of 2D materials when it comes to interlayer interaction, in-plane bonding configuration, Fermi energy level, electronic band structure and spin–orbit coupling.[5]

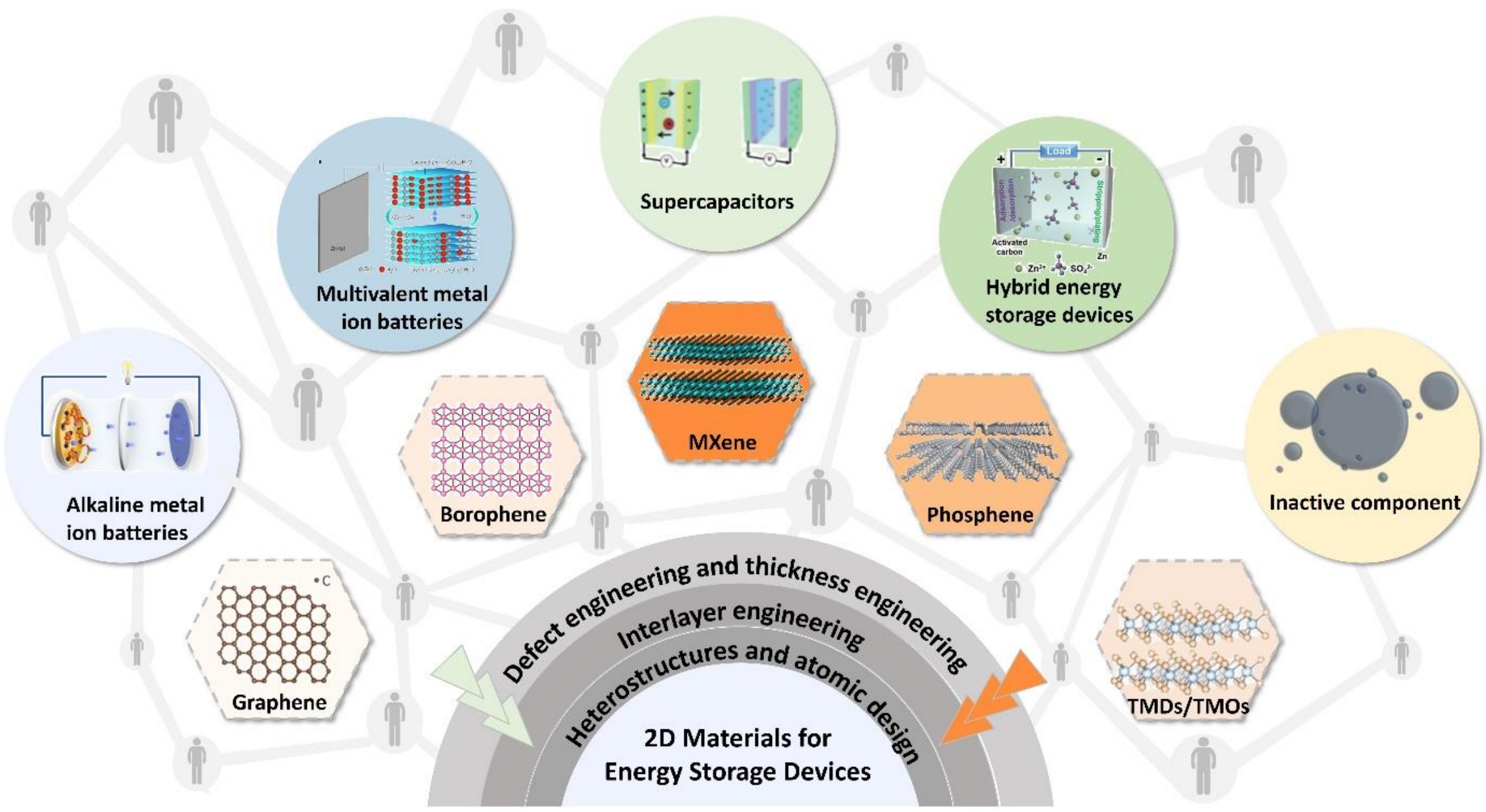

**Figure 1.** Status of 2D materials for energy storage.

## Current and Future Challenges

Some issues must be addressed before the properties of 2D materials could be used to the fullest extent:[6, 7] (i) Instability of nanosheets morphology and limited electrochemical performance of single 2D materials for electrode materials. Due to the high surface free energy of 2D materials, tendency to restack in no order restrict the utilization rate of edge and in-plane active sites. Additionally, although 2D materials with expanded interlayer distance contribute to the improved electrochemical performance for electrode materials, the larger interlayer distance means the decrease of volumetric energy density of batteries. (ii) Unclear specific mechanisms for energy storage devices. Previous investigations report that graphene and other 2D materials could be combined to improve the energy storage performance via different methods, such as ample geometric defects, surface winkles/corrugations, but most of them rarely clarify clearly the synergistic effect. (iii) Lack of precise application-driven design. Notably, the general permutations and combinations of different 2D materials cannot meet the requirements from different scenarios. The greater the complexity and multifunctionality demanded by the intended applications, the more sophisticated the synthesis strategies should be to exercise control over the structure, atomic arrangement and morphology.[8] Specially, the atomic-level microstructure and electrochemical performance for energy storage devices, such as catalysts for metal-air batteries, needs further explorations. (iv) Difficult scalability and reproducibility.[9] It is still difficult to control dimensions and structures of 2D nanosheets with high quality and high production rate, considering the high cost and environmental concerns. Future generations of printable energy storage devices require the processability of the hybrid materials.

## Advances in Science and Technology to Meet Challenges

To deal with the above challenges, ingenious strategies and technologies need to be explored to enable accurate synthesis of 2D materials and deep understanding of structure-property relationships. Recent advances in fabrication technologies, such as solid lithiation and exfoliation, atomic layer deposition and molecular beam epitaxy, allow precise control over thickness and composition, especially for the scalable production. To improve the electrochemical performance for energy storage devices, surface chemistry of 2D materials is adjusted through chemical doping and surface functionalization for pseudocapacitance and charge carrier concentration improvement, interlayer enlargement for large ion storage and fast kinetics, Janus structure for improving binding energy with electrolyte ions and metals and other controlling methods. Since Janus structure is not cost-effective, alternative 2D materials-based heterostructures with amorphous architecture could alleviate structural stress in charging/discharging process, devoting to the high-rate capability and long cyclability. From this point of view, defect engineering and interface engineering are promising to modulate the physicochemical characteristics of 2D materials-based heterostructures. Furthermore, regulation of chemical affinity of 2D interfaces associated with the surface charge density and distribution could be an interesting topic for metal-air batteries. To further figure out the synergistic effects, specific model system and further DFT investigations are supposed to be carried out. Unique technologies that go beyond conventional characterization techniques, such as scanning electrochemical microscopy or electrochemical quartz crystal microbalance, appropriately supplement our understanding of such systems by offering direct insights into the reaction behaviour of samples at the nanoscale level, including the interfacial charge transfer and ionic transfer. Furthermore, advanced computational techniques like high-throughput computational screening and machine learning to guide more accurate and efficient synthesis is a promising solution.

**Concluding Remarks**

Since it is in the early stage of 2D materials development and their industrialization, the promising pathways of 2D materials in the future could be considered in the following aspects: (1) Cost-effective and precise synthesis technologies with high performance. To overcome the traditional methods with cumbersome and complex feature, solution-based methods, such as solid lithiation and exfoliation, become alternatively innovative method, which provide a way of modifying the characteristics of 2D materials during the growth process. Specifically, interlayer engineering could be a complement way to host the ions with large radii. (2) Advanced characterization and computational simulation. To ensure quality control and clarify the material growth mechanisms, it is essential to record real-time observations during the growth of 2D materials and device fabrication processes. Additionally, comprehensive understanding of the promotion mechanisms in energy storage devices (e.g., binding ability, charge transfer number) would require further developments on the in-situ probe techniques and DFT approaches. (3) All-2D system construction. Capitalizing on the advantages of 2D materials in terms of size reduction, planar electrochemical energy storage systems are of great importance for further practical application and model system for characterization, which have been achieved through complete integration of electronics and energy storage on a single substrate. (4) Selective and specific applications. For all-solid-state batteries and future generation of printable energy storage devices, 2D materials provides a feasible way for constructing a highly conductive network or highly ion conductive interphase. Close collaboration and coordination among different fields could facilitate smooth development of 2D materials for next-generation energy storage.

**Acknowledgements**

The authors acknowledge support from the National Natural Science Foundation of China (Grants. 22125903, 22439003, and 22209173), DICP (DICP I202471), Energy Revolution S&T Program of Yulin Innovation Institute of Clean Energy (Grant E411050316), Natural Science Foundation of Liaoning Province (2023BS006).

## 15. 2D materials for high-performance electrocatalysis

**Bilu Liu[1] and Yi Xie[2]**

[1] Tsinghua Shenzhen International Graduate School, Tsinghua University, P. R. China.
[2] University of Science and Technology of China, P. R. China.

**Status**

Electrochemically converting a chemical into another more value added one via electrocatalysis is a promising technology, especially when the reaction is driven by green electricity from solar, wind, hydropower, etc, which can realize the goal of ‘kill two birds with one stone’. On one hand, it starts with cheap resource like water or even unwanted harmful reactants like carbon dioxide, contaminates in water, etc. One the other hand, it can produce valuable products without consumes much fossil resource. Different types of electrocatalysis reactions have been studied extensively, from water electrolysis to produce hydrogen and oxygen discovered in 1800, to other reactions including electrochemical carbon oxide reduction, nitrogen reduction, oxygen reduction, water oxidation, small organic molecules oxidation, and different coupled reactions. The field is becoming increasingly important and urgent these years due to global mission of carbon neutrality.

The use of an efficient electrocatalyst is key for industrial implementation of these techniques, because a suitable electrocatalyst can decrease reaction barrier and consequently electricity consumption during reaction. This in turn will reduce the cost of the conversion process, to make it economically competitive over traditional conversion process like thermal catalysis or others. Among various catalysts, two-dimensional (2D) materials including single-element ones like graphene, thin layer metal (metallene) or non-metals (e.g., phosphorus, silicene), as well as compounds like transition metal dichalcogenides/carbides/nitrides/phosphides, carbon nitrides, are particularly promising.[1-2] First, they have large surface area which will help adsorb reactants and catalytic reaction. Second, 2D materials have a flat and well-defined structure without surface reconstructions, making them ideal model catalysts for mechanism studies (Figure 1). Third, there are many types of 2D materials with widely tunnable compositions and electronic structures, and 2D materials can feasibly assemble into a whole structure with finely engineerable micro-and macro-structures,[3] which are advantages for improving intrinsic activity as well as mass and charge transport. Other advantages of 2D materials include low-cost mass production capability, thin thickness induced easier charge separation, etc. All these features suggest that 2D materials are promising in various electrochemical conversion technologies, as witnessed by the achievement of using 2D material electrocatalysts in the past two decades.

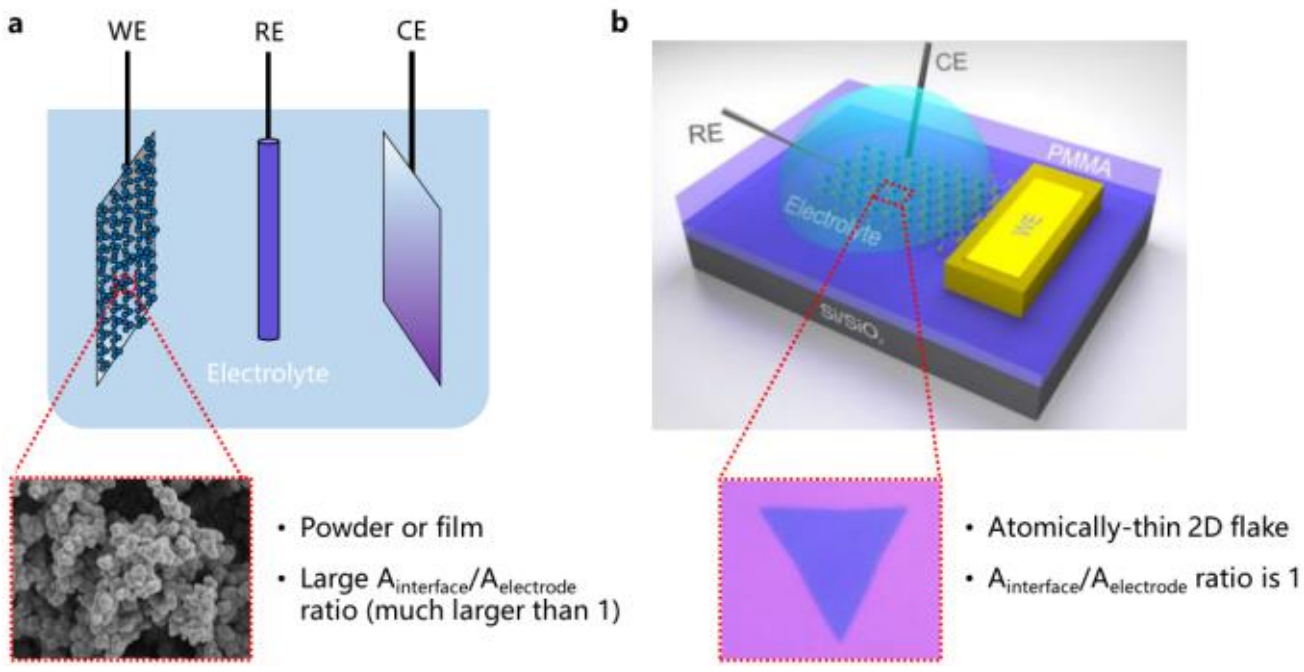

**Figure 1.** A comparison of powder catalysts versus 2D flat catalysts.

**Current and Future Challenges**

Despite with great success, there are several challenges in 2D material electrocatalysis. The first challenge is about the structure of real active site. Of course, one needs a catalytic active site with the best intrinsic activity, meaning suitable and balanced adsorption energy to different intermediate species. In reality, several factors may affect the structure of real active site, including electrolyte environment, pH, pressure, voltage, etc. Meanwhile, catalyst usually undergoes surface and sub-surface reconstruction, especially at oxidative potential, making accurate identification of real catalytic active site under operation conditions challenging.[4] The second challenge is about high-current-density (HCD) performance over a long period. To make various electrocatalysis technologies practical useful, it is a must to go to HCD. The overall performance of a catalyst at HCD are jointed determined by many factors including intrinsic activity of active site, amount of active site and their exposure, mass transport, electron transport, microstructure of catalysts, etc. In addition, when translocate the performance of an electrocatalyst from three-electrode cell to a membrane electrode assembly device, other factors will play rules, for example, interface between catalyst and support, chemical and mechanical robustness of the electrode, etc. The third challenge is about selectivity to a certain product. For some electrocatalytic conversion processes like carbon dioxide reduction, there are multiple products generated simultaneously, including C1, C2, C3, and higher.[5] Separation of different products could be energy and time extensive, and therefore, developing catalyst which show high selectivity to certain products and can operate with a partial HCD is important. The fourth challenge is about economic considerations. On one hand, we need catalysts with high activity and high performance to reduce electricity cost during electrocatalysis conversion in produce valuable products. On the other hand, catalysts themselves can be quite expensive because noble metals are common catalysts for many electrocatalysis technologies, and therefore we need to develop catalysts made of cheap and abundant elements.

**Advances in Science and Technology to Meet Challenges**

New techniques and understandings are needed to address the above challenges. First, among all types of catalysts, 2D materials might be the ideal platform to identify the real active site, mainly because the dangling bond free and surface state free feature of 2D materials. For example, nanoparticle catalysts may have surface reconstruction even in vacuum, which becomes more complicated when putting into liquid electrolyte and applying a voltage. From this angle, researchers can pay more attention to use 2D materials as model catalysts for mechanism studies, including compare intrinsic activity of different sites, dopants, point defects, edge configurations, etc.[6-7] Of course, in situ techniques like spectroscopy and high-resolution imaging will be necessary. Note that micro-cell devices are interesting to study 2D material electrocatalysis and correlate activity with material structures. Second, 2D materials can assemble into a monolith with designed structures. By selecting 2D materials with high activity, and optimizing macro- and micro-structures, one can obtain catalysts with high performance at HCD. Besides that, for such high performance monolithic catalytic electrodes, one needs to pay attention to their stability, mainly including chemical stability and mechanical stability, which is a prerequisite for long term use. For chemical stability, the formation of a stable reconstructed layer may be important. For mechanical stability, a strong covalent or metallic or ionic bonding at interface of catalyst and support is important. Meanwhile, such a strong bonding should not induce a large charge transfer resistance across the catalyst-support interface. Third, the selectivity of hydrogen or oxygen production in water electrolysis under pure, alkaline, or acidic water is high enough. Meanwhile, the selectivity of C1 or certain C2 products in carbon oxide reduction is also high to meet applications. However, for water electrolysis in complicated electrolyte like contaminated water or sea water, selectively generation of hydrogen and oxygen still needs to be improved. For carbon dioxide reduction, obtaining C3 or higher carbon products with high selectivity and Faraday efficiency are important tasks, which can be solved by using highly active and selective catalysts well positioned on a stable support. Also, there are various new electrochemical reactions aiming for synthesizing products by electricity, in which selectivity towards one certain produce is important. Fourth, increasing energy efficiency, i.e., produce more target products under certain electricity is an important criterium for economic considerations, which calls for high performance and highly selective catalysts. Reducing amount of noble metals by alloying, single atom catalysis, exposing as many as possible noble metals to reactants, are useful ways to reduce catalysts cost. Meanwhile, using earth abundant resources especially minerals like molybdenite made of 2D $MoS_2$,[8] is promising to pursuit further.

**Concluding Remarks**

To conclude, 2D materials have shown promising as high performance durable catalysts for various electrochemical reactions. Several 2D materials based water electrolysis catalysts have shown potential to run at HCD over long period in three-electrode cells and membrane electrode assembly electrolyzers, while more efforts are needed to move forward for large scale industrial implementation. The use of 2D materials in other electrochemical reactions are mostly in the fundamental study stage, in which more efforts are needed to elucidate the reaction active site, improve selectivity, and make it works robust over long term operation at industrial required current density conditions.

**Acknowledgements**

This work was supported by the National Science Fund for Distinguished Young Scholars (No. 52125309), Guangdong Basic and Applied Basic Research Foundation (No. 2022B1515120004), Innovation Team Project of Department of Education of Guangdong Province (2023KCXTD051), Shenzhen Science and Technology Program (ZDSYS20230626091100001), the Shenzhen Basic Research Project (No. JCYJ20220818101014029), and Tsinghua Shenzhen International Graduate School-Shenzhen Pengrui Young Faculty Program of Shenzhen Pengrui Foundation (No. SZPR2023002).

## 16. Filtration & separation

**Wanqin Jin[1] and Rahul Raveendran Nair[2]**

[1] Nanjing Tech University, China
[2] University of Manchester, UK

### Status

Membranes are key components in systems as diverse as batteries, fuel cells, controlled delivery devices, solar cells and large-scale energy-efficient separation and purification systems. Wherein the membranes act as a selective barrier that selectively lets some components (e.g., gases, liquids and ions) transport while restricts the movement of others. Current membranes, especially polymeric membranes, generally suffer from the trade-off relationship between permeability and selectivity. New membrane materials are need to be developed to surpass the limitation.

The discovery of graphene opens a new era of two-dimensional (2D) materials[1]. Featuring with unique atomic thickness and micro-meter lateral dimensions, 2D materials offer new opportunities for the next generation of high-performance (high permeability, high selectivity, and stability) separation membranes. There are two main types of 2D materials membranes (2DMMs): a single-layer (porous nanosheet) membrane and a laminar membrane. The former one is composed of monolayer or a few layers of 2D materials with intrinsic or perforated in-plane nanopores. The latter one is fabricated by assembling nanosheets.

2DMMs represented by graphene and graphene-derived materials have seen rapid progress in the past decade. Figure 1 displays a historic timeline and highlight some breakthrough in 2DMMs field. The first 2DMMs were produced between 2008 and 2012 with single layer graphene[2] and laminar graphene oxide membranes[3]. In the following years, more attention is paid to the theoretical study of molecular transport[4], the precise construction and regulation of nanochannels in 2DMMs (e.g., tuning interlayer spacing[5] and intrinsic pore size[6], creating in-plane pores[7]), and the improvement of stability [8]. The booming of graphene membranes also inspire researchers develop other 2DMMs, such as transition-metal phosphorus trichalcogenides nanosheets[9], $MoS_2$[10] membranes. To meet the standard of industrial application, the scale-up in 2DMMs[8] is underway.

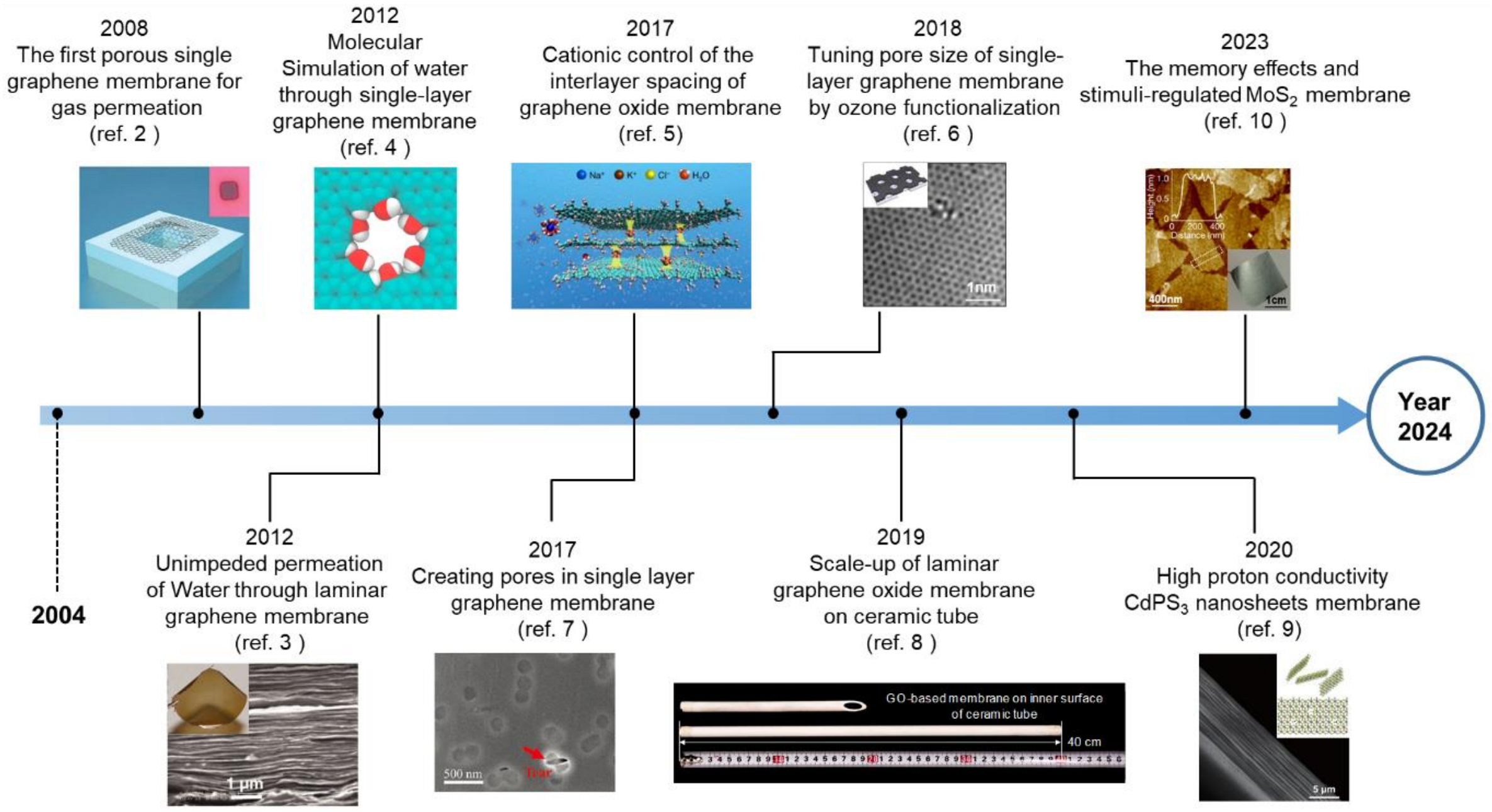

**Figure 1.** Timeline of the milestones for 2DMMs.

## Current and Future Challenges

Since the discovery of graphene in 2004, a great success for 2DMMs has been achieved. Nevertheless, there are still some issues and challenges to be addressed. For the single-layer (porous nanosheet) membrane, the major challenge lies in that the synthesis and subsequent transfer of high-quality single or few layer 2DMMs with controllable features (e.g., pore size and its size distribution, in-plane porosity, and functionalization degree),and their insufficient reproducibility for stable membrane separation, which generally are essential to unlock the potential of scale-up of 2DMMs towards practical application. For laminar membrane, the key point lies in how to precise control the interlayer spacing and channel surface chemistry especially at the angstrom scale. The stability for real-world applications is another concern. The membrane swelling especially in aqueous solution will enlarge the interlayer channels and degrade the size-sieving property, even lead to the membrane disintegration. At present, it remains challenging to accurately characterize sub-nanometer transport channels and their interaction with various ions and molecules, in addition, there is a lack of theoretical models for the description of mass-transport behaviour within these channels in 2DMMs. The application of 2DMMs towards challenging mixtures separation (e.g., olefin/paraffin, xylene separations) has not been fully demonstrated yet. More importantly, the 2DMMs synthesis is only on the laboratory scale, more efforts should be devoted on the scale-up of defect-free and uniform 2DMMs to meet the requirement of real industry applications.

## Advances in Science and Technology to Meet Challenges

To meet the above mentioned challenges, the following aspects should be considered:

(i) Nanochannel construction and application for challenging separation. Owing to the uneven distribution of functional groups or intercalators in 2DMMs, the interlayer nanochannels or in-plane pores are undesirable for separating challenging mixtures with very close molecular sizes, such as olefin/paraffin, $O_2/N_2$, and xylene separations. Thus, it needs to develop novel 2D materials with uniform functional groups and well-defined pore structures, in addition, novel strategies to regulate the nanochannels to sub-angstrom-scale precision are also highly necessary.
(ii) Nanochannel characterization and transport mechanism. Advanced characterization techniques at atomic-scale resolution, such as positron annihilation lifetime spectra and in situ high-resolution transmission electron microscopy, need to be developed to reveal the structures of the nanochannels of 2DMMs and probe the real-time molecular/ionic transport behaviors in nanochannels, which can be combined with theoretical calculation to establish the theoretical model for guiding the rational structural design and performance optimization of 2DMMs.
(iii) Large-scale fabrication and performance evaluation under practical conditions. Novel synthesis methods for 2D materials and defect-free, large-scale 2DMMs need to be developed, and the yield of high-quality nanosheets needs to be enhanced to satisfy production demands. The performance evaluation should consider the real application conditions, such as high temperature, high pressure, and steam-complicated contamination.

**Concluding Remarks**
The field of 2DMMs is rapidly growing, high-quality single and few-layered nanosheets with diverse choice of materials have been developed for high-performance separation membranes. Future efforts should be devoted to  atomic-scale characterizations, mass transport mechanism understanding, and innovative construction strategies that integrate principles from mathematics, physics, chemistry, materials science, and artificial intelligence as well as the scalingup these technologies for real-world industry application.

**Acknowledgements**
This work was supported by the National Natural Science Foundation of China (22038006, 21921006, 22278210) and a project funded by the Priority Academic Program Development of Jiangsu Higher Education Institutions (PAPD).

## 17. 2D materials for thermal management of electronic devices

**Yan Xu[1]*, Qing Zhang[2], Wencai Ren[2]***

[1] Huawei Technologies Co., Ltd.
[2] Shenyang National Laboratory for Materials Science, Institute of Metal Research, Chinese Academy of Sciences, 72 Wenhua Road, Shenyang 110016, P. R. China.
Emails: mullich.xuyan@huawei.com, wcren@imr.ac.cn

### Status

The integration of chiplets into system-level packaging, driven by advancements in artificial intelligence, high-performance computing, the Internet of Things, and 5G/6G communications, has been formally incorporated into the International Roadmap for Devices and Systems (IRDS). However, the high degree of integration leads to significant overheating issues, which have become a primary factor limiting the performance, reliability and longevity of electronic devices. This urgently calls for the development of efficient thermal management materials and technologies to ensure the optimal performance and reliability of electronic devices. A typical thermal management system in integrated circuit (IC) chips, as illustrated in Figure 1, consists of a heat sink, an integrated heat spreader (HS), and two thermal interface materials (TIMs), designated as TIM 1 and TIM 2 [1]. TIMs reduce thermal resistance by filling air gaps between the IC and heat sink, while the HS dissipate heat from hotspots to broader areas.

Two-dimensional (2D) materials, characterized by atomic-scale thickness and extended lateral dimensions, exhibit exceptional in-plane thermal conductivity due to their highly ordered lattice structures and strong covalent bonds, offering promising solutions for next-generation thermal management challenges. For instance, graphene demonstrates an ultrahigh in-plane thermal conductivity of up to 5,300 $W \cdot m^{-1} \cdot K^{-1}$, while hexagonal boron nitride (h-BN), an electrical insulator, also exhibits high in-plane thermal conductivity up to 700 $W \cdot m^{-1} \cdot K^{-1}$ [1,2]. However, 2D materials (nanosheets) typically need to be assembled into macroscopic assemblies for thermal management applications, such as films for HS and arrays or networks for TIMs. The thermal conductivity of these 3D assemblies is influenced not only by lattice defects in the nanosheets, such as vacancies, heteroatoms, and grain boundaries, but also by macroscopic structural disorders like disordered nanosheet arrangements [3]. These challenges can be effectively mitigated through high-temperature treatments and alignment assembly techniques that orient the nanosheets along a specific direction [3-4]. In addition, these 3D assemblies must also meet stringent requirements for mechanical properties and stability to ensure their practical applicability in thermal management systems.

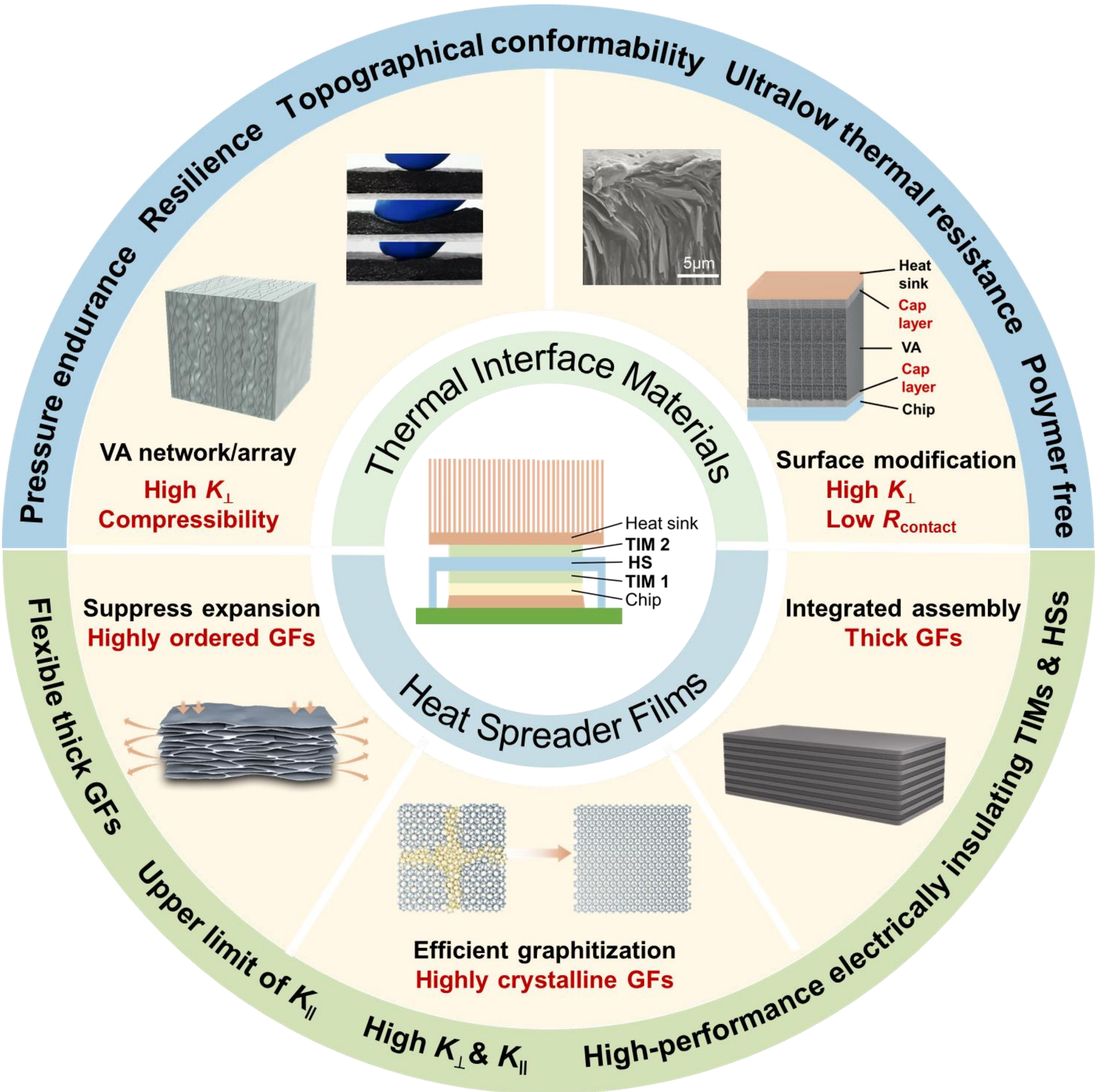

**Figure 1. Status and challenges of 2D materials for thermal management of electronic devices. VA means vertically aligned.**

Advanced TIMs must simultaneously exhibit high through-plane thermal conductivity and excellent compressibility. The mostly widely used TIMs are polymer-based composites, which are fabricated by incorporating highly thermally conductive fillers, such as graphene and h-BN, into polymer matrices [5]. However, a trade-off exists between thermal conductivity and compressibility. While higher filler content enhances thermal conductivity, it often compromises compressibility, making the composite less effective at accommodating assembly tolerances. To meet practical application requirements, TIMs should demonstrate good compressibility under low pressure, specifically achieving a thickness reduction of 30% to 60% under pressures ranging from 0.2 to 0.5 MPa. To fully utilize the intrinsic thermal properties of 2D materials, researchers are focusing on creating continuous and efficient thermal pathways, such as constructing high-crystalline vertically aligned graphene and h-BN networks or arrays [6-8]. Furthermore, to optimize heat dissipation efficiency under packaging conditions, surface structure optimization of these vertically aligned 3D assemblies is crucial for minimizing interface thermal resistance. This can be achieved by applying a cap layer to the surface of TIMs in contact with the chip/heat sink to match the rough topography of heat sinks and increase the effective heat transfer area. Strategies include modifying the surface with low-melting-point metals [9] and constructing arc-like arrangements on vertically aligned networks or arrays [10].

Compared to traditional HS technologies, including metal foils, heat pipes, and vapor chambers, graphene film (GF) offers superior advantages, including high thermal conductivity, tunable thickness, good flexibility, remarkable stability, low density, and exceptional tolerance to extreme environmental. These properties make GF highly promising for applications in portable electronics and spacecraft. For these applications, achieving an in-plane thermal conductivity as high as possible is crucial, which requires GFs with ultra-low defect density, large grain size, and highly ordered crystalline structures. Current strategies to enhance the thermal conductivity of GFs focus on two key areas: (1) Suppression of structural expansion. Graphene oxide (GO) is a widely used precursor for fabricating GFs owing to its exceptional processability. However, during thermal annealing, the decomposition of functional groups into gaseous products causes structural expansion in the film, disrupting the ordered structure. Strategies such as pre-reduction of GO precursors [11], construction of gas escape channels within GO films [12], and multi-step hot pressing [13] have proven effective in fabricating compact and ordered GFs. (2) Efficient graphitization strategies. Based on in-depth studies of the structural evolution and graphitization behavior of GFs, strategies such as introducing high-temperature lattice defects can significantly boost their graphitization [14]. In addition, thick GFs are necessary to ensure superior thermal transfer capabilities. Advanced integrated assembly methods, such as continuous centrifugal casting [15]and water-induced self-fusion [16], can produce highly aligned thick GFs. Stacking and adhering thin GFs through metal welding or binders has also been demonstrated as an effective approach [17]. The repaid development of fabrication technologies of GFs has also facilitated the development of h-BN films [18-19]. Both films now have been commercialized and are now used in various high-tech fields such as smartphones and satellites.

**Current and Future Challenges**

The development of high-performance TIMs still faces several technical challenges. First, considering that the gap between the heatsink and IC can increase by several tens of micrometers at elevated temperatures, TIMs must exhibit self-recovery properties with a recovery ratio exceeding 50% to maintain optimal thermal contact and prevent interface delamination. Second, compression in vertically aligned 2D materials-based TIMs might cause sheet tilting and interface separation, reducing thermal conductivity under high packaging pressures. Consequently, the fabrication of high-performance TIMs with resilience and pressure endurance capabilities represents an essential technical requirement. Third, traditional thermally conductive silicone greases, which are widely used as TIMs for CPU cooling due to their adequate compressibility and topographical conformability, face limitations such as volatility, poor thermal shock resistance, and inadequate long-term durability. Addressing these challenges is crucial for developing TIMs that meet the demands of diverse and extreme application scenarios.

Significant opportunities remain for advancing 2D materials-based HS. First, the upper limit of thermal conductivity for GFs is still unclear from both fundamental and technological perspectives. Key challenges include achieving high thermal conductivity at large film thicknesses while maintaining good flexibility. It is also highly desired to improve the out-of-plane thermal conductivity of the film without significantly compromising in-plane performance. Moreover, the currently used high-temperature graphitization (2,500 °C~3,200 °C) results in significantly elevated equipment maintenance requirements and substantially increased energy consumption levels. Second, the progress in thermally conductive yet electrically insulating h-BN based HS significantly lags behind that of GFs, primarily due to the difficulties in mass-production of h-BN nanosheets and their chemically inert surface. A stable, additive-free h-BN dispersion is still lacking. Further research is needed to explore the potential of heat treatment to enhance the thermal conductivity of h-BN film and to understand h-BN's high-temperature fusion and recrystallization behavior.

**Advances in Science and Technology to Meet Challenges**

Advances in 2D materials-based TIMs are fundamentally dependent on the rational structural design of 3D assemblies. The construction of interlocking architectures is crucial for stabilizing vertically aligned graphene arrays. Optimization of the polymer matrix and the interfacial characteristics between 2D materials and the polymer matrix is also needed. Furthermore, designing vertically aligned crumpled graphene arrays or networks would be helpful for achieving excellent resilience capability. As a high-performance alternative to traditional thermally conductive silicone greases, all-graphene or h-BN-based 3D assemblies can be strategically designed to fully exploit their unique advantages, including good flexibility, low thermal expansion coefficient, and outstanding thermochemical stability, to function simultaneously as both an efficient thermal filler and a flexible matrix. Artificial intelligence (AI) is anticipated to play an important role in the design of 3D assemblies for TIMs.

To approach the limit of thermal conductivity in GFs, a comprehensive understanding of the structural and property evolution throughout the whole fabrication process from graphene into GF is crucial. This includes systematic investigation of precursor material, film fabrication process, and graphitization parameters, particularly through advanced in situ characterization techniques combined with theoretical simulations. The formation of multiple covalent bonds between adjacent graphene layers would be essential for achieving high out-of-plane thermal conductivities. To address the challenges of high-temperature graphitization, developing efficient catalysts to reduce graphitization temperatures below 2,000 °C represents a critical research direction with broad implications. For practical applications, low-cost and high-yield fabrication methods of 2D materials-based films and their integration with electronic components need to be developed.

The realization of thermally conductive yet electrically insulating TIMs or HSs requires more research on h-BN, including the scalable production, surface functionalization, uniform dispersion, and controlled assembly. The chemical conversion from graphene to h-BN, together with exploration of other thermally conductive yet electrically insulating 2D materials, offers a promising alternative approach.

**Concluding Remarks**

In summary, 2D materials including graphene and h-BN have emerged as pivotal components in TIMs and HS for advanced thermal management applications. By leveraging the assemblability and scalability of 2D materials, structural designs such as vertically aligned graphene networks or arrays and highly ordered GFs have been developed to achieve optimal thermal conductivity. Although significant efforts are still need to approach the upper limit of the thermal conductivity, practical applications require a careful balance between thermal conductivity and mechanical properties, including tensile strength, flexibility, compressibility, and resilience. Additionally, the structural and performance stability of these materials needs to be further evaluated under extreme conditions, such as high temperature, high humidity, vacuum, and strongly corrosive environments, to ensure reliability in some real-world scenarios. Furthermore, the adaptability of these materials for seamless integration with other components remains a critical consideration for their widespread adoption in next-generation electronic devices. Last but not least, cost-effectiveness is a key factor that must be prioritized for large-scale applications.

**Acknowledgements**
Wencai Ren acknowledges financial supports from the National Key R&D Program of the Ministry of Science and Technology of China (2022YFA1205300 and 2022YFA1205301) and the Guangdong Basic and Applied Basic Research Foundation (2020B0301030002).

## 18. Flexible electronics with 2D materials: current research and challenges

**Ajit K Katiyar[1] and Jong-Hyun Ahn[1]**

[1] School of Electrical and Electronic Engineering, Yonsei University, Seoul 03722, Republic of Korea.

**Status**

Since the discovery of the world's first transistor in 1947, the solid-state electronics has undergone a remarkable development and currently achieved tremendous success. However, the next generation of everyday electronic gadgets and wearable healthcare devices is aiming for a light-weight design with mechanical flexibility which is hard to achieve with conventional bulk semiconductors. Originally, flexible devices were created using soft organic materials or by using unconventional device architectures, where rigid bulk semiconductors were embedded into an elastomer with stretchable interconnects. However, the rigid nature of bulk semiconductors, combined with the weak biological compatibility and poor stability of organic materials, impeded further advancements and led to a need for alternative ultra-thin functional materials that are cost-effective and energy-efficient.

After the experimental demonstration of graphene flakes two decades ago, two-dimensional (2D) materials attracted significant research attention and have been extensively explored for use in flexible electronics. The exceptional electrical, optical and sensing characteristics of 2D materials make them a suitable choice for developing high-performance electronics. Particularly, their mechanical superiorities associated with ultrathin nature and thickness controllability, make them excellent candidates for developing flexible devices on soft substrates. Various 2D materials such as graphene, transition-metal dichalcogenides (TMDs), black phosphorous, hexagonal boron nitride, MXenes, transition-metal oxides, and 2D perovskites, have been explored so far and used according to their characteristics and target applications [1]. The 2D materials based ultrathin devices possess specific mechanical capabilities, such as bendability, flexibility, conformability, and stretchability (figure 1). Exploiting these features, 2D materials-based flexible devices have demonstrated several applications including smart wrist band [2], bendable integrated circuits [3], flexible/stretchable displays [4], and printed flexible electronics [5,6]. In addition, origami-inspired unconventional devices [4], flexible bioelectronics [7], conformable and implantable healthcare devices [8-10] have also been demonstrated (figure 2).

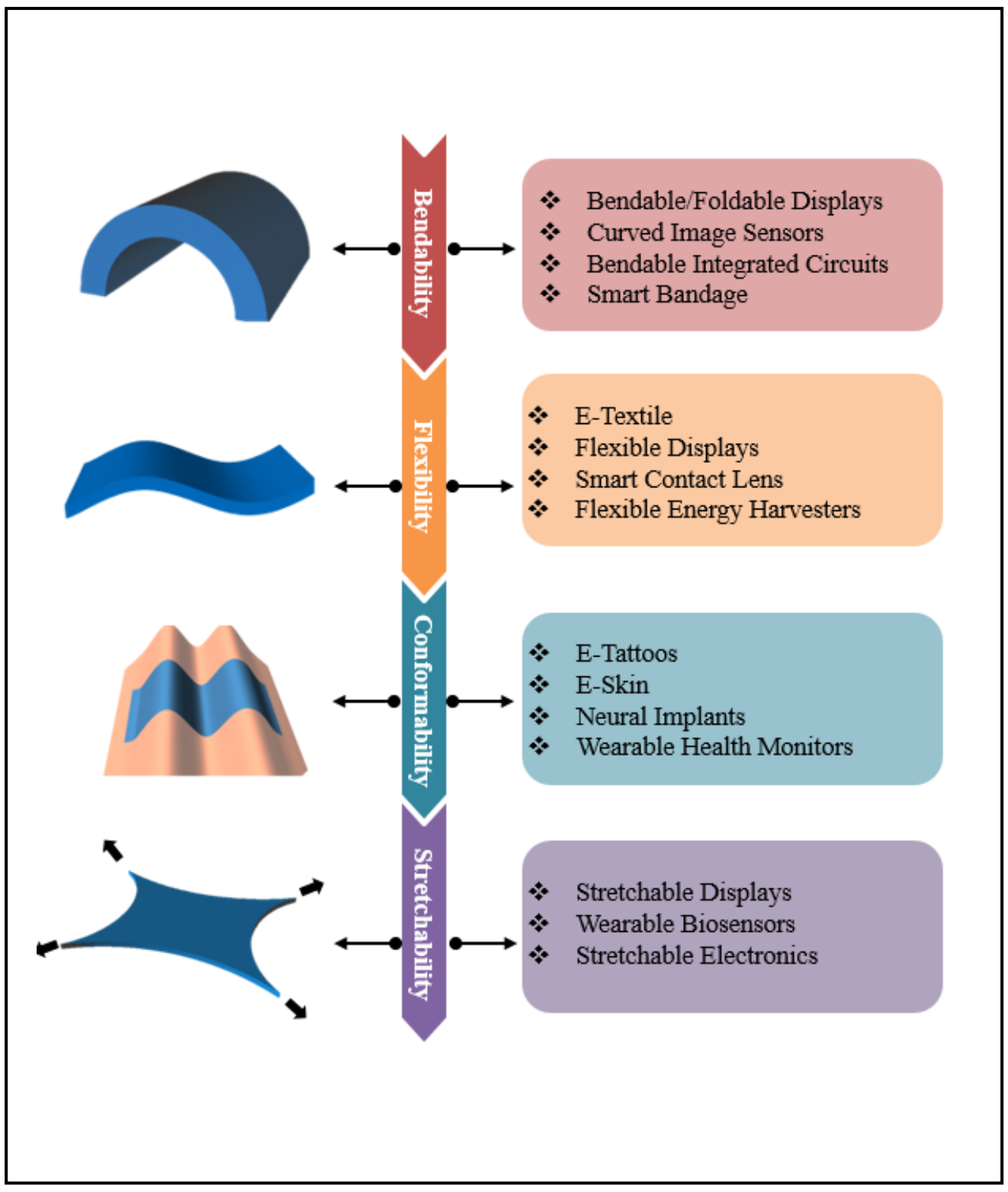

**Figure 1.** Typical mechanical deformations required during the operation of 2D materials-based flexible electronics and representative applications. Reprinted with permission from [10].

**Current and Future Challenges**

Due to ultralow thickness and dangling bonds-free surfaces, 2D materials hold significant potential for developing high-performance flexible electronics. However, 2D materials face several inherent limitations and challenges that hinder their deployment at commercial level. For example, the existence of high contact resistance, difficult deposition of high-quality dielectric material on dangling bonds-free 2D material's surface and complicated selective doping for local p-or n-type conversion are the major challenges of 2D electronics. Additionally, the defect-free 2D materials are desirable to develop high performance electronic devices, but the production of ultrapure large-area 2D materials is still a challenge. Apart from the aforementioned intrinsic issues, the flexible devices developed with 2D material experience several additional challenges. For example, the high surface roughness and low thermal budget of the commonly employed flexible polymeric substrates are the detrimental issues. Additionally, the poor water vapor transmission ratio (WVTR) of the polymeric substrates can accelerate the oxidation induced degradation of 2D materials and subsequently leads to instability and performance deprivation in the fabricated device. The 2D materials based large area flexible electronic devices such as flexible/stretchable micro-displays, flexible image sensors, and wearable/attachable healthcare devices usually consist of a multitasked structure including a substrate, backplane electronics, a front plane of active units, and the encapsulation layer. During the function, these devices repeatedly experience various mechanical deformations such as bending, twisting, shearing, and stretching which incorporate an unsolicited strain into the device components leading to performance degradation and even mechanical failure. The recurrent mechanical deformations can cause sliding, buckling or delamination in the device layers, resulting in a loss of mechanical integrity and device functionality. Additionally, applied strain can alter the intrinsic properties of 2D materials, which can affect the device's performance and stability. Therefore, to effectively integrate 2D materials into practical flexible applications, the material-related intrinsic issues along with other challenges associated with flexibility such as limited processing temperature, strain-induced failure, and inferior performance must be addressed appropriately.

**Advances in Science and Technology to Meet Challenges**
Significant advances via extensive research and investigations have been made aiming to establish the appropriate solutions to existing challenges. To tackle the most common issue of doping in 2D materials, various nonconventional surface doping strategies have been developed to obtain the desired tuning in charge carrier polarity while preserving the lattice structure and other intrinsic features. Treating 2D material's surface with specific organic compounds and chemicals offers an easy way of tuning the doping type through electronic charge transfer. However, these methods have a serious limitation of unstable doping. Depositing a thin layer of an oxide dielectric on 2D material’s surface, particularly on TMDs, is another strategy to achieve stable doping and p-/ n-type conversion via oxygen vacancies-mediated electronic charge transfer at the interface. Advancements in local doping techniques for 2D materials not only enabled the fabrication of electronic circuits but also helped in mitigating contact resistance issues. For example, the existence of an ultra-thin oxide layer between the metal electrode and 2D materials helps to decrease contact resistance by preventing physical damage associated with the Fermi level pinning issue that occurs in the 2D materials during the metal deposition. To address the high contact resistance issue in 2D materials, several other approaches, such as the use of transferred metal electrodes, the use of metallic or semi-metallic phases of 2D materials, edge contacts, and the use of semi-metals such as Bi and Sb have also been attempted. Depositing a thin oxide buffer layer on top of the polymeric substrates not only extends device lifespan by reducing moisture permeation but also mitigates performance degradation caused by high surface roughness. The numerous advantages of the atomic layer deposition (ALD) method, including low-temperature processing, conformal surface coverage, uniform thickness, and precise thickness control over large areas, make ALD the most preferred technique to deposit high-k dielectrics, such as $Al_2O_3$ and $HfO_2$ on the 2D material’s surfaces as a gate dielectric or as a buffer layer for reducing surface roughness and moisture permeation of flexible polymeric substrates. The limited temperature tolerance and huge mismatch in the thermal expansion coefficients of commonly employed polymeric substrates lead to the search for alternative substrates. To this end, mica and ultrathin glass (UTG) (UTG ∼400 °C, mica ∼700 °C) have been explored as non-polymeric alternatives. However, their brittleness and limited strain tolerance provide only modest mechanical advantages. To fabricate flexible devices, 2D materials are typically synthesized on a hard $SiO_2$/Si or sapphire substrate followed by transferring them to the desired polymeric substrates via solution- or laminator/hot press-based dry transfer method for the device fabrication. These transfer processes usually incorporate several defects including surface contaminations, wrinkles, tears and micro-cracks, leading to degradation in material quality and thereby device performance. The recently developed adhesive wafer bonding based large area dry transfer method can be a possible alternative to the conventional methods.

Advancing the low-temperature synthesis of 2D materials would revolutionize the development of 2D material-based flexible devices on polymeric substrates. Additionally, the low-temperature synthesis can enable the monolithic-three-dimensional (M3D) hetero-integration of various 2D materials on flexible platforms, paving the way for novel flexible 3D electronics with high integration density and unparalleled multifunctionality. Recent research has demonstrated a substantial achievement with the wafer scale synthesis of high quality $MoS_2$ monolayers at a temperature as low as 150 ˚C on the UTG and polymer substrate [11]. An alternative approach is the all-solution-based processing of device fabrication, which includes the synthesis and printing of 2D materials and other device components on flexible substrates at low temperatures. Although 2D materials with ultrathin thickness exhibit high flexibility and are ideal for flexible electronic applications, they are also susceptible to strain-induced damage. Therefore, to avoid mechanical failure in 2D materials-based flexible devices, it is crucial to adopt a suitable device architecture with proper strain management. Integrating the active 2D materials near the neutral mechanical plane (NMP) in the device architecture substantially decreases the risk of mechanical failure. Furthermore, ensuring consistent performance under repeated deformations is crucial, particularly for the applications of wearable healthcare devices, which require stretchability comparable to that of human skin (~60%). To achieve such mechanical virtues in the devices, several non-conventional architectures, such as serpentine, buckling, origami, and kirigami, have been employed in interconnects and active 2D materials. As the field advances, 2D materials-based flexible electronics devices have been integrated with wireless communication capability, particularly flexible healthcare devices for efficient data recording and transfer while maintaining human comfort. Additionally, with the rapid advancement in artificial intelligence (AI) technology, machine learning-based data analysis techniques can dramatically decrease the time required for analysing physiological data and can provide real-time clinical information to users.

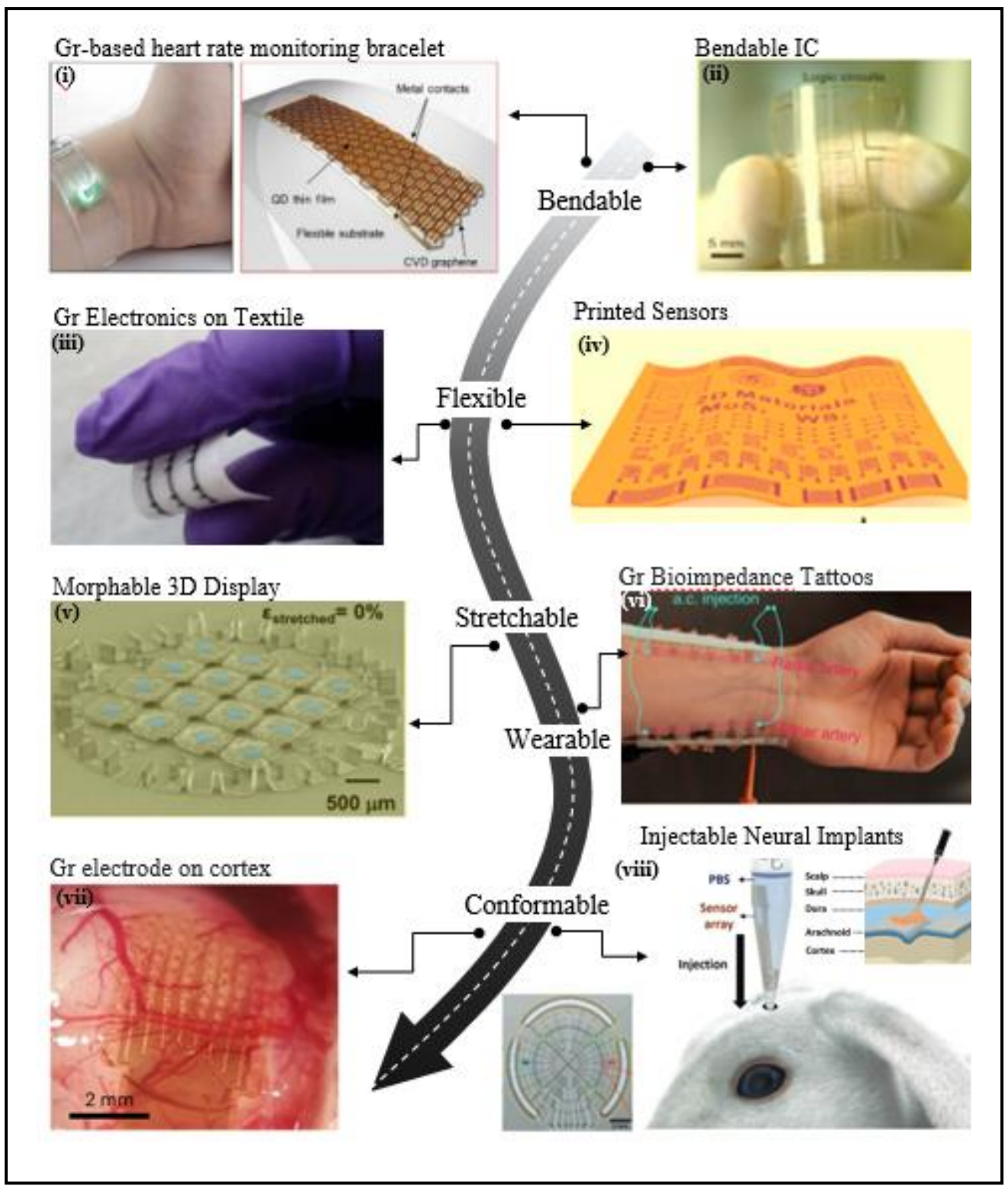

**Figure 2.** A technological roadmap representing the development of 2D materials based flexible electronic applications. (i) Graphene based bendable smart wrist band. Reprinted with permission from [2] (ii) Bendable integrated circuit. Reprinted with permission from [3] (iii) Graphene based flexible electronics on textile. Reprinted with permission from [4] (iii) Printed sensor on flexible polymeric substrate. Reprinted with permission from [5] (v) Morphable 3D display. Reprinted with permission from [6] (vi) Graphene based bioimpedance tattoo. Reprinted with permission from [7] (vii) Graphene electrodes attached on rat cortex. Reprinted with permission from [8] (Viii) Graphene and $MoS_2$ based Injectable neural implants. Reprinted with permission from [9].

## Concluding Remarks

The advent of 2D materials as alternatives to traditionally brittle materials marked a pivotal moment for the field of flexible electronics. Although substantial progress has been made toward the utilization of 2D materials in flexible electronics at the laboratory scale, their commercial implementation is still at the early stages and an in-depth understanding and technological development to overcome various challenges associated with 2D technology is needed. Specially, it is essential to establish a production and fabrication line specifically designed for the manufacturing of devices utilizing 2D materials similar to that of conventional Si foundry.

**Acknowledgements**
This work was supported by the National Research Foundation of Korea (NRF) funded by the Korean Government (RS-2024-00435661).

## 19. Next-generation neuromorphic, quantum, and spintronic computing

**Igor Aharonovich[1], Mark C. Hersam[2] and Stephan Roche[3]**

[1]University of Technology Sydney, Australia
[2] Northwestern University, USA
[3] ICREA and Catalan Institute of Nanoscience and Nanotechnology, Spain

### Status

The design of Van der Waals (vdW) heterostructures, integrating two-dimensional materials (2DMs), offers fascinating opportunities for material innovation and the integration of novel charge, spin and optical features in functional devices. Thanks to a large 2DMs family portfolio, including metallic, magnetic, superconducting, semiconducting to insulating structures, unprecedented possibilities for material design in computing architectures are made possible, especially in the fields of neuromorphic computing, spintronics and quantum information processing. Particularly specific is their atomically thin nature which paves the way for ground-breaking band-structure engineering through "proximity" effects, opening new avenues for exploration in nanotechnology and ultracompact device design.

The recent proliferation of artificial intelligence and machine learning (AI/ML) has led to significant demands on conventional von Neumann computing architectures. In particular, the need for training on massive datasets for AI/ML results in uncomfortably high-power consumption due to the need to move large amounts of data between spatially separated memory and logic blocks. To overcome this von Neumann bottleneck, alternative neuromorphic (i.e., brain-like) computing concepts are being explored. Neuromorphic computing draws inspiration from the brain (Fig. 1) such as the co-location of memory and information processing, thus minimizing the movement of data and reducing power consumption by orders of magnitude compared to a digital computer. Other neuromorphic characteristics include highly nonlinear spiking voltage profiles that are analogous to the action potentials in biological neurons and hyper-interconnectivity that attempts to mimic the hundreds to thousands of synaptic connections per neuron in the brain. With high gate-tunability that facilitates nonlinear charge transport and unconstrained vdW bonding that enables integration into complex geometries, 2DMs have emerged as leading candidates for next-generation neuromorphic computing [1].

In what concerns quantum technologies, 2D materials and various heterostructures have emerged in full force. Vital components of scalable quantum chips such as quantum light sources, photon detectors and nanoscale sensors have been realised. In addition, availability of myriad of strong excitonic systems, coupled with the ability to optically select the valley state, may pave the way towards many body physics and exploration of exotic phases of matter through quantum simulations. In a more detailed approach, realisation of quantum technologies from layered 2D vdW systems can stem from either exploring excitonic phenomena (limited to transition metal dichalcogenides, phosphorene, and related 2D semiconductors) or defects in wide band gap systems (hBN)[2]. The final application will dictate the hardware selection. As an example, valleytronics and many body exciton physics would be pursued with the former, while quantum sensing might be realised with the latter. It is important to add that both TMDCs and hBN can be interfaced with traditional cavities and photonic components – fabricated both from the host crystal, as well as from an external material (e.g. silicon nitride). This diversity of structures and properties truly makes the library of 2D materials a unique and rich playground for quantum information technologies for years to come [3].
In the realm of spintronics, the demonstration of proximity effects between graphene and strong spin-orbit coupling materials or magnetic materials have revealed ground-breaking phenomena, such as the room-temperature manipulation of pure spin currents through the spin Hall effect [4], or the imprinted exchange coupling in graphene from layered antiferromagnets that have achieved values on the order of 170 Tesla on length scales of hundreds of nanometres [5]. In a recent roadmap, scientific leaders and representatives from large industries such as Samsung, Global Foundry, and Thales have established the main advantages of 2DMs for spintronic applications and discussed the way for the further development of 2DM-based non-volatile memory technologies [6]. Furthermore, horizons for developing topological spintronics have been proposed theoretically [7].

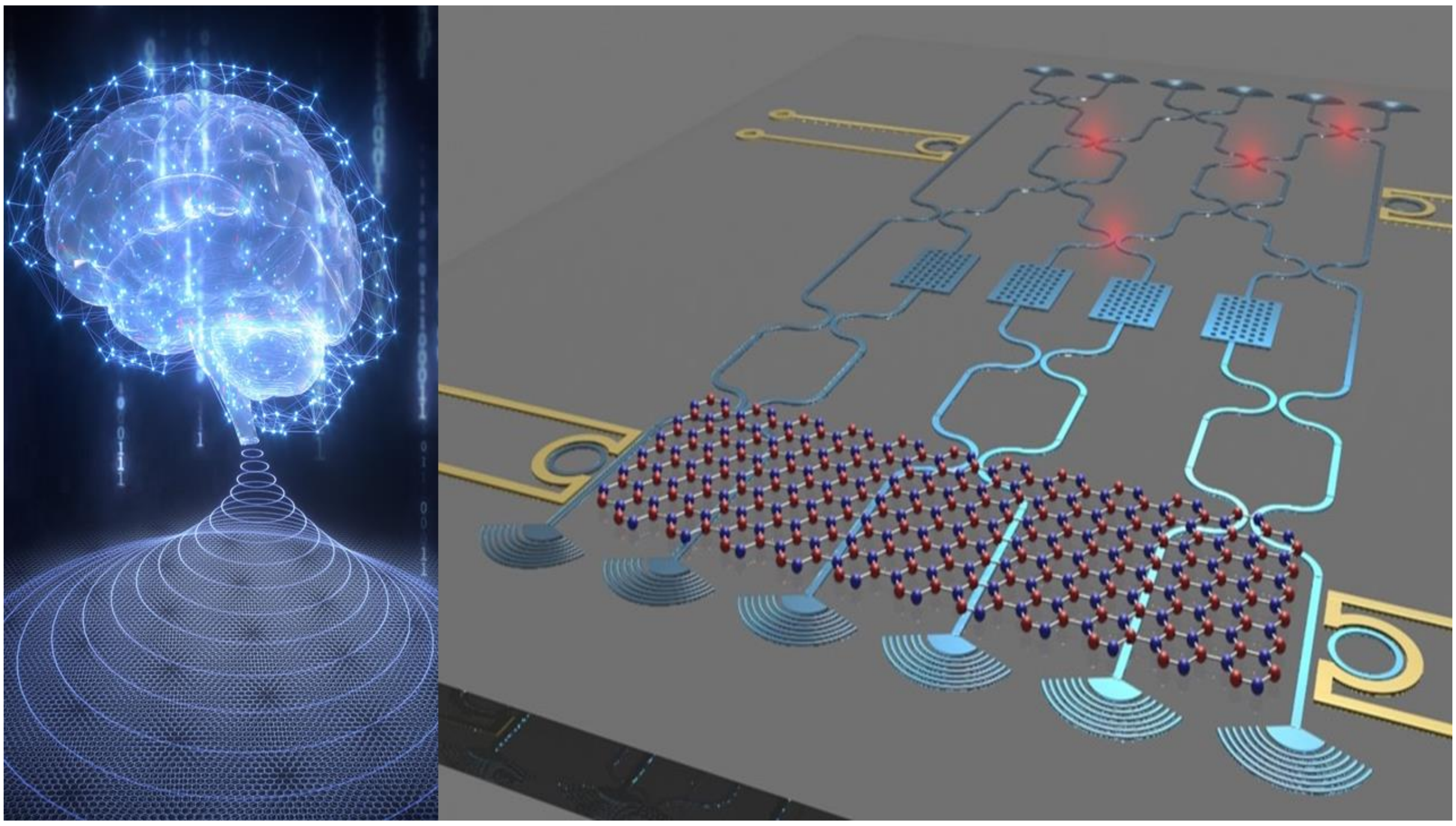

**Figure 1.** Left panel: With high gate tunability and unconstrained vdW integration possibilities, 2DMs are leading candidates for next-generation neuromorphic (i.e., brain-like) computing. Rigth panel: Schematics of a coupled cavity–emitter system using hexagonal boron Nitride two-dimensional materials

## Current and Future Challenges

Due to weak dielectric screening in 2DMs, nearly any 2D device can be made gate-tunable, thus presenting opportunities for achieving highly nonlinear charge transport characteristics that mimic biological neurons and synapses. For example, the carrier concentration on each side of 2DM p-n heterojunctions can be tailored with applied gate potentials, enabling nonlinear anti-ambipolar characteristics that can be harnessed for voltage spiking in addition to energy-efficient platforms for AI/ML classification [8]. While early demonstrations have achieved voltage spiking on biologically relevant timescales (i.e., kHz frequencies), opportunities remain for significantly increasing the frequency response to timescales characteristic of modern digital computers (i.e., GHz frequencies). In another example, by exploiting the low barrier to defect motion at grain boundaries in polycrystalline 2DMs, doping profiles can be reconfigured laterally with modest electric fields. These reconfigurable doping profiles provide non-volatile changes in conductance (i.e., memristor responses) that are analogous to synaptic responses in biological neural networks. By combining this memristive behaviour with gate-tunability of the overall carrier concentration (i.e., transistor responses), memory and information processing can be achieved at the same point in space in a hybrid memristor/transistor device that is referred to as a memtransistor. Moreover, the planar geometry of memtransistors implies that multiple contacts can be applied to the channel region to realize high degrees of interconnectivity in addition to bio-realistic heterosynaptic phenomena [9]. While 2DM memtransistors achieve many desirable neuromorphic features, their reliance on defective, polycrystalline materials presents challenges for realizing device-to-device uniformity over large areas, especially as the individual device sizes are scaled to nanometre dimensions. Most recently, the unique physics of twisted moiré heterostructures has presented additional opportunities for 2DM neuromorphic computing including room-temperature excitonic ferroelectricity and ratcheting charge transport that enable hardware implementation of higher order cognitive function such as associative learning with input-specific adaptation (Fig. 2) [10]. However, these moiré synaptic devices have only been realized through exquisite micromechanical exfoliation and sequential stamping at the single flake level, which presents significant challenges for scalable manufacturing. In this regard, direct growth and/or large-area sequential transfer with precise angular control will be required to fully exploit moiré phenomena in practical neuromorphic computing systems.

For spintronics, as discussed in [6], there are many avenues for reducing the write current for practical magnetic random-access memory (MRAM) technologies (especially the STT and SOT-MRAM devices) while largely improving compactness of the technology and lowering power consumption. However, many challenges remain to be tackled. The main hurdle lies in the co-integration of 2DMs in heterostructures in a scalable way, limiting the use of dry or wet transfer and stacking by material assembly. Direct growth of heterostacks using chemical vapor deposition (CVD) or molecular beam epitaxy (MBE) methods would be ideal, but it requires ultimate control of gas flow, substrate quality, and homogeneous temperature monitoring. On the other side, if advanced modelling is essential to anticipate on the upper limits of device performance, the variety of structural properties (number of layers in the stack, twist angle, substrate or encapsulation materials, disorder forms, etc.) is too large for realistic modelling to be implemented through “human” action. Consequently, artificial Intelligence-based approaches have become a cornerstone to accelerate material design and access to predictive modelling of their properties.

In the realm of quantum technologies, a key outstanding issue is understanding the photophysical properties of the quantum light source. Within the TMDC communities, there is a growing debate whether these are strain induced or localised to a particular defect. On the other hand, within the hBN community, the crystallographic nature of the majority of the visible emitters is still unknown, with a growing proposition that these are originating from donor–acceptor pairs or a unique localisation within the N antibonding orbitals [11]. In addition, it is yet to be seen what the most promising architecture is to combine quantum light sources based on 2D systems with integrated photonics on chip. One can envision a fully integrated, wafer-scale vdW quantum photonic chip, given the high refractive index of some of the TMDCs, and the ability to grow those over large areas.

**Advances in Science and Technology to Meet Challenges**

Advances in material fabrication and integration are definitely at the core of future improvement and commercialization of emerging technologies integrating 2DMs in neuromorphic computing, quantum technologies and spintronics. Possible twist angle control would enlarge the possibility for optimising a given property but that would also require robotic assembling to be operational at wafer scale. Despite proof-of-principle demonstrations, significant efforts are needed to develop more systematic approaches and tools able to upscale device fabrication in a manner compatible with industrial-scale manufacturing.

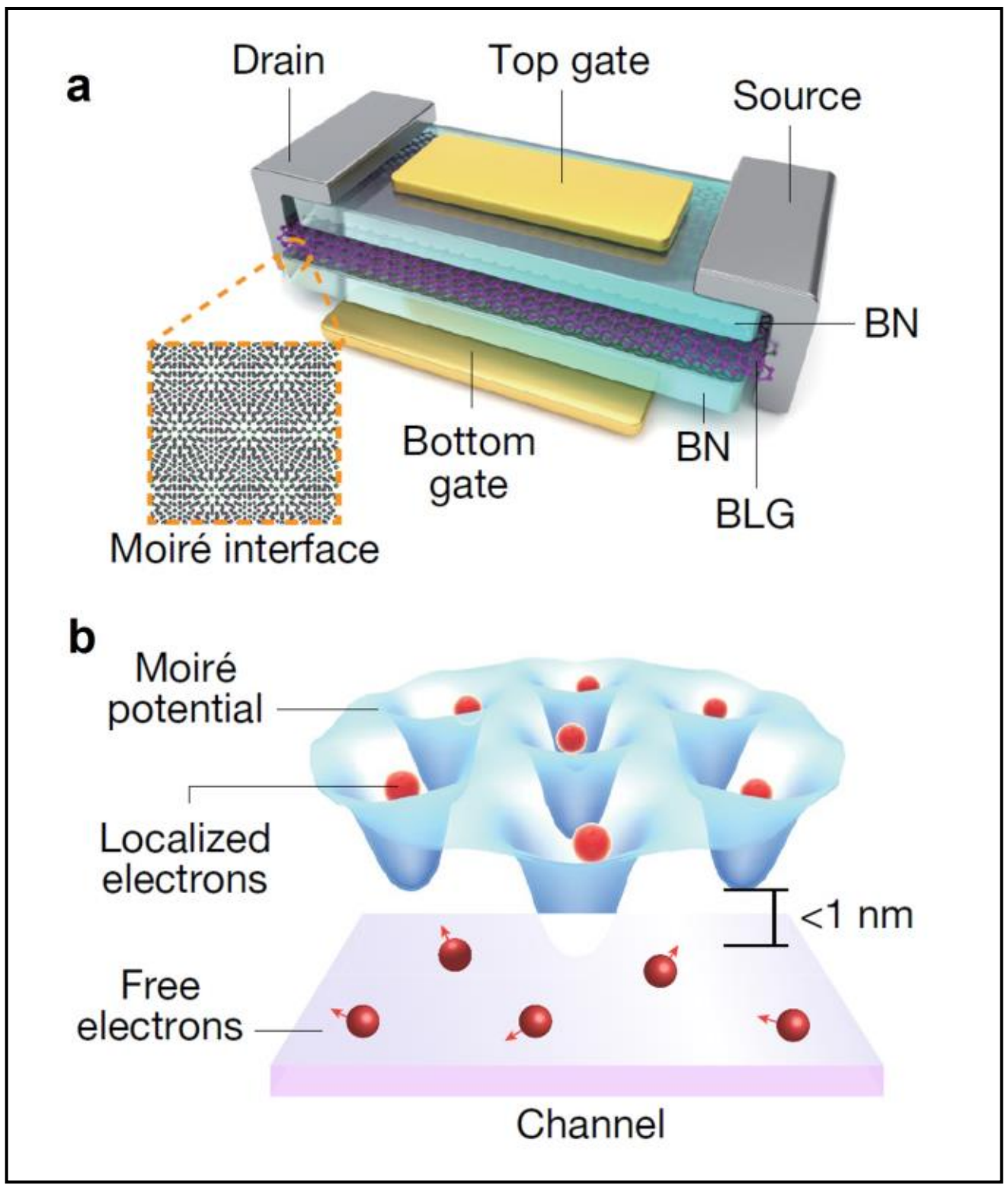

**Figure 2.** (a) Schematic of a moiré synaptic transistor produced from a twisted vdW heterostructure based on hexagonal boron nitride (BN) and bilayer graphene (BLG). (b) The moiré potential between the top BN and top BLG graphene results in a localized electron layer in direct proximity to a free electron channel, thus enabling ratcheting charge transport for advanced neuromorphic functionality. Panels a and b are adapted from Ref [10].

2D spintronics will also need to extract from the everlasting increasing family of 2DMs, the way to benchmark the best combination of materials for targeted metrics (e.g., larger spin-orbit torque efficiency with free field switching operation at the lowest possible current for magnetization switching). Again, this quest will demand the massive use of AI tools for, material fabrication and exploitation of measurement data.

## Concluding Remarks

After twenty years of intense research following the discovery of graphene, the enlarged family of 2DMs has brought high value for active control of charge and spin degrees of freedom, opening possibilities for more energy efficient and compact technologies able to solve current limitations of standard technologies in computing. However, despite the large investment and efforts worldwide, the lack of synergies between industry and academia and a suitable platform to network the concerned partners is a limiting factor for faster transfer from fundamental research to applications and commercialization, especially in the deep tech sector. International collaboration and large-scale public programs (such as the Graphene and Quantum Flagships in Europe) have initiated incubators and accelerators of knowledge-to-technology developments, but challenges remain in proper alliance design for overcoming the great challenges of implementation and deployment of breakthroughs to society as a whole.

## Acknowledgements

M.C.H. acknowledges support from the National Science Foundation EFRI BRAID Program under award number NSF EFMA-2317974 and the National Science Foundation Materials Research Science and Engineering Center at Northwestern University under award number NSF DMR-2308691. S.R

acknowledge grant PCI2021-122035-2A-2 funded by MCIN/AEI/10.13039/501100011033 and European Union ``NextGenerationEU/PRTR" and the support from Departament de Recerca i Universitats de la Generalitat de Catalunya. I.A. acknowledges the financial support from the Australian Research Council (CE200100010, FT220100053) and the Office of Naval Research Global (N62909-22-1-2028).

## 20. 2D Materials-Based Sensors

**Qilin Hua[1], Guozhen Shen[1*] and Tianling Ren[2**]**

[1] School of Integrated Circuits and Electronics, Beijing Institute of Technology, China
[2] School of Integrated Circuits, Tsinghua University, China

E-mail: *gzshen@bit.edu.cn **rentl@tsinghua.edu.cn

**Status**

Since its discovery in 2004, graphene, a single layer of carbon atoms arranged in a hexagonal lattice, has marked a breakthrough, and paved the way for research in the emerging field of 2D materials (2DMs). Due to their unique electronic, optical, and mechanical properties, the 2DM systems, such as transition metal dichalcogenides (TMDs), black phosphorene (BP), and MXene, have been extensively studied from material properties to emerging applications. Notably, 2DM exhibits high surface-to-volume ratios, excellent mechanical flexibility, and remarkable sensitivity to external stimuli, making them highly suitable for sensing applications.[1, 2] 2DM sensors offer exceptional capabilities in detecting a wide range of signals, including temperature, humidity, light, strain, gas, and biomolecules. Moreover, their atomic thinness enables the development of flexible and transparent sensors that can be seamlessly integrated on diverse substrates.

Over the past decade, significant progress has been made in the emerging field of 2DM sensors. Graphene and other 2D materials, ranging from MXene to h-BN (Figure 1a), have been extensively investigated in sensing applications. Such new layered materials, including metallic, semiconducting, and insulating types, have been utilized as active sensing elements. For instance, graphene-based thermometers have been developed for sensing temperature, MXene-based photodetectors for sensing infrared (IR) light, BP-based gas sensors for detecting volatile organic compounds, $MoS_2$-based strain sensors for monitoring human blood pressure, and rGO-based biosensors for sweat diagnostics. [1-4] Specifically, the sensing mechanisms in 2DM sensors rely on the interaction between the analytes or target molecules and the 2DM itself, affecting sensor design and optimization. [2] Common sensing mechanisms include piezoelectric, electrical, optical, or electrochemical effects (Figure 1b). The piezoelectric mechanism involves changes in the band structure and electrical conductivity of 2DM (e.g., $MoS_2$) under strain, enabling tactile sensing. The electrical mechanism relies on modifications in charge carrier concentration, mobility, or bandgap of 2DM (e.g., semiconducting TMDs) in field-effect transistors (FETs)[5] due to the presence of analyte molecules in the channel region. The optical mechanism exploits changes in the absorption, emission, or scattering properties of 2DM when analyte molecules bind, facilitating sensitive and label-free detection. The electrochemical mechanism involves the quantitative detection of electron transfer or oxidation/reduction reactions between the 2DM-modified electrode surfaces and analytes for environmental or health monitoring.[1] Ongoing research aims to uncover new mechanisms and refine sensor designs to enhance the sensitivity, selectivity, and reliability of 2DM sensors.

As FET being the fundamental sensor structure, 2DM FET sensors have the capability of converting physical, chemical, or biological quantities into electrical signals and can modulate performance using diverse gates (Figure 2c), enabling fast and sensitive detection.[2] To parallel and simultaneous monitor multiple analytes or parameters, 2DM sensor arrays have been developed, offering enhanced selectivity, sensitivity, and multiplexing capabilities. In addition, the associated circuit is designed to allow for on-chip signal processing, calibration, and data analysis. With the integration of 2DM sensors and CMOS peripheral circuits (Figure 1d), the 2DM sensor arrays and circuits offer promising avenues for advanced sensing technologies. Moreover, multifunctionalities in sensor array systems can be implemented by a monolithic 3D (M3D) integration of 2DMs and Si or III−V layers via van der Waals growth and transfer techniques.[6] Figure 1e illustrates the M3D architecture, which enables vertically stacking multiple chiplets in 3D packaging. This stacking integration allows for the combination of various layers, such as transmitters, 2DM sensors, memory, and Si CMOS, thereby providing additional functionality. Incorporating artificial intelligence (AI) algorithms, such as artificial neural networks, enhances the performance of 2DM sensors, addressing challenges related to cost, speed, and power consumption.[7] Additionally, in contrast to rigid substrate-based sensor chips, flexible and stretchable sensors based on 2DMs can offer disruptive solutions for emerging applications such as healthcare monitoring, humanoid robotics, and human-machine interface technologies.[8]

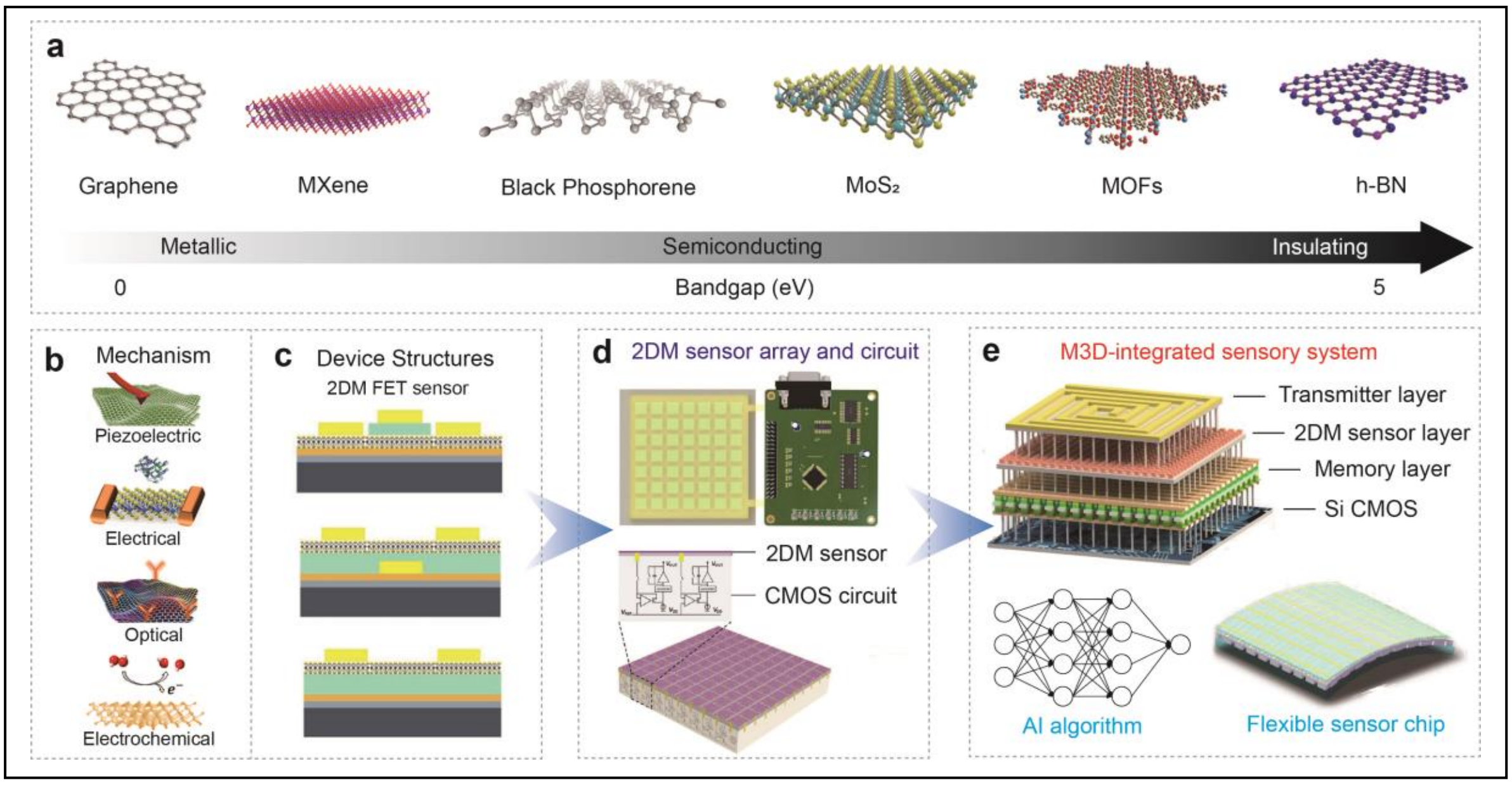

**Figure 1**. Roadmap of 2D materials (2DM)-based sensors. (a) Typical 2D materials for sensing applications, including metallic, semiconducting, and insulating types. (b) Illustration of sensing mechanisms in 2D materials. Reproduced from [1], with permission from Elsevier. (c) Schematic of device structures of 2DM sensors based on FET configurations. (d) Current progress in 2DM sensor arrays and the CMOS peripheral circuits. (e) Future development trends of 2DM sensors, including M3D-integrated sensory systems, advanced AI algorithms, and flexible sensor chips. Reproduced from [2], with permission from Springer Nature.

**Current and Future Challenges**

While 2DM sensors offer great promise, there are several challenges that researchers are currently addressing and will continue to face in the future.

**Material quality and scalability**: Achieving large-scale, high-quality production of wafer-scale 2DMs with consistent properties is essential to ensure uniformity and reproducibility across specific sensor devices. Controlling the growth, synthesis, and transfer processes is crucial in this regard. Van der Waal integration of 2DM in both n-type and p-type configurations is highly desirable for extending functionalized sensing capabilities.

**Sensing performance**: A deep understanding sensing mechanisms at the atomic and molecular level is essential for optimizing 2DM sensor design. Enhancing the sensor's ability to accurately discriminate between different substances or stimuli (i.e., selectivity) is crucial for accurate and reliable sensing. Strategies to mitigate degradation caused by environmental factors such as humidity, temperature, and chemical exposure are necessary to ensure long-term stability. Additionally, integrating multiple sensing modalities or functionalities into a single 2DM sensory system presents a complex task.

**Integration and compatibility**: Seamless integration with existing Si-based electronics and systems necessitates compatibility with substrates, encapsulation materials, and fabrication processes. Addressing challenges related to coordinating different sensing elements, signal processing, and data fusion in system-level integration is crucial for implementing advanced multimodal sensing systems.

**Power efficiency**: Striking a balance between high performance and low power consumption is crucial for battery-powered applications. Strategies such as reducing power requirements, enhancing energy efficiency, or exploring self-powered configurations are necessary to optimize power efficiency in 2DM sensors.

**Cost-effectiveness**: Advancements in manufacturing processes, material synthesis techniques, and economies of scale are needed to reduce the cost of 2DM sensors. Improving cost-effectiveness enhances accessibility and commercial viability, making 2DM sensors more widely available.

**Advances in Science and Technology to Meet Challenges**

To address the challenges in 2DM sensors, a range of key techniques and methodologies are presented, encompassing material-level design, device-level fabrication, and system-level integration, as shown in Figure 2.

**(1) Material-level design**. Advanced techniques in layered material synthesis, such as optimized growth recipes, substrate engineering, and doping, are being developed to enable large-scale and high-quality production of wafer-scale 2DMs with uniform and defect-free attributes. In addition, advanced techniques in materials characterization, such as high-resolution in-situ electron microscopy and spectroscopy, would provide insights into the structure, composition, and properties of 2DMs, facilitating material quality control and optimization. Advancements in theoretical modeling, computational simulations, and experimental techniques are essential to deepen the understanding of sensing mechanisms in 2DM sensors. By studying the interactions between analytes or stimuli and 2DMs at different length and time scales, valuable insights into the electronic and chemical processes involved in sensing can be obtained. Theoretical analysis with techniques such as atomistic simulations, density functional theory calculations, and quantum mechanical modelling contribute to this understanding and verification. Overcoming challenges related to selectivity, stability, and integration of 2DM sensors requires innovative sensor designs. Exploring novel architectures like heterostructures and hybrids enhances sensing performance, while integration with complementary materials (e.g., polymers or nanoparticles) can add functionalities and improve selectivity. Additionally, strategies, including encapsulation techniques, protective coatings, and interface engineering, play a crucial role in enhancing the stability and reliability of 2DM sensors.

**(2) Device-level fabrication.** Leveraging their compatibility with CMOS technology, 2DM sensors can adopt micro-electro-mechanical system (MEMS) processes that are highly desirable for making the integration of sensing and processing functions on a single chip. To ensure seamless integration of 2DM sensors into system-level architectures, compatible fabrication processes, interconnectivity strategies, and interface engineering are crucial to achieve excellent compatibility with existing electronics. Advancements in material synthesis methods, such as chemical vapor deposition, and solution-based preparation are essential for enhancing the quality and scalability of 2DM sensors. Scalable manufacturing processes like roll-to-roll or large-area fabrication techniques enable high-throughput production at reduced costs, facilitating the widespread adoption of 2DM sensors. Furthermore, the utilization of reliable interfacial materials and flexible manufacturing approaches (e.g., 3D printing and deformation techniques) enable the integration of 2DMs, including van der Waals heterostructures, into flexible and wearable sensory electronics.

**(3) System-level integration.** Optimizing the power efficiency of 2DM sensors can be achieved through energy-efficient sensor designs and low-power peripheral circuits. The utilizing of piezoelectric, thermoelectric, or triboelectric materials enables the development of self-powered sensory systems, while advanced signal processing methodologies enhance power efficiency for high-performance sensing applications. Moreover, as the complexity and volume of data generated by sensors increase, the development of advanced AI algorithms becomes crucial to improve the reliability and accuracy for real-time data processing, pattern recognition, and data fusion in 2DM sensory systems. AI-based approaches enable autonomous sensor operation, adaptive calibration, and intelligent decision-making. Leveraging the inherent parallelism and low-power operation of 2DMs, novel architectures,[9, 10] such as neuromorphic computing or in-sensor computing, enable real-time processing of sensor data and the implementation of deep learning algorithms directly within the sensory system. These advancements revolutionize data processing and analysis, leading to more efficient, intelligent, and autonomous sensing applications.

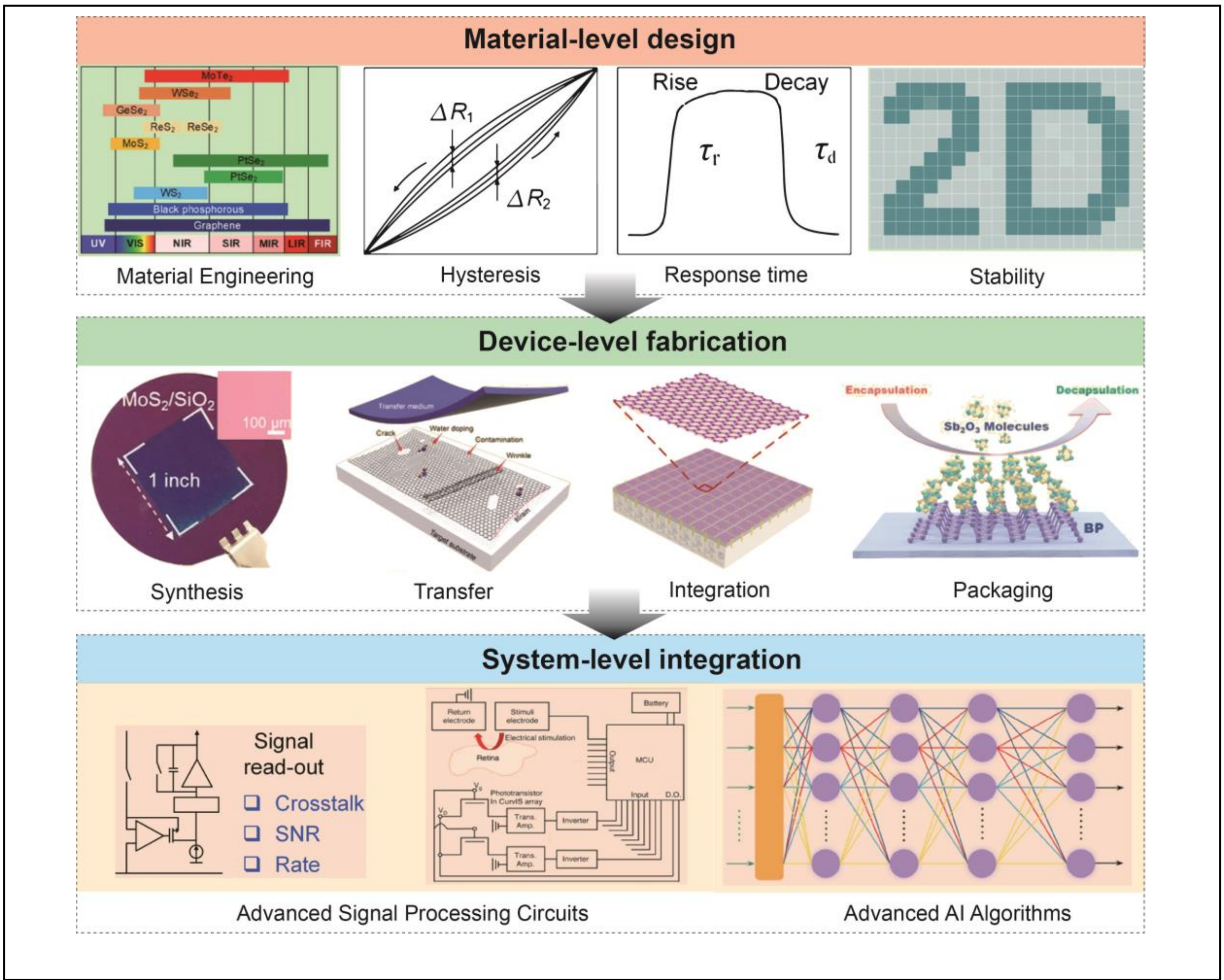

**Figure 2**. Challenges in developing 2DM sensor chips from the perspective of material-, device-, and system-level considerations. Reproduced from [2], with permission from Springer Nature.

## Concluding Remarks

In the post-Moore era, as chip manufacturing approaches physical size limits, the "More than Moore" paradigm has emerged. 2DMs, possessing high surface-to-volume ratios, excellent mechanical flexibility, and remarkable sensitivity to external stimuli, have highlighted significant progress in transformative sensing applications. The ongoing development of 2DM sensors with unique characteristics focuses on enhancing sensitivity, selectivity, response time, stability, and scalability through extensive studies in material engineering, integration techniques, and data analysis. These advancements will unlock novel intelligent applications, improve existing sensor technologies, and contribute to the development of energy-efficient advanced sensory systems (even in M3D architecture) for diverse fields such as healthcare, environmental monitoring, industrial process control, and beyond.

## Acknowledgements

The authors are thankful for the support from the National Natural Science Foundation of China (62374018, and 61904012), the National Key Research and Development Program of China (2023YFC3603501), Beijing Natural Science Foundation (L223006), and Beijing Institute of Technology Research Fund Program for Young Scholars.

## 21. 2D nanomaterials for electromagnetic interference shielding

**Hao-Bin Zhang[1] and Chong Min Koo[2]**

[1] State Key Laboratory of Organic-Inorganic Composites, Beijing University of Chemical Technology, Beijing 100029, China

[2] School of Advanced Materials Science and Engineering, School of Chemical Engineering, Sungkyunkwan University, 16419, Republic of Korea

**Status**

Electromagnetic interference (EMI) occurs when electromagnetic waves cause interference among various circuits and devices operating in proximity. With the proliferation of telecommunication devices and microelectronics, the detrimental effects of EMI have become more serious, necessitating stringent solutions. EMI shielding materials are employed to protect sensitive electronic devices and components from undesirable electromagnetic radiation through reflection, absorption, and multiple reflection mechanisms, which are primarily defined by the electrical conductivity, magnetic permeability, frequency and thickness and structural makeup of the materials, described by Simon's formulae shown in equation 1 and Figure 1a [1].

$$SE_T = 50 + 10\,log\left(\frac{\sigma}{f}\right) + 1.7t\sqrt{\sigma f} + SE_M \qquad (1)$$

where $\sigma$ ($S{\cdot}cm^{-1}$), $t$ (cm), and $f$ (MHz), and $SE_M$ refer to electrical conductivity, sample thickness, frequency, and shielding due to multiple reflection, respectively. Therefore, advanced EMI shielding materials with high conductivity, magnetic permeability, and controlled structural design are needed for efficient EMI shielding in applications such as electronic communication, consumer electronics, flexible and wearable devices, autonomous vehicles, medical equipment, and aerospace.

The emergence of two-dimensional (2D) nanomaterials, especially graphene and two-dimensional transition carbides/nitrides (MXenes), presents tremendous opportunities for EMI shielding material research and applications due to excellent electrical conductivity, 2D structural feature, and microstructure designability through solution-processability [2-4]. Since the pioneering work on graphene EMI shielding materials, studies on 2D nanomaterial and their related EMI shielding composites have attracted great interest from both academic and industrial communities. The number of research papers on 2D materials-based EMI shielding has surged, with graphene-based materials showing an overwhelming contribution compared to other competing materials (Figure 1b). In 2016, MXenes emerged as state-of-the-art EMI shielding materials, outperforming other synthetic materials of comparable thickness due to their superb metallic conductivity, tunable composition and surface chemistry, and solution processability. This triggered another wave of EMI shielding research into 2D nanomaterials [4]. The unique combination of high electrical conductivity, large surface area, high aspect ratio, and abundant surface functional groups allows for the production of efficient EMI shielding polymer composites/additives at low filler loadings, particularly when compared to traditional metal- and carbon-based shielding materials. Additionally, their nanometre-thick 2D structure enables them to assemble into ultrathin films, lightweight 3D architectures, and high-performance fibres/fabrics, adapting to various application scenarios (Figure 1c). While some graphene-based products are being commercialized, and their potential for scalable applications is optimistically expected, further in-depth and systematic research is still required to replace traditional EMI shielding materials fully.

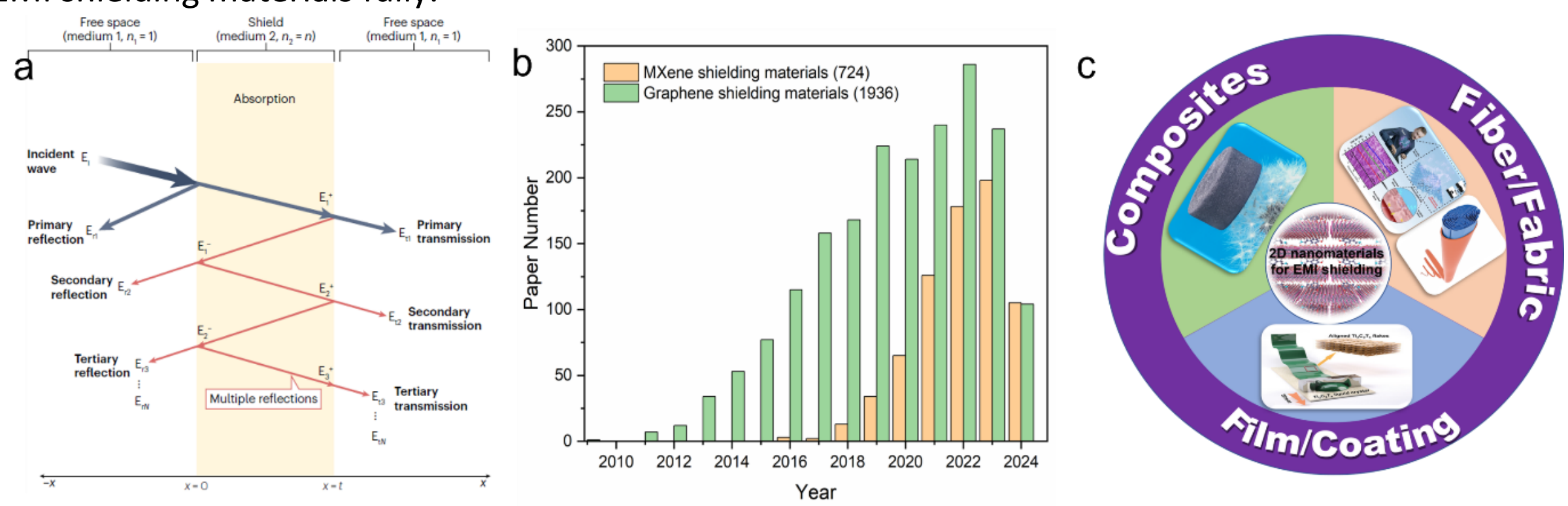

Figure 1. (a) EMI shielding mechanism in a film geometry; (b) The statistic paper number of graphene and MXenes in the past 15 years; (c) The forms of 2D nanomaterial-based EMI shielding materials.

## Current and Future Challenges

The research advances in 2D nanomaterials, notably Graphene and MXenes, have leveraged their intriguing physical characteristics and EMI shielding performances, positioning them as leaders in current and future lightweight EMI shielding applications. However, several challenges impede their practical applications, summarized as follows: (i) Developing controllable synthesis methods for 2D nanomaterials with well-defined nanostructures and intrinsic properties are demanded. Understanding the intrinsic electrical, dielectric, and electromagnetic characteristics of 2D nanomaterials is essential to meet various application requirements. While some scalable production methods have been developed for graphene-based EMI shielding materials, the precise fabrication of diverse MXene materials still requires significant efforts. (ii) The general and scalable production methods of 2D nanomaterial-based polymer composites and macroscopic assemblies with customized microstructures and performances are needed. Rational microstructure design of 2D materials can modulate EMI shielding performance and other characteristics, such as lightweight feature, mechanical flexibility/stretchability, visible transparency, and required mechanical properties, making 2D nanomaterial-based shielding materials suitable for various specific applications. (iii) The fundamental interaction mechanisms of electromagnetic waves with 2D nanomaterials and their macroscopic structures/composites are still not well understood. It is essential to explore how the hierarchical structures of macroscopic shielding materials affect the electromagnetic parameters, internal multiple reflections, attenuation mechanisms and broadband EMI shielding performances. Understanding the performance stability and evolution mechanisms under extreme conditions (e.g. high/low temperatures, moisture, UV exposure, and environmental oxygen) is crucial for the practical applications in complex scenarios. (iv) The relatively higher costs of 2D materials may be a significant obstacle to their practical applications compared to that of currently dominant metallic shielding materials. Scalable and cost-effective production technologies for 2D nanomaterial-based shielding materials are needed. Alternatively, identifying critical and irreplaceable applications for 2D materials could justify their use over the conventional EMI shielding materials. (iv) The reflection-dominant EMI shielding mechanism of current 2D materials, due to their high electrical conductivity, leads to undesirable secondary EM pollution. Developing absorption-dominant EMI shielding materials requires significant attention. (v) Current 2D EMI shielding materials are mostly investigated within a narrow frequency range. However, the operating frequency of highly integrated electronic systems is continuously increasing. Therefore, it is essential to explore these materials across a broader frequency spectrum, including terahertz, infrared, and beyond.

**Advances in Science and Technology to Meet Challenges**

Significant progress in the fabrication, property modulation, and applications of graphene- and MXene-based shielding materials has provided valuable insights for the research of 2D nanomaterials (Figure 2). However, further advancements are still needed to address current and future challenges. The EMI shielding research initially focused on polymer composites of conductive graphene, with tremendous efforts aimed at improving EMI shielding performance by optimizing matrix-filler interfaces, distribution/orientation, or forming efficient 3D networks. Importantly, free-standing 3D structures not only act as conductive networks for electron conductance in composites, but also serve as multifunctional lightweight aerogels, delivering both EMI shielding and thermal insulation. Techniques such as CVD, electrostatic assembly, emulsion-templated method, and 3D printing show great potentials for producing scalable and multifunctional EMI shielding materials.

MXenes offers versatile elemental compositions, nanolayer structures, surface terminations, and physical properties, presenting exciting opportunities for developing EMI shielding materials [4]. Among the available MXenes, $T_3C_2T_x$ is considered as the most promising candidate for EMI shielding applications [5]. The porous structure with rich interfaces in MXene foam significantly enhances EMI shielding capability [6]. Additionally, thermal annealing has been shown to enhance the microwave absorption properties of $Ti_3CNT_x$ films [7]. Precise regulation of MXene terminations and flake size will help clarify the fundamental shielding mechanisms and enable the preparation of high-performance films and fibres [8]. The deliberate design of interfacial interactions integrates high tensile strength, electrical conductivity, and shielding performance in 2D nanomaterial films [9]. Furthermore, processing approaches including drop-casting, spray-coating, and scanning centrifugal casting facilitate the scalable production and practical applications of 2D nanomaterial EMI shielding materials.

Fabricating fibres and fabrics using 2D nanomaterial, either by loading them onto existing fabric substrates or through wet-spinning, is also a promising approach for developing efficient EMI shielding [10]. Although graphene-based fibres have exhibited high electrical, thermal, and mechanical performances, significant work remains to be done for MXene fibres, despite some encouraging advances.

Developing predictive models to understand the EMI shielding mechanism of millimetre-wavelength EM waves interacting with nanometre-thin 2D materials will aid in designing superior shielding material for next generation electronics. Traditional electromagnetic theories mainly address the EMI shielding properties and mechanisms of uniform materials, making them not perfectly suitable for 2D nanomaterial composites. Machine learning methods and theoretical simulations can provide valuable insights and guidance for the rational design and mechanism understanding of 2D nanomaterial-based EMI shielding materials.

Developing magnetic MXenes will not only improve EMI shielding properties but also mitigate the secondary reflections and EMI pollution. Exploring novel precursors, including Cr, Mn, or other magnetic elements in single, bimetallic, or multi-metallic compositions, within ordered or solid-solution structures of MAX phases, could impart magnetic properties to the respective MXene structures. Currently, only Cr-based magnetic MXene has been reported at low temperatures, necessitating further developments to achieve room-temperature magnetic properties for absorption dominant and multispectral EMI shielding.

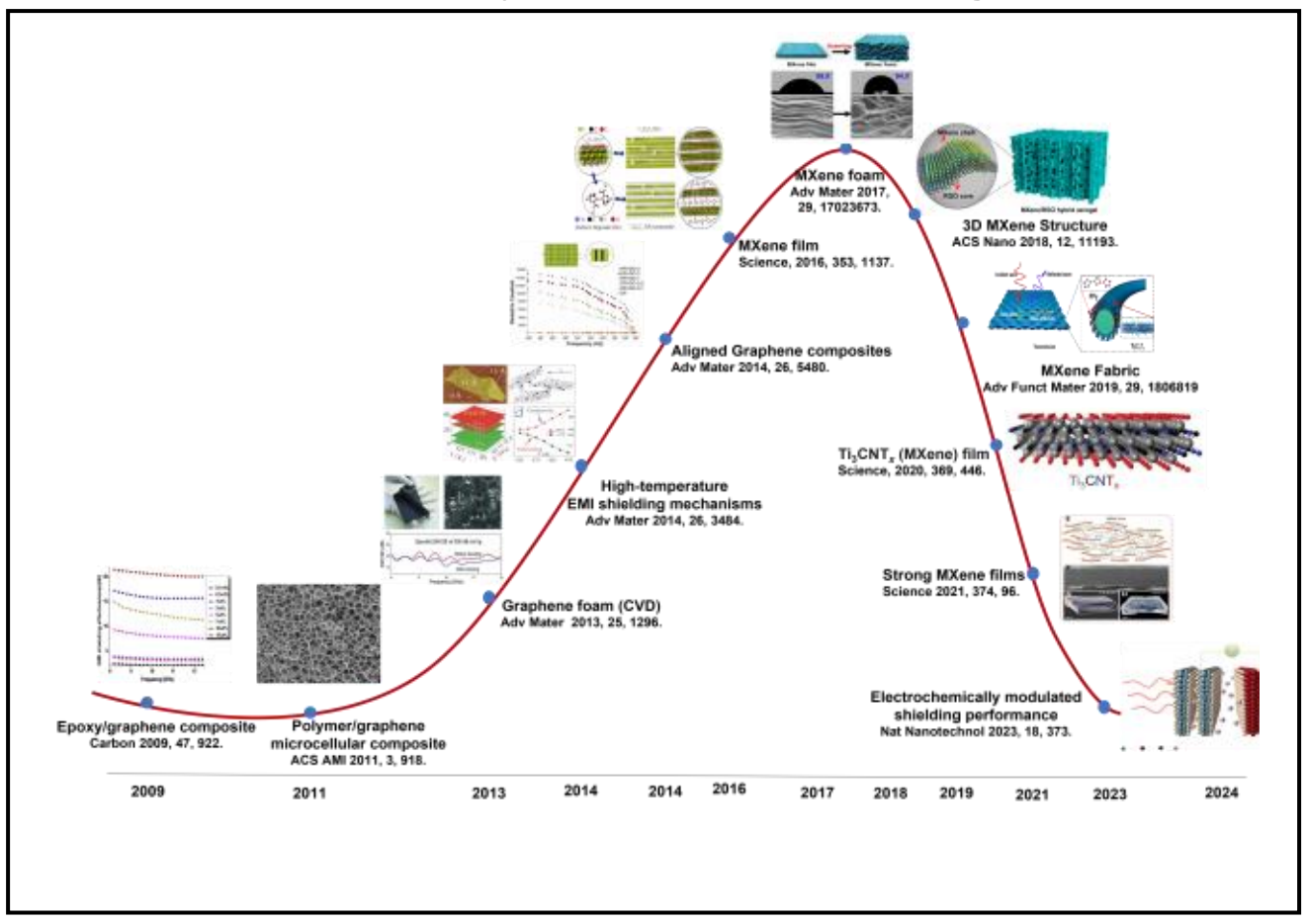

Figure 2. The significant advances in 2D nanomaterial-based EMI shielding materials.

## Concluding Remarks

Over the past 15 years, significant advancements have been made in the field of 2D nanomaterial-based electromagnetic shielding materials. This progress includes leveraging the intrinsic electromagnetic properties of 2D nanomaterials, developing polymer composites and hybrids, flexible films, and multifunctional fibres/fabrics, and understanding of EMI shielding properties and underlying mechanisms. However, critical issues and challenges remain, such as precisely modulating the structure and performance of 2D nanomaterials, integrating efficient electromagnetic shielding properties with other material characteristics, and developing large-scale production and application technologies for these materials. Future research should address these challenges to overcome barriers to the widespread adoption of 2D nanomaterials for EMI shielding applications in advanced electronics.

## Acknowledgements

This research was supported by the National Natural Science Foundation of China (52273064, 51922020). This study was also partially supported by grants (2021M3H4A1A03047327, 2022R1A2C3006227, CRC22031-000) funded by the Ministry of Science, ICT, and Future Planning, Republic of Korea.

## 22. Composites (2D Materials as Additives in Nanocomposites)

**Nikhil Koratkar[1], Vittorio Pellegrini[2] and Robert J Young[3]**

[1] Rensselaer Polytechnic Institute, USA
[2] BeDimensional, Italy
[3] University of Manchester, UK

**Status**

A range of 2D materials has been discovered over the past two decades that includes graphene, hexagonal boron nitride (h-BN), transition metal dichalcogenides (TMDs) and transition metal carbides, nitrides or carbonitrides (i.e., MXenes) to name just a few. Such 2D materials are typically derived from 3D van der Waals solids and can be exfoliated from the bulk by mechanical, chemical or electrochemical methods to isolate their 2D form [3]. Table 1 lists the basic physical, mechanical, electrical and thermal properties of typical 2D materials. The values in the table are combination of both experimental and theoretical studies. A review of Table 1 indicates that 2D materials such as graphene and h-BN possess low densities and high specific surface areas, which maximizes interfacial contact with the matrix material into which these fillers are dispersed. In terms of elastic properties, 2D materials offer Young's moduli in the range of 133 to 1000 GPa, with graphene having the highest value. For fracture properties, graphene offers the highest tensile strength (~130 GPa), h-BN displays the highest fracture toughness (~5.5 MPa√m), while MXenes exhibit exceptionally high strain-to-break (up to ~25%). Graphene is by far the most electrically and thermally conductive member of the 2D materials family, while h-BN is nearly electrically insulating. TMD materials such as $MoS_2$ have the highest dielectric constants. From the above discussion, it is evident that 2D materials offer an impressive range of physical, mechanical, electrical and thermal properties, and show outstanding potential to influence the properties of polymer, ceramic and metal matrix materials into which they are dispersed. The challenge is to what extent can this potential be realised in nanocomposites.

| | Physical properties | | Mechanical Properties | | | | Electrical/Thermal properties | | |
|---|---|---|---|---|---|---|---|---|---|
| | Density ($g/cm^3$) | Specific surface area ($m^2/g$) | Young's Modulus (GPa) | Ultimate tensile strength (GPa) | Fracture Toughness (MPa√m) | Strain to failure (%) | Electrical conductivity (S/cm) at RT | Dielectric constant | Thermal conductivity (W/m.K) |
| **Graphene** | 2.267 | 2630 | 1000 | 130 | 4 | 13-19 (armchair) 6-11 (zigzag) | $10^6$ | ~1.8 (in plane ~3 (out of plane) | 1500-2500 |
| **Molybdenum disulphide** | 5.06 | 5.28-27.82 | 270 ± 100 | 23 | 1.8 | 6 – 11 | $10^4$ | 15.5-15.9 (in plane) 6.2 – 6.9 (out of plane) | 23.2–155 |
| **MXenes** | 2.6 | 49.5-61.4 | 133-517 | 14-70 | - | 9-24 (armchair) 11-23 (zigzag) | 5000-15000 | - | 55.8 |
| **Hexagonal Boron Nitride h-BN** | 2.1 | 198.41 - 964.3 | ~860 | ~70 | 5.25 - 5.76 | 12.5 - 13.3 | $10^{-13}$ – $10^{-6}$ | 6.82-6.93 (in plane) 3.29-3.76 (out of plane) | 340 |

Table 1: Typical properties of two-dimensional (2D) materials that show promise as additives in nanocomposite materials [4].

**Current and Future Challenges**

Composites are materials that are constitute of at least two components and have properties superior to those of the component materials in isolation. The main properties of interest for nanocomposites are mechanical, thermal, electrical and barrier although perhaps the most exciting prospect is the ability of nanocomposites to excel in more than one of these properties, i.e. displaying multifunctionality. The addition of a filler with a high Young's modulus to a matrix with a low modulus leads to a composite with a Young's modulus higher than that of the matrix [5]. Table 1 shows that 2D materials have very high levels of intrinsic stiffness and strength that were anticipated to lead to nanocomposites containing 2D materials reflecting these properties. In the event it has been shown [6] that the main factor limiting the improvement in mechanical properties is the aspect ratio (length/thickness) of the 2D material (Figure 1(a)) controlling stress transfer from the matrix to the reinforcement. Other factors controlling stiffness include the orientation and degree of dispersion of the nanofillers. Figure 1(b) shows that there is an increase in Young's modulus upon adding graphene nanoplatelets (GNPs) to an epoxy resin and that the level of reinforcement increases with increasing *s*. Aspect ratio is found to have similar effect upon toughness, since higher levels of *s* lead to tougher nanocomposite materials through mechanisms such as interfacial debonding and pull-out [1]. On the other hand, the spectacular values of strength of 2D materials (Table 1) are never realised in nanocomposites as the interface will generally fail before the 2D reinforcement undergoes fracture.

**Addition of 2D Materials to Control Physical Properties**

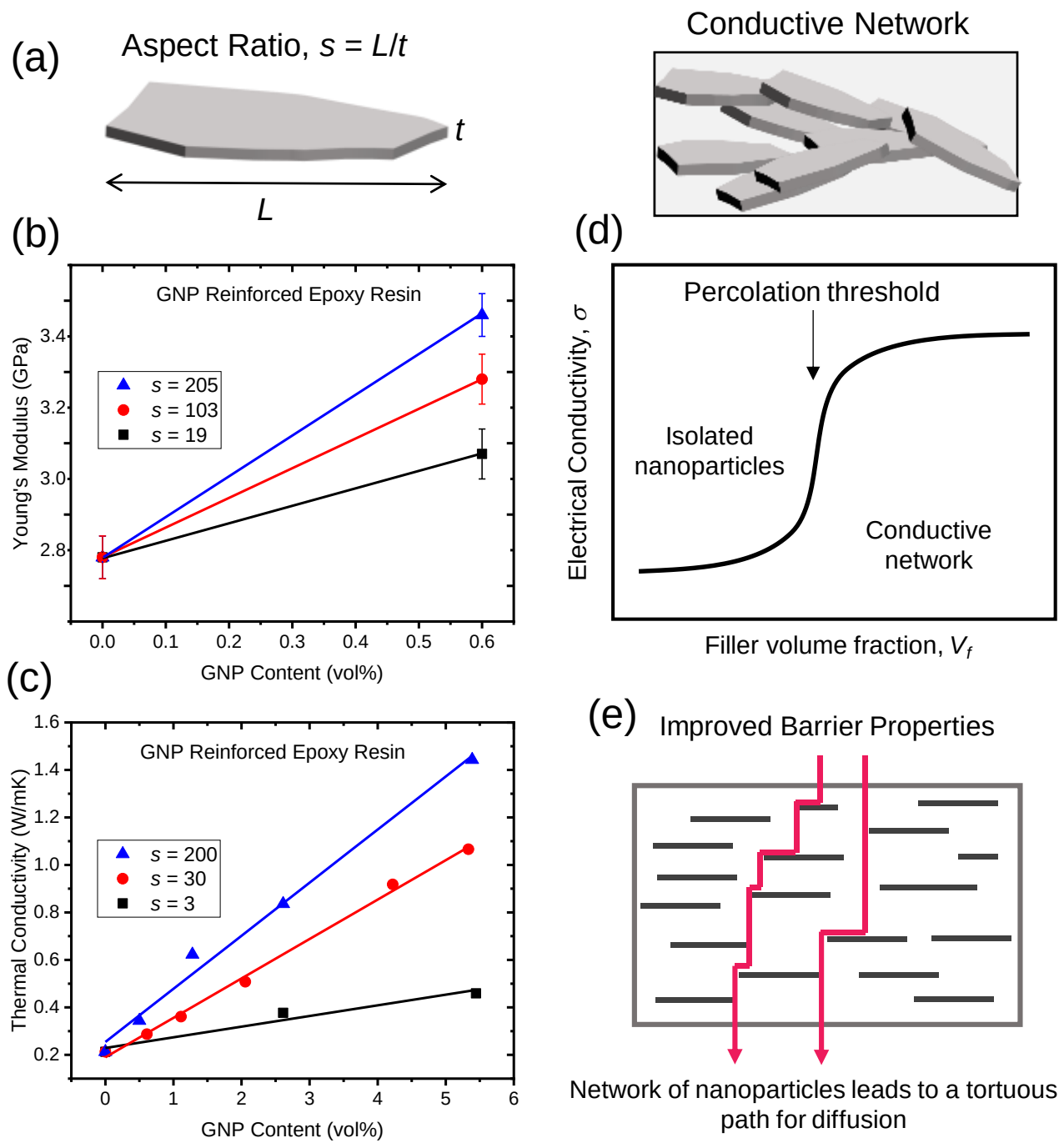

**Figure 1.** Control of the properties of nanocomposites by the addition of 2D materials (a) Definition of aspect ratio. (b) Dependence of the Young's modulus of an epoxy resin upon the addition of GNPs (Data taken from [1]). (c) Dependence of thermal conductivity of an epoxy resin upon the addition of GNPs (Data taken from [2]). (d) Increase in electrical conductivity with the addition of a conductive nanofiller showing the percolation threshold and conductive network. (e) Improvement in barrier properties through the formation of a tortuous path.

One general feature of the behaviour is that if the property of the nanofiller is orders of magnitude different from that of the matrix, the scope for improvement is limited as the result of the mechanisms involved. This is shown in Figure 1(c) for the addition of GNPs to an epoxy resin where although a higher *s* leads to better enhancement of the thermal conductivity of the resin, there is only a factor of 5 improvement, despite the thermal conductivity of graphene being up to $10^4\times$ higher (Table 1). Although the values of Young's modulus and thermal conductivity generally increase linearly with nanofiller loading, electrical conductivity shows quite different behaviour as seen schematically in Figure 1(d). The conductivity remains similar to that of the insulating matrix at low loadings and then undergoes a rapid rise by many orders of magnitude at the percolation threshold once a conductive network is formed. One advantage of 2D nanofillers, as opposed to nanoparticles or nanotubes, is that they have they have the potential of being used in barrier applications as they have the possibility of forming networks that lead to a tortuous path for the diffusion of a gas through a polymer matrix as shown in Figure 1(e).

**Advances in Science and Technology to Meet Challenges**

One of the main challenges in the development of nanocomposites reinforced with 2D materials is by obtaining the nanofillers with good control of nanoparticle microstructure (size, shape and aspect ratio) and the ability to produce them reproducibly in large quantities at a reasonable cost [3]. On the other hand, there is little need to develop any new nanofillers as we now have a broad palette of materials available with a whole range of impressive physical properties (Table 1).

One of the attractions of polymer-based nanocomposites is that they are generally easy to process and can be moulded into complex shapes using conventional techniques such as injection moulding. There is plenty of scope for developments in production processes such as additive manufacturing [7] and using inks based upon 2D materials for the ink-jet printing of components [8]. It is well established that the properties of nanocomposites depend strongly upon their microstructure [6] as shown in Figure 2(a). The different processing techniques could be employed to control and optimise microstructure of nanocomposites. For example, the best mechanical properties are generally obtained with uniform distributions of nanofillers whereas recent studies have shown that improved barrier properties can be obtained with the 2D materials concentrated near the surface [9]. In addition, processing methods such as fibre spinning can be used to align the 2D nanofillers in a particular direction that may again produce materials with both enhanced mechanical properties and anisotropic electrical and thermal conductivities.

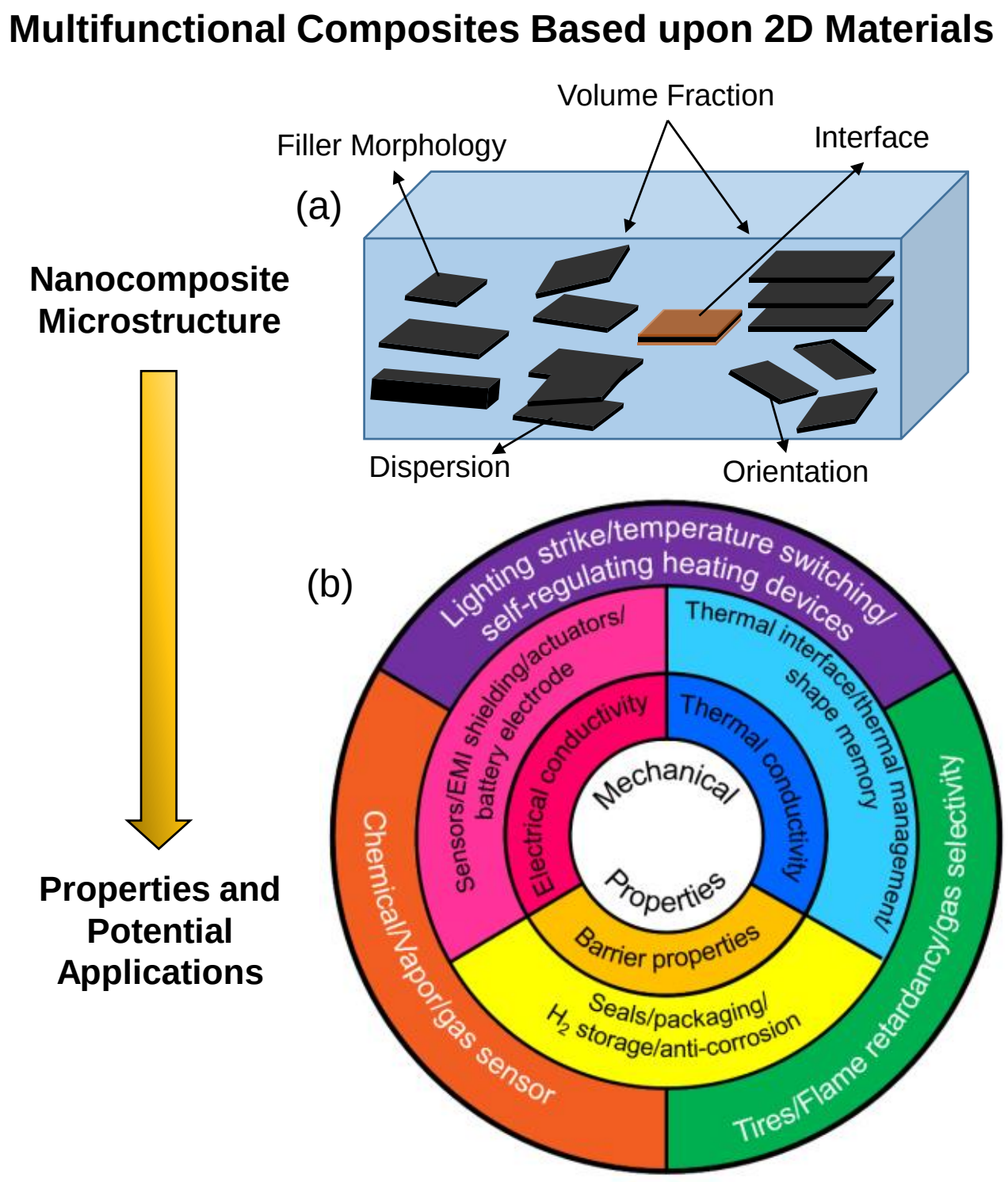

**Figure 2.** Development of multifunctional nanocomposites based upon the addition of 2D materials. (a) Microstructural features that control properties. (b) Potential multifunctional applications of 2D nanocomposites based upon a combination of enhanced physical properties.

Most of the original studies upon using 2D materials in nanocomposites have employed the nanofillers in their native states without any special surface treatments. As it is now well established that mechanical properties depend strongly upon stress transfer across the filler-matrix interface, there is scope for more systematic studies being undertaken of methods of functionalising interfaces in 2D materials to optimise properties. In particular, the filler-matrix interface strength must be carefully tuned to concurrently enhance both the elastic and fracture properties. One of the most exciting prospects for nanocomposites based upon 2D materials is being able to exploit their multifunctional properties [10] as shown schematically in Figure 2(b). There is further scope to control multifunctionality through the use of hybrid systems with two or more types of 2D materials or the inclusion of other geometries of nanofiller.

**Concluding Remarks**

It is clear that there have been spectacular developments in the field of 2D materials over the past 20 years. Such materials have been shown to have a whole spectrum of impressive intrinsic physical properties that could be exploited in nanocomposites. As the mechanisms whereby the properties of the composite matrix are enhanced are becoming better understood, we are also learning the limits for property modification. Nevertheless, there is plenty of scope for us to use our imagination to develop new ways in which 2D materials can be employed in nanocomposites, particularly for exciting multifunctional applications.

**Acknowledgements**

NK acknowledges funding support from the USA National Science Foundation (Award 2015750). The authors are also grateful to Dr Mufeng Liu for help with the preparation of the Figures.

## 23. Commercialization and standardization

**Bill Qu[1], Max Lemme[2] and Andrew J. Pollard[3]**

[1] The Sixth Element Inc.
[2] RWTH Aachen University & AMO GmbH, Germany
[3] National Physical Laboratory (NPL), UK

### Status

Possessing outstanding properties, graphene related 2D materials (GR2Ms) and 2D materials (2DMs) more generally have demonstrated their potential in completely different application areas. There are now hundreds of companies reported to sell 'graphene' worldwide, with many companies now able to offer many tons of material per annum or even large-scale deposition machines. This has been achieved after several years of the industry increasing the scale of production, but to make an economic impact with a material of such potential, enabling the commercialization process is key. Although new products containing 2DMs are now on the market, more effort is still required to achieve significant commercialization (see Figure 1). Nevertheless, the graphene industry has greatly progressed over the last 15 years, from a few companies trying to convince mainstream end users of a wonder material that would revolutionize their sector, to many companies working within different international supply chains for real-world products.

Currently the word 'graphene' is informally used within the community as a generic name for a broad range of carbon-based 2D-materials differing in structure, morphology and/or chemical composition, more accurately referred to as 'graphene-related 2D materials' or GR2Ms (as defined in ISO/TS 80004-13). There are also many cases of materials being sold commercially that are not actually 'graphene' or GR2Ms, instead being nano-graphite [1]. Therefore, setting accurate international standards for the different forms of GR2Ms is crucial for efficient communication across the supply chain, from producers to customers. Both terminology and measurement standards are being developed or have been published through joint working groups of two international standardization bodies [2], namely the International Organization of Standardization (ISO) and the International Electrotechnical Commission (IEC), within their corresponding nanotechnology technical committees, ISO/TC229 and IEC/TC113. Through a holistic approach, different GR2Ms can be measured and defined using a set of complementary international standards, as described in Figure 2, that are already published or due to be published in 2025. Furthermore, these standards are based on verified processes using international interlaboratory studies [3] within the Versailles Project on Advanced Materials and Standards (VAMAS), established by the G7 in 1982 to promote world trade by innovation and adoption of advanced materials. There are also other standards published or underdevelopment within IEC/TC113 WG8, that are focused on more specific measurements taken from academic literature, for example IEC/TS 62607-6-6: "Graphene – Strain uniformity: Raman spectroscopy". However, there are absences of reference materials or specific standards within the technological areas being targeted by graphene-producing or graphene-enabling companies.

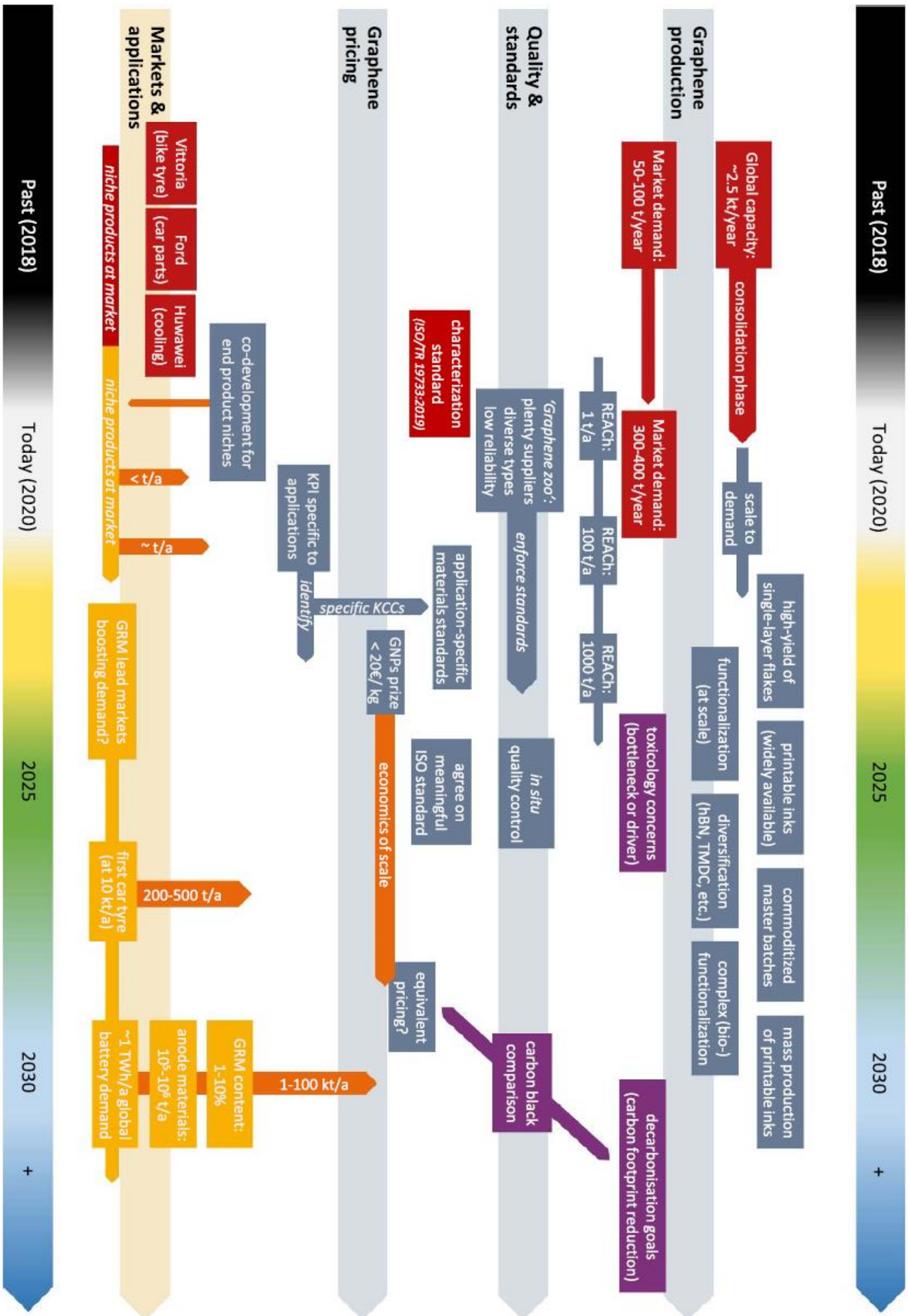

**Figure 1.** Graphene and related materials Roadmap 2020, from ref. [4]

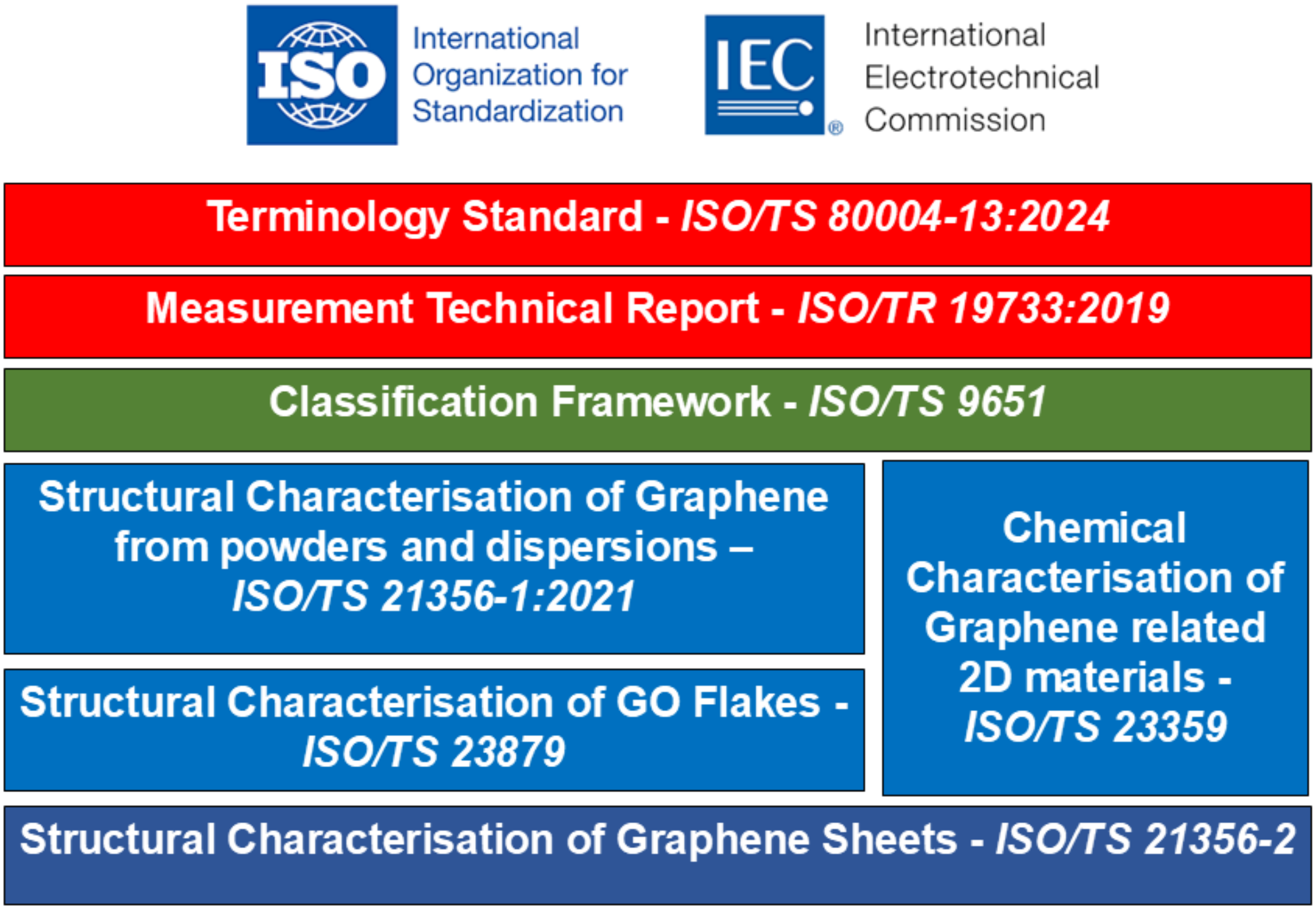

**Figure 2.** Current landscape of joint ISO and IEC 2DM standards published or under development and their associated designations (including publication date where applicable), showing both published (red) and unpublished (green) enabling standards, as well as measurement standards for materials consisting of flakes or particles in a powder or liquid dispersion form (light blue) or produced as sheets (dark blue), for example through chemical vapor deposition (CVD).

### Current and Future Challenges

The commercialization of such a versatile material begins with applications that are generally less technically challenging or belong to more mature areas, typically as an additive. As illustrated in Figure 1, some of the first commercial application of graphene nanoplatelets are thermally-conductive films, functional coatings, composite materials, and conductive additives in lithium-ion batteries. These areas are useful examples in understanding the commercialization process and the challenges of moving new materials from the laboratory and into real-world products.

In thermally-conductive films, graphene nanoplatelets (GNPs) are assembled into films for use in microchip cooling and in 2018, Huawei launched a high-performance cellphone, Mate 20X, claiming the world's first application equipped with such technology. Since then, many other cellphone manufacturers have adopted these films, for example, Xiaomi, Oppo, and OnePlus. Over 300 million devices equipped with these films have now been sold. The early adoption of graphene in this case is due to this industry's short supply chains, as well as the rapid feedback process from end users to materials producers.

Unlike thermal conductive films where GNPs are used as the primary material, most other applications involve GNPs as functional additives. In functional coatings, the GNPs are used to increase the electrical conductivity and/or protection against electrochemical corrosion. Functional coatings also have a relatively short supply chain to reach their final customers, and the industry has many tools, dispersing agents and experience in handling nanoparticles. However, the qualification time required is much longer than in consumer electronics, hindering progress. For example, salt spray tests take thousands of hours to assess corrosion, with the NORSOK M-501 standard requiring a 4200-hour cyclic corrosion test. This slows down the feedback cycle for product improvements and therefore the commercialization process. Another issue is that these coatings are used in large construction projects, such as bridges or mega-structures and there is therefore significant regulation that must be adhered to in these areas. Without well accepted standards and design guidelines, companies hesitate to use innovative materials, unless their use has been thoroughly demonstrated in such mega-structures. This can become a “chicken and egg” problem, as development of a standard that can be the basis of regulation requires users and application cases to support its necessity.

Adding GNPs into polymers to improve their electrical and/or thermal conductivity, and mechanical properties, has been a focus of the graphene community for many years. An early commercial application using these nanocomposites is polyamide (PA6) fiber, used for textile materials requiring an increased wear resistance, modified through the incorporation of GNPs dispersed into the organic compound caprolactam. GNP-modified PA6 does not have long testing periods nor standards as hurdles for its commercialization, however, it does involve many different manufacturing steps before reaching end users; 1) dispersing GNPs into caprolactam, 2) the use of a PA6 polymerization plant to manufacture PA6 chips, 3) a spinning company to then produce the fiber, 4) stock dyeing, 5) weaving the fiber into fabric and 6) tailoring products such as apparel or bed linens. This makes it relatively difficult for commercialization, affecting the time to market. However, after more than 10 years of development, MLILY Inc. has launched products with GNP- modified PA6 fiber. The products can be purchased in mainstream outlets in over 73 countries and regions around the world.

For a final example, the use of GR2Ms in energy storage has been of great interest from the very beginning of the commercialization of graphene. Many studies have shown that the conductivity and chemical stability of graphene brings advantages in various aspects of lithium-ion batteries, typically using GNPs as a conductive additive. The supply chain in this application is very short. The GNPs are dispersed into NMP (N-Methyl pyrrolidone), a solvent commonly used in both the production/supply of GNPs and in lithium-ion batteries electrode coating. Then GNPs are added into a slurry (mostly $LiFePo4$) for cathode production. Thus, very rapidly lithium-ion battery manufacturers such as BYD and Gotion in China have one or more product lines that benefit from GNP additives, in a similar way to how carbon nanotubes have been used in this application.

**Advances in Science and Technology to Meet Challenges**

There are several aspects of standardisation that still need to be addressed in the future, on top of the publication of the current ISO/IEC standards underdevelopment (see Figure 2). This will require the completion of VAMAS studies, as well as the continued revision of the standards already published, to continually improve and update standards in this fast-paced industry. Another consideration is the measurement of other inorganic 2D materials, through the inclusion of 2D materials as a whole into the current measurement standards, as well as the consideration of new measurement standards (for example, measuring the chemical properties).

Meanwhile, new international or sector-specific standards may be required for new products containing 2D materials, as they come to markets where no standards are already in place. At the same time, standards will typically be required only for the final product and not be specific for any involved 2D materials. For example, there are already standards for assessment of the mechanical properties of composites, as this is a mature area, and so nanocomposites containing GR2Ms, will not require a new standard.

The use of GNPs as an additive in concrete for construction is a good example of standardization needs. Many benefits and substantial improvements such as improved mechanical properties have been demonstrated by companies, which would lead to less concrete used and therefore a significant benefit to net zero carbon emission targets. This material is already in use in the real world with hundreds of tons of product poured to date. However, there is understandably significant regulation in place in construction due to safety. This can hinder this type of concrete being used in load-bearing applications and ensure there is no removal of the steel reinforcement, even if the mechanical properties still meet expectations. The removal of steel would simplify the construction process as well as further reduce associated carbon emissions. As regulation is typically based on standards, which are in turn based on data and verification, more work in this area is required if the commercialization of this product is going to be able to meet the regulation requirements.

Focussing on the beginning of the supply chain to achieve the reproducible production of material by a company, well-defined procedures for mass production are necessary, whilst also considering the application-specific requirements. The standardization of product definitions between different manufacturers are also helpful to avoid confusion for downstream users and are being investigated within ISO and IEC through a Classification Framework (Figure 2). Not knowing product characteristics have led to many inconsistent and even contradictory results in graphene applications. Finally, quality control processes will be required that are more rapid and cost effective than the currently standardized processes. But these quality control methods must be verified, through using the more accurate and precise quality control standards.

**Concluding Remarks**

A coherent, hierarchical structure of international standards is required for industry, allowing the comprehensive methodology of controlling materials properties as well as the performance of the final products. Strong communication between academia and industry will be key in order to achieve more scientific understanding of the roles of 2DMs in particular products, moving away from 'graphene' just being used as a gimmick in consumer products. Forming an ecosystem involving the scientific community, business community, government and the general public, will lead to the significant contribution of these 2D materials to the development of human civilization, as many other materials have in the past, creating a step change in technology.

**Acknowledgements**

AJP would like to acknowledge the National Measurement System (NMS) of the Department for Science, Innovation and Technology (DSIT), UK (projects #127931 and #128807, Enabling innovation to meet UK Net Zero targets through Physicochemical Metrology of Nanoscale Advanced Materials) for funding.